\documentclass{cernyrep}
\usepackage[T1]{fontenc}
\usepackage[scale=.78]{geometry}

\usepackage{xspace}
\usepackage{graphicx}
\usepackage{epsfig}
\usepackage{float}
\usepackage[nointegrals]{wasysym}
\usepackage{hyperref}
\usepackage{url}
\usepackage{subfig}
\usepackage{amsmath}
\usepackage[dvipsnames,usenames]{xcolor}
\usepackage{amssymb,array}
\usepackage{multirow}

\hypersetup{
  linkcolor  = black, 
  citecolor  = blue,
  urlcolor   = blue,
  colorlinks = true,
  breaklinks=true
}

\usepackage{placeins}
\usepackage{subfig}
\usepackage{adjustbox}
\usepackage{blindtext}
\usepackage{verbatim}
\usepackage{bm}
\usepackage[normalem]{ulem}
\usepackage[T1]{fontenc}

\newcommand{\OO}{\ensuremath{\mathcal{O}}}

\newcommand{\pdp}{\ensuremath{\phi^\dagger\phi}}
\renewcommand{\phi}{\ensuremath{\varphi}}
\newcommand{\sss}{\scriptscriptstyle}

\newcommand{\Op}[1]{\OO_{\sss #1}}

\def\lra#1{\overset{\text{\scriptsize$\leftrightarrow$}}{#1}}

\makeatletter\g@addto@macro\bfseries{\boldmath}\makeatother

\def\equationautorefname~#1\null{Eq.\,(#1)\null}
\newcommand{\fullref}[2]{\hyperref[#2]{#1\,\ref*{#2}}}
\makeatletter

\newcommand{\rcite}[1]{\hyper@@link[cite]{}{cite.#1}{Ref.\,\cite*{#1}}}
\makeatother
\newcommand*{\eqsref}[2]{\hyperref[#1]{Eqs.\,(\ref*{#1}--}\hyperref[#2]{\ref*{#2})}}

\allowdisplaybreaks[4]

\makeatletter
\newcommand{\parenbar}{\mathpalette\p@renb@r}
\def\p@renb@r#1#2{%
	\vbox{%
		\ifx#1\scriptscriptstyle\dimen@.7em\dimen@ii.20em\else%
		\ifx#1\scriptstyle	\dimen@.8em\dimen@ii.25em\else%
					\dimen@.9em\dimen@ii.35em\fi\fi%
		\offinterlineskip%
		\ialign{%
			\hfill##\hfill\cr
			\vbox{\hrule width\dimen@ii height .35pt}\cr
			\noalign{\vskip-.35ex}%
			\hbox to\dimen@{$\mathchar300\hfil\mathchar301$}\cr
			\noalign{\vskip-.35ex}%
			$#1#2$\cr%
		}%
	}%
}%
\makeatother

\newcommand*{\tmp}[4]{\ensuremath{%
	{#4%
	\ifx\empty#3\empty\ifx\empty#1\empty\else^{#1}\fi\else^{#1(#3)}\fi%
	\ifx\empty#2\empty\else_{#2}\fi}%
}}

\newcommand*{\ccc}[4][]{\tmp{#2}{#3}{#4}{#1{c}}}

\newcommand*{\qq }[4][]{\tmp{#2}{#3}{#4}{#1{\mathcal{O}}}}

\newcommand{\rien}[1]{}

\newcommand{\be}{\begin{equation}}
\newcommand{\ee}{\end{equation}}
\newcommand{\bea}{\begin{eqnarray}}
\newcommand{\eea}{\end{eqnarray}}
\newcommand{\bi}{\begin{itemize}}
\newcommand{\ei}{\end{itemize}}
\newcommand{\ben}{\begin{enumerate}}
\newcommand{\een}{\end{enumerate}}

\newcommand{\lc}{\left[}
\newcommand{\rc}{\right]}
\newcommand{\lp}{\left(}
\newcommand{\rp}{\right)}

\def\frac#1#2{{{#1}\over {#2}}}

\def\lra#1{\overset{\text{\scriptsize$\leftrightarrow$}}{#1}}

\usepackage{booktabs}
\usepackage{pifont}
\usepackage{cleveref}

\newcommand*\rot{\rotatebox{90}}

\begin{document}

\preprint{CERN-LHCEFTWG-2022-001\\ CERN-LPCC-2022-05}

\title{LHC EFT WG Report: Experimental Measurements and Observables}

\author{Nuno Castro\aff{LIP},
Kyle Cranmer\aff{Wis},
Andrei V. Gritsan\aff{JH},
James Howarth\aff{SUPA},
Giacomo Magni\twoaff{NIK}{VU},
Ken Mimasu\aff{King},
Juan Rojo\twoaff{NIK}{VU},
Jeffrey Roskes\aff{JH},
Eleni Vryonidou\aff{Man},
Tevong You\threeaff{TH-CERN}{AMTP}{Cav}
}

\institute{
\naff{LIP}{LIP, Departamento de F\'{i}sica, Escola de Ci\^{e}ncias, Universidade do Minho, 4710-057 Braga, Portugal}
\naff{Wis}{Department of Physics, University of Wisconsin, Madison, WI 53706, USA}
\naff{JH}{Department of Physics and  Astronomy, Johns Hopkins University, Baltimore, MD 21218, USA}
\naff{SUPA}{SUPA - School of Physics and Astronomy, University of Glasgow, Glasgow, UK}
\naff{NIK}{Nikhef, Science Park 105, 1098 XG Amsterdam, The Netherlands}
\naff{VU}{Department of Physics and Astronomy, Vrije Universiteit, NL-1081 HV Amsterdam, The Netherlands}
\naff{King}{Department of Physics, King's College London, Strand, London WC2R 2LS, UK}
\naff{Man}{Department of Physics and Astronomy, University of Manchester, Oxford Road, Manchester M13 9PL, UK}
\naff{TH-CERN}{Theoretical Physics Department, CERN, CH-1211 Geneva 23, Switzerland}
\naff{AMTP}{AMTP, University of Cambridge, Wilberforce Road, Cambridge CB3 0WA, UK}
\naff{Cav}{Cavendish Laboratory, University of Cambridge, J.J. Thomson Avenue, Cambridge CB3 0HE, UK}
}

\date{November 15, 2022}

\begin{abstract}
The LHC effective field theory working group gathers members of the LHC experiments and the theory community 
to provide a framework for the interpretation of LHC data in the context of EFT. In this note we discuss experimental 
observables and corresponding measurements in analysis of the Higgs, top, and electroweak data at the LHC. 
We review the relationship between operators and measurements relevant for the interpretation 
of experimental data in the context of a global SMEFT analysis.
One of the goals of ongoing effort is bridging the gap between theory and experimental communities 
working on EFT, and in particular concerning optimised analyses.
This note serves as a guide to experimental measurements and observables leading to EFT fits 
and establishes good practice, but does not present authoritative guidelines how those measurements 
should be performed. 
\end{abstract}

\maketitle

\setcounter{tocdepth}{3}
\tableofcontents

\clearpage


\section{Introduction}
\label{sect:area3_intro}

The LHC effective field theory working group (LHC EFT WG)~\cite{lhceftwg} gathers members of the LHC experiments 
and the theory community to provide a framework for the interpretation of LHC data in the context of 
effective field theories (EFTs). The LHC EFT WG studies the physics requirements needed to facilitate 
an interpretation commensurate with the available measurements performed in a wide range of different 
processes with Higgs bosons, top quarks, and electroweak bosons. It provides recommendations 
for the use of EFT by the experiments to interpret their data, and a forum for theoretical discussions of 
EFT issues. This includes recommendations on the theory setup as well as Monte Carlo simulation and 
other tools needed for EFT analyses. It focuses on recommendations, developments, 
and combinations that require coordination between the existing WGs (Higgs, Top, Electroweak), in order 
to allow global EFT analyses inside and outside experimental collaborations. 
The following six areas of activity have been identified:
\begin{enumerate}
\item EFT Formalism;
\item Predictions and Tools;
\item Experimental Measurements and Observables;
\item Fits and Related Systematics;
\item Benchmark Scenarios from UV Models;
\item Flavour.
\end{enumerate}

This note is focussed on the topics from Area 3 devoted to {\it experimental measurements and observables}.
This activity area covers how observables relate to operators, which measurements are important 
for a given operator or set of operators, differential/fiducial measurements vs. dedicated ones, identification
of optimal observables, machine learning, re-interpretation vs. static, presentation of results: covariance, 
multi-D likelihood, etc., compatibility with global fits (i.e. assumptions used in deriving measurement and 
reporting results).

The goal is to study observable, channel, process sensitivities and complementarities:
\begin{itemize}
\item Experimental targets: survey of the sensitive channels and corresponding operators;
\item Differential distributions, optimal observables, including machine learning, and dedicated EFT measurements, 
         spin density matrices, EFT-optimized fiducial regions, amplitude analyses, angular distributions (e.g. for $CP$), pseudo observables...
\item Agreement across experiments (for fiducial regions in particular);
\item What observables are most sensitive to new physics? Exploit energy growing effects, non-interferences, and other theoretical knowledge;
\item Expected uncertainties: systematics or statistics dominated.
\end{itemize}

The goal is also to study analysis strategies and experimental outputs, also with a view at legacy measurements and their possible reinterpretation:
\begin{itemize}
\item Dedicated EFT extractions by collaborations;
\item Differential measurements and the best choice of observables for re-interpretation;
\item Presentation of measurements: cross sections, correlations/covariance, multi-D likelihood...
\item Experimental systematics related to EFT (e.g. accounting for detector effects);
\item Detector effects: unfolding, forward folding, efficiency maps, recasting through re-weighting...
\item EFT in backgrounds: final-state driven instead of signal-background, statistical model.
\end{itemize}


This note serves as a guide to experimental measurements leading to EFT fits, but does not establish authoritative 
guidelines how those measurements should be performed. There is a spectrum of experimental approaches. 
Each approach has its own stronger and weaker sides, and none of the approaches has been established as the
universally best approach to perform the measurements. Therefore, one of the goals of this note is to survey these
approaches and identify their key features. 

In order to discuss experimental approaches, we will use the following notation.
We will denote a {\it channel} to be a process used to perform a measurement, for example production
of a top-antitop pair in association with the Higgs boson $t\bar{t}H$ in pp collisions at the LHC. 
We will denote an {\it observable} to be an experimentally defined quantity in such a process, 
for example transverse momentum of the Higgs boson $p_T^H$. 
We will denote a {\it measurement} to be an experimentally delivered quantitative result, 
for example differential cross section in bins of $p_T^H$ in the $t\bar{t}H$ process.  
Experimental {\it measurements} using the {\it observables} sensitive to EFT effects in a given set of 
{\it channels} at the LHC become the input to EFT fits which provide constraints on the EFT operator coefficients.

\section{Experimental observables and corresponding measurements}
\label{sect:area3_observables}
A typical particle physics analysis requires construction of a likelihood function ${\cal L}$ 
(or other means of inference), which describes a sum of contributing {\it processes} defined for
a set of {\it observables} $\vec{x}_\mathrm{reco}$ reconstructed in an experiment 
as a function of parameters of interest~$\vec{\theta}$. 
In application to an EFT analysis, for example, $\vec{\theta}$ could represent the EFT parameters, 
which describe the truth-level process, characterized by the quantities $\vec{x}_\mathrm{truth}$, 
which could be the four-momenta of all partons involved in the process (incoming and outgoing particles
in the hard process). The challenges in experimental analysis are related to the interplay of the
reconstructed $\vec{x}_\mathrm{reco}$ and truth-level $\vec{x}_\mathrm{truth}$ quantities,
where the parton shower, detector response, and reconstruction algorithms stand in between. 

The probability density function (pdf) for a reconstructed process can be obtained with the transfer function
from the truth-level process pdf as  
\begin{equation}
{\cal P}(\vec{x}_\mathrm{reco}|\vec{\theta}) = 
\int \mathrm{d}\vec{x}_\mathrm{truth} ~p(\vec{x}_\mathrm{reco}|\vec{x}_\mathrm{truth}) {\cal P}(\vec{x}_\mathrm{truth}|\vec{\theta}) \,,
\label{eq:P} 
\end{equation}
where $p(\vec{x}_\mathrm{reco}|\vec{x}_\mathrm{truth})$ is the transfer function. 
If ${\cal P}(\vec{x}_\mathrm{truth}|\vec{\theta})$ represents the matrix element squared of the hard process, 
including the parton luminosities, propagators, and phase-space factors, 
then the transfer function reflects the parton shower, detector response, and reconstruction algorithms. 
In this case, the \emph{truth level} corresponds to the \emph{parton level}. 
Alternatively, ${\cal P}(\vec{x}_\mathrm{truth}|\vec{\theta})$ could represent the \emph{particle-level} process 
including the parton shower, which needs to be clarified based on the use case. 
When treated as the probability density, ${\cal P}$ is normalized to unit area. 
The assumption that $p(\vec{x}_\mathrm{reco}|\vec{x}_\mathrm{truth})$ does not depend on EFT parameters
$\vec{\theta}$ is often valid, but this assumes, for example, that QCD effects in parton shower factorize. 
Let us also define $p(\vec{x}_\mathrm{reco},\vec{x}_\mathrm{truth}|\vec{\theta})=
p(\vec{x}_\mathrm{reco}|\vec{x}_\mathrm{truth}) {\cal P}(\vec{x}_\mathrm{truth}|\vec{\theta})$
for future use in Section~\ref{sect:ml}. 

More generally, the \emph{truth level} can refer 
to any set of observables that are defined based on information in the Monte Carlo (MC) event record, however, two definitions 
dominate most results; the \emph{particle level} and \emph{parton level}. At particle level, observables are 
constructed using objects that closely resemble those at reco level and are built using only particles in the event record, usually 
those with a mean lifetime $\tau \geq 3\cdot 10^{-9}$ s. For example, a particle level jet could be defined using the same 
anti-k$_T$ algorithm as is used on reco level, with stable particles as input instead of calorimeter objects. Particle-level objects 
and observables are agnostic to the choice of MC generator. The parton level is a more nebulous concept and seeks 
to mimic the results of a fixed-order calculation using undecayed objects such as quarks and bosons. However, as each MC 
generator calculates and stores the four vectors of such objects differently, it is challenging to establish a generator-agnostic 
definition. In the area of top quark physics, for example, the parton-level definition of a top quark is commonly taken to be the 
last top quark in the decay chain which decays to a $b$ quark and $W$ boson.

The distinction of the \emph{particle level} and \emph{parton level} will become important when we discuss unfolding 
techniques in Section~\ref{sect:unfold}, which are designed to invert the transformation in Eq.~(\ref{eq:P}). 
It is not always possible to reverse the relationship in Eq.~(\ref{eq:P})
and recover ${\cal P}(\vec{x}_\mathrm{truth}|\vec{\theta})$ from the observed reconstructed distributions,
due to loss of information in the transfer. This loss could happen either because of loss in the detector, 
e.g. particles lost along the beampipe, or because $\vec{x}_\mathrm{reco}$ does not match the full 
set~$\vec{x}_\mathrm{truth}$. 
However, if the truth-level observables are closely related to those at reco-level,
the truth-level distributions might be recovered with good accuracy.

Modeling of the transfer encoded in Eq.~(\ref{eq:P}) is often performed with MC techniques.
The hard process ${\cal P}(\vec{x}_\mathrm{truth}|\vec{\theta})$ is usually modeled at parton level
with an event generator, such as POWHEG~\cite{Frixione:2007vw}, MadGraph\_aMC$@$NLO~\cite{Alwall:2014hca}, 
or another program (which are discussed in Area 2 of the group effort); 
parton shower with e.g. Pythia~\cite{Sjostrand:2014zea}; detector response with e.g. GEANT~\cite{GEANT4:2002zbu}; 
and the reconstruction software specific to an experimental approach is used for calculation of $\vec{x}_\mathrm{reco}$. 
Data-driven techniques can be employed either to calibrate or to substitute for MC, but the general idea 
remains the same: some approximation is used to predict ${\cal P}(\vec{x}_\mathrm{reco}|\vec{\theta})$. 
This predicted distribution should describe the observed distribution in experiment 
${\cal P}_\mathrm{obs}(\vec{x}_\mathrm{reco})$, which, in general, 
is represented by an unbinned distribution of observed events. 
Based on distributions of observables, a quantitative {\it measurement} is derived, which could be 
confidence intervals set on cross sections, parameters $\vec{\theta}$, or any other quantities, 
which could later be used as input to a global~fit. 

One particular application in EFT is a set of parameters $\vec{\theta}$ that consists of $(K+1)$ couplings, 
including the standard model (SM) $\theta_0$ and $\vec{\theta}=(\theta_0,\theta_1,..,\theta_K)$.
Let us consider the models $\vec{\theta}_i$, each corresponding to a non-zero coupling $\theta_i$.
For simplicity in the equations and discussion, we assume $\theta_i$ are real, but
complex couplings do not significantly change the situation.
The full probability density is proportional to the sum of the amplitudes squared and can be written as
\begin{eqnarray}
{\cal P}(\vec{x}_\mathrm{reco}|\vec{\theta})  \propto
{\cal P}_0(\vec{x}_\mathrm{reco}) 
~~~~~~~~~~~~~~~~~~~~~~~~~~~~~~~~~~~~~~~~~~~~~~~~~~~~~~~
~~~~~~~~~~~~~~~~~~~~~~~~~~~~~~~~~~~~~~~~~~~~~~~~~~~~~~~
\nonumber  \\
+  \sum_{1\le k \le K} \left(\frac{2\theta_k}{\theta_0}\right) {\cal P}_{0k}(\vec{x}_\mathrm{reco}) 
+ \sum_{1\le k \le K} \left(\frac{\theta_k}{\theta_0}\right)^2 {\cal P}_k(\vec{x}_\mathrm{reco}) 
+ \sum_{1\le i<j\le K} \left(\frac{2\theta_i\theta_j}{\theta_{0}^2}\right) {\cal P}_{ij}(\vec{x}_\mathrm{reco}) \,, 
\label{eq:probreco} 
\end{eqnarray}
where the single index of the probability density indicates the model type $\vec{\theta}_k$ and
double index indicates interference between the two models $\vec{\theta}_i$ and $\vec{\theta}_j$,
and where we do not carry the overall normalization factor, which corresponds to the total cross section 
measurement of a process. 
Equation~(\ref{eq:probreco}) allows parameterization of the pdf as a function of parameters 
of interest $\vec{\theta}$, and we will come back to this equation below.

\subsection{Approaches to experimental observables}
\label{sect:obs}

An ideal analysis would reconstruct the full set of observables $\vec{x}_\mathrm{reco}^\mathrm{\,full}=\vec{x}_\mathrm{truth}$,
which completely describe the process of interest in Eq.~(\ref{eq:P}), without any loss of information. 
For example, this may be the complete set of four-vectors describing the incoming and outgoing partons,
or an equivalent set of observables which contains the same information. In reality, this often becomes either 
impossible or impractical, and at the very least some information is lost or obstructed due to detector effects 
or parton shower. 
Such an approach is often considered impractical because it requires a highly multi-dimensional space of
observables which would be hard to analyze. Therefore, $\vec{x}_\mathrm{reco}$ often becomes a
reduced set of observables, often just one or two, characterizing an event. This could
be illustrated with the following transformation, where $\vec{x}_\mathrm{reco}^\mathrm{\,full}$ represents
the most information about the process that can be reconstructed from an event
\begin{equation}
\vec{x}_\mathrm{reco} = f(\vec{x}_\mathrm{reco}^\mathrm{\,full})  \,.
\label{eq:xreco} 
\end{equation}
In the case of such reduction of information, it is important to choose observables in such a way that their
differential distributions are sensitive to variation of the model parameters $\vec{\theta}$.

We loosely group observables sensitive to EFT effects in the following categories, without necessarily 
clear boundaries between those:
\begin{itemize}
	\item typical SM observables, such as invariant masses, transverse momenta $p_T$ of reconstructed particles, etc...
	\item EFT-sensitive observables: angular information, four-momenta squared $q^2$ of the propagators, etc...
	\item optimized observables: matrix-element-based calculations, machine learning, special construction, etc...
	\item full information $\vec{x}_\mathrm{reco}^\mathrm{\,full}$ in any of the above forms. 
\end{itemize}
The distinction between the above is usually historical. The SM observables were often chosen for 
the best background rejection and isolation of signal without necessarily targeting EFT effects. 
The EFT-sensitive observables are typically individual calculations which emphasize certain features
of higher-dimension operators in a simple and transparent manner. The optimized observables
usually do not seek transparency but are targeting the maximal information possible. Observables
which carry full information are meant to be used by advanced methods for statistical inference. 
We note that even if the shapes of certain observables are not sensitive to EFT effects, 
the rates measured using such observables still provide some sensitivity. 

The choice of observables in experimental analysis depends on many factors. Ideally, there should be a minimal number
of observables sensitive to the maximal information contained in a certain process, with minimal correlation 
between these observables. This leads to the idea of optimized observables targeted to given operators. 
Such an approach provides better sensitivity. At the same time, generic observables may have wider application 
and usage, without being tuned to a particular set of operators. Reproducibility of observables is another important 
consideration. For example, while an observable obtained with machine-learning techniques may appear most
optimal, the differential distribution of such an observable may become hard to interpret. 
Historically, many EFT measurements evolved from analyses which did not target EFT operators specifically and are 
therefore based on generic observables.

\subsubsection{Observables sensitive to EFT effects}
\label{sect:obs-eft}

One particular consideration in building EFT-sensitive observables is the fact that higher-dimension operators 
typically lead to enhancement at the higher values of four-momenta squared $q^2$ distributions of the particles 
appearing in the propagators. 
Therefore, observables based on the $q^2$ calculations or correlated with those quantities become sensitive probes 
of deviations from the SM. An example of such an observable correlated with $q^2$ could be the transverse momentum of 
reconstructed objects. At the same time, such generic probes of $q^2$ may not be sensitive to distinguish multiple 
operators that all lead to the same $q^2$ enhancement. One example of such a situation is the study of 
$CP$-even and $CP$-odd operators, which may require special $CP$-sensitive observables to differentiate them. 

\begin{figure}[t!]
  \begin{center}
    \captionsetup{justification=centerlast}
    \includegraphics[width=0.32\linewidth]{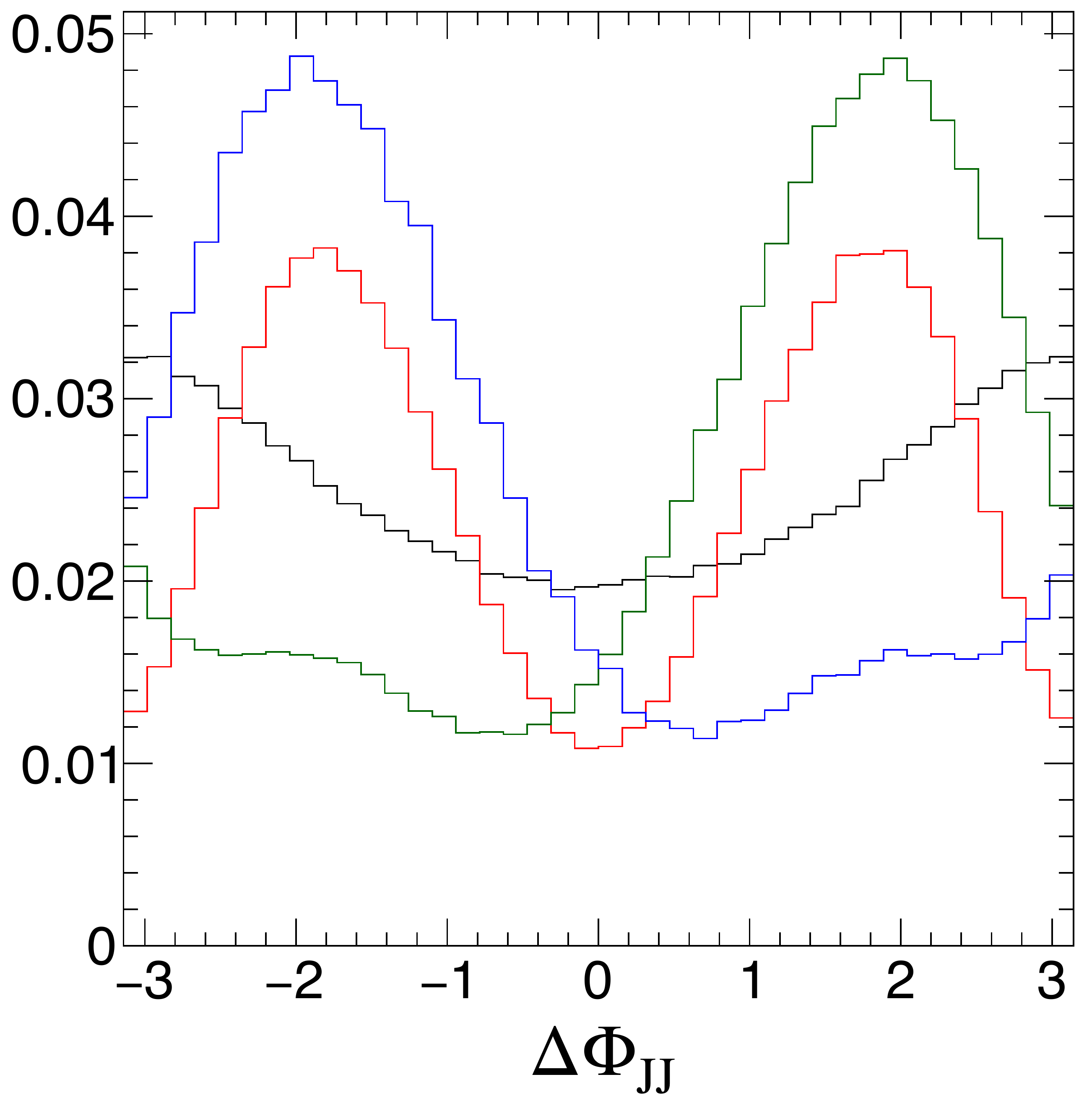}
    \includegraphics[width=0.32\linewidth]{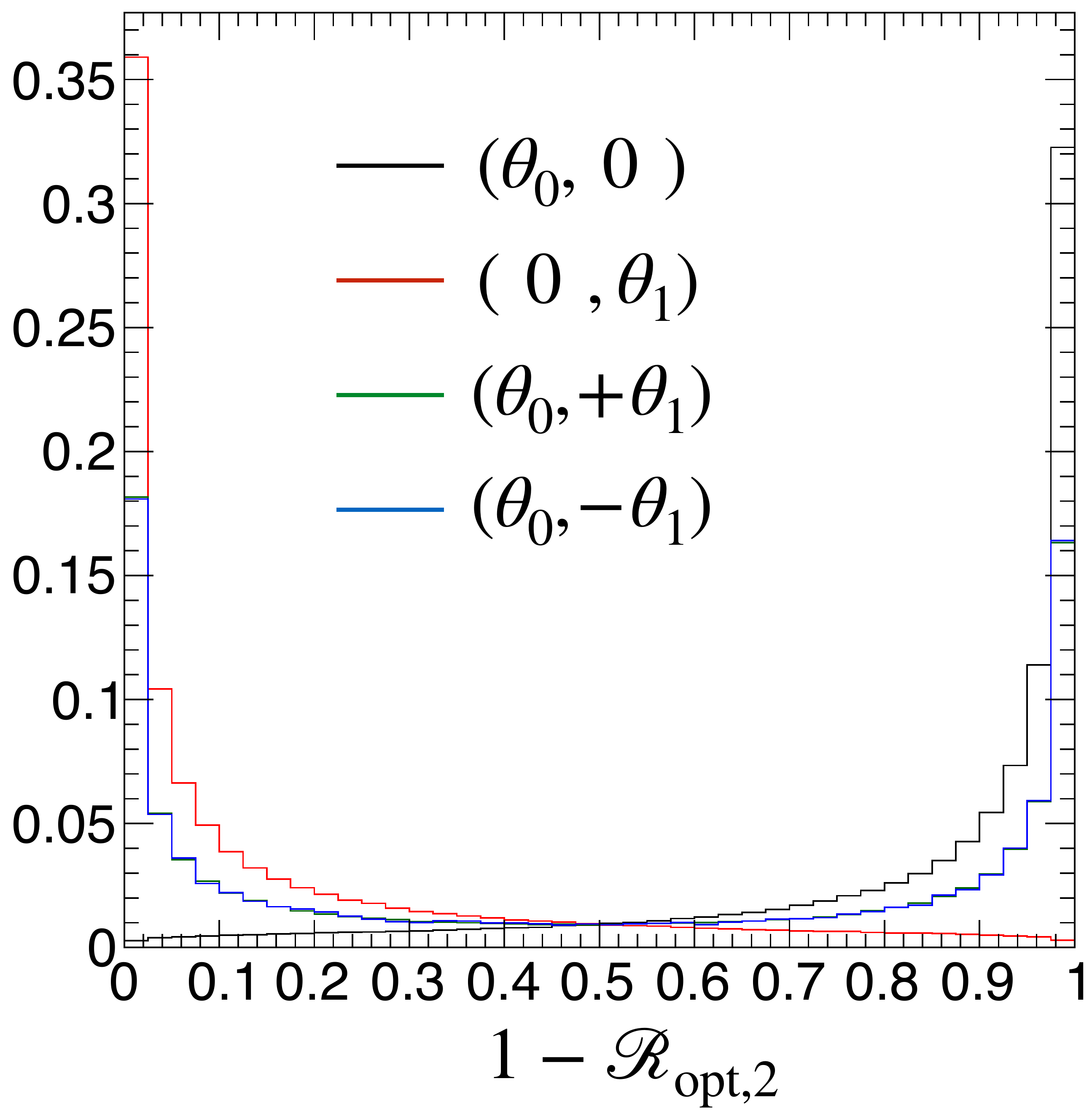}
    \includegraphics[width=0.32\linewidth]{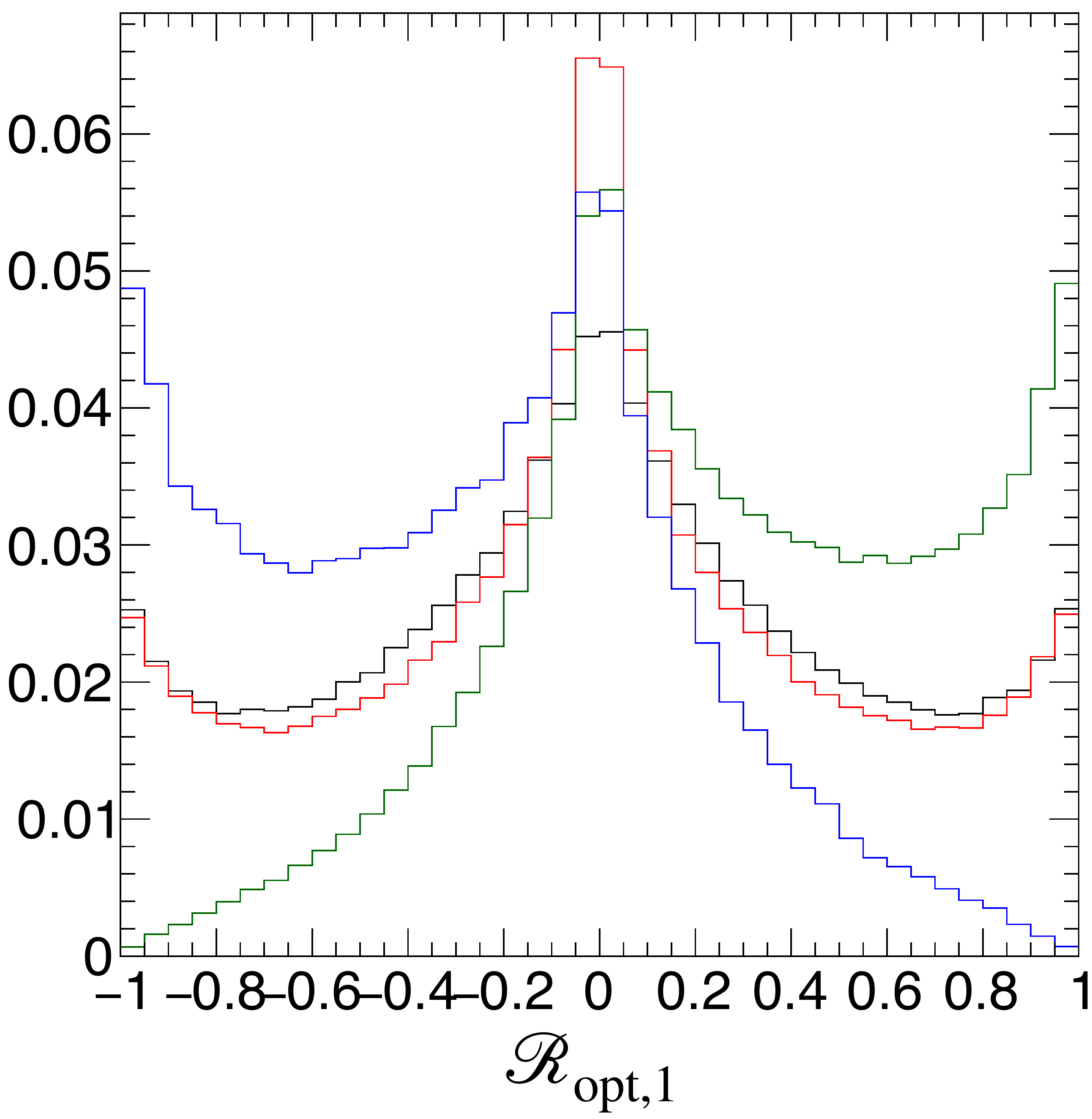}
    \caption{
    Observables reconstructed in production of the on-shell Higgs boson in the $ZZ$ and $WW$ fusion:
    azimuthal angle between two associated jets $\Delta\Phi_\mathrm{JJ}$ (left);
    observable ${\cal R}_\mathrm{opt,2}$ optimized for the quadratic term $\theta_1^2$ (middle);
    observable ${\cal R}_\mathrm{opt,1}$ optimized for the linear term $\theta_1$ (right). 
    Four distributions are shown: SM ($\theta_0$, black), $CP$-odd operator ($\theta_1$, red), 
    and 50\% mixture with positive (green) and negative (blue) relative sign of $\theta_0$ and $\theta_1$. 
    The study is inspired by Ref.~\cite{Gritsan:2020pib}. 
     }
    \label{fig:obs}
  \end{center}
\end{figure}

Building EFT-sensitive observables usually follows certain features of the higher-dimension operators. 
As an example of such a construction, let us consider the 
vector boson scattering process with creation of an on-shell Higgs boson. The operators
$C_{H\widetilde{W}B}$, $C_{H\widetilde{W}}$, and $C_{H\widetilde{B}}$ give rise to the $CP$-odd
amplitudes in the $HVV$ interaction. The azimuthal angle between the two associated jets 
$\Delta\Phi_\mathrm{JJ}$ is an observable sensitive to these amplitudes~\cite{Plehn:2001nj}, 
as illustrated in Fig.~\ref{fig:obs}.

\subsubsection{Observables optimized with matrix-element calculations}
\label{sect:obs-me}

The idea of optimized observables was introduced in particles physics in Refs.~\cite{Atwood:1991ka,Diehl:1993br}. 
For a simple discrimination of two hypotheses, the Neyman-Pearson lemma~\cite{Neyman:1933wgr} 
guarantees that the ratio of probabilities 
${\cal P}_1(\vec{x}_\mathrm{reco}^\mathrm{\,full})/{\cal P}_0(\vec{x}_\mathrm{reco}^\mathrm{\,full})$ for the two hypotheses 
0 and 1 provides an optimal discrimination power.
For a continuous set of hypotheses with an arbitrary quantum-mechanical mixture of two states, one could apply the 
Neyman-Pearson lemma to each pair of points in the parameter space, but this would require a continuous, and 
therefore infinite, set of probability ratios. 
However, equivalent information is contained in a combination of only two probability ratios, 
which can be chosen as two optimized observables~\cite{Anderson:2013afp}
\begin{eqnarray}
{\cal R}_\mathrm{opt,2} =
\frac{{\cal P}_1(\vec{x}_\mathrm{reco}^\mathrm{\,full})}
{{\cal P}_0(\vec{x}_\mathrm{reco}^\mathrm{\,full})+c\cdot{\cal P}_1(\vec{x}_\mathrm{reco}^\mathrm{\,full})} \,,
\label{eq:optimized2} 
\\
{\cal R}_\mathrm{opt,1} =
\frac{2{\cal P}_{01}(\vec{x}_\mathrm{reco}^\mathrm{\,full})}
{{\cal P}_0(\vec{x}_\mathrm{reco}^\mathrm{\,full})+c\cdot{\cal P}_1(\vec{x}_\mathrm{reco}^\mathrm{\,full})} \,,
\label{eq:optimized1} 
\end{eqnarray}
where the index of the probability density ${\cal P}$ is discussed in application to Eq.~(\ref{eq:probreco}). 
The constant $c$ is introduced for convenience and could be set to $c=1$ when symmetric appearance 
is desired, and these observables can also be defined with $c=0$.
The information content of the two observables used jointly is the same for any fixed value of $c$,
and the observable optimized for discrimination between two arbitrary hypotheses 
can be written as a function of ${\cal R}_\mathrm{opt,1}$ and ${\cal R}_\mathrm{opt,2}$.
In the case of a small contribution of $\theta_1$ to differential cross section, Eq.~(\ref{eq:optimized1})
with $c=0$ reproduces the optimal observable defined in Ref.~\cite{Diehl:1993br}.

Equation~(\ref{eq:probreco}) allows us to make several important observations. 
When interference terms ${\cal P}_{ij}$ are absent, as for example in the case of non-interfering background,
the optimized observables are of the type defined in Eq.~(\ref{eq:optimized2}).
There is one optimized observable to separate signal from background, assuming background 
does not depend on EFT parameters. 
When interference is present and there are only two types of couplings with $K=1$, there are only
two optimized observables defined in Eqs.~(\ref{eq:optimized2}) and~(\ref{eq:optimized1}),
as stated above. This illustrates the power of the multivariate techniques when the 
full information contained in the high-dimensional space of 
$\vec{x}_\mathrm{reco}^\mathrm{\,full}$ can be preserved in just two observables. 
This power is limited to the measurement of one parameter $\theta_1/\theta_0$, though. 
However, when multiple couplings are present with $K>1$, the number of optimized observables 
grows significantly as $(K+2)!/(2K!)-1$. 
At the same time, one can observe that for the purpose of an EFT 
measurement, the last two terms in Eq.~(\ref{eq:probreco}) are quadratic in non-SM couplings and
can be neglected. This leaves $K$ linear terms, 
describing interference with the SM amplitude, in the coupling expansion. 
The $K$ observables, one for each term of the type defined in Eq.~(\ref{eq:optimized1}) with $c=0$,
can be picked for optimal separation in an EFT analysis.
Alternatively, the $2K$ observables, one for each term of each of the types defined in 
Eqs.~(\ref{eq:optimized2}) and~(\ref{eq:optimized1}) with any fixed value of $c$, 
will define the complete set of optimized observables. 

Calculation of optimized observables as probability ratios in Eqs.~(\ref{eq:optimized2}) and~(\ref{eq:optimized1}) can be 
performed with the help of the matrix-element calculations in Eq.~(\ref{eq:P}). The transfer functions 
introduced in Eq.~(\ref{eq:P}), $p(\vec{x}_\mathrm{reco}|\vec{x}_\mathrm{truth})$, are required 
for proper probability calculations. However, as opposed to the matrix element method
discussed in Section~\ref{sect:mem} where the transfer functions have to be modeled fully,
in the case of probability ratios certain effects cancel, and in any case, any imperfection does not bias
the result, but only reduces optimality somewhat. Therefore, the transfer functions 
can often be omitted, with only minor effect on optimality, in cases where the process 
is fully reconstructed with good enough resolution. 

While multiple operators usually contribute to each process, it is often possible to isolate a small 
number $K$ of operators that are preferably constrained in a given process. Other operators may 
be much better constrained elsewhere. It is also important to pick the correct basis, or rotation, 
of the operators, in which blind or weak directions are removed. When such a situation is possible, 
the matrix-element calculations provide a prescription for calculating the observables optimal for 
EFT measurements. Therefore, before a measurement can be attempted, one must obtain 
a map of processes and operators with the weak directions resolved.

\begin{figure}[t!]
  \begin{center}
    \captionsetup{justification=centerlast}
    \includegraphics[width=0.32\linewidth]{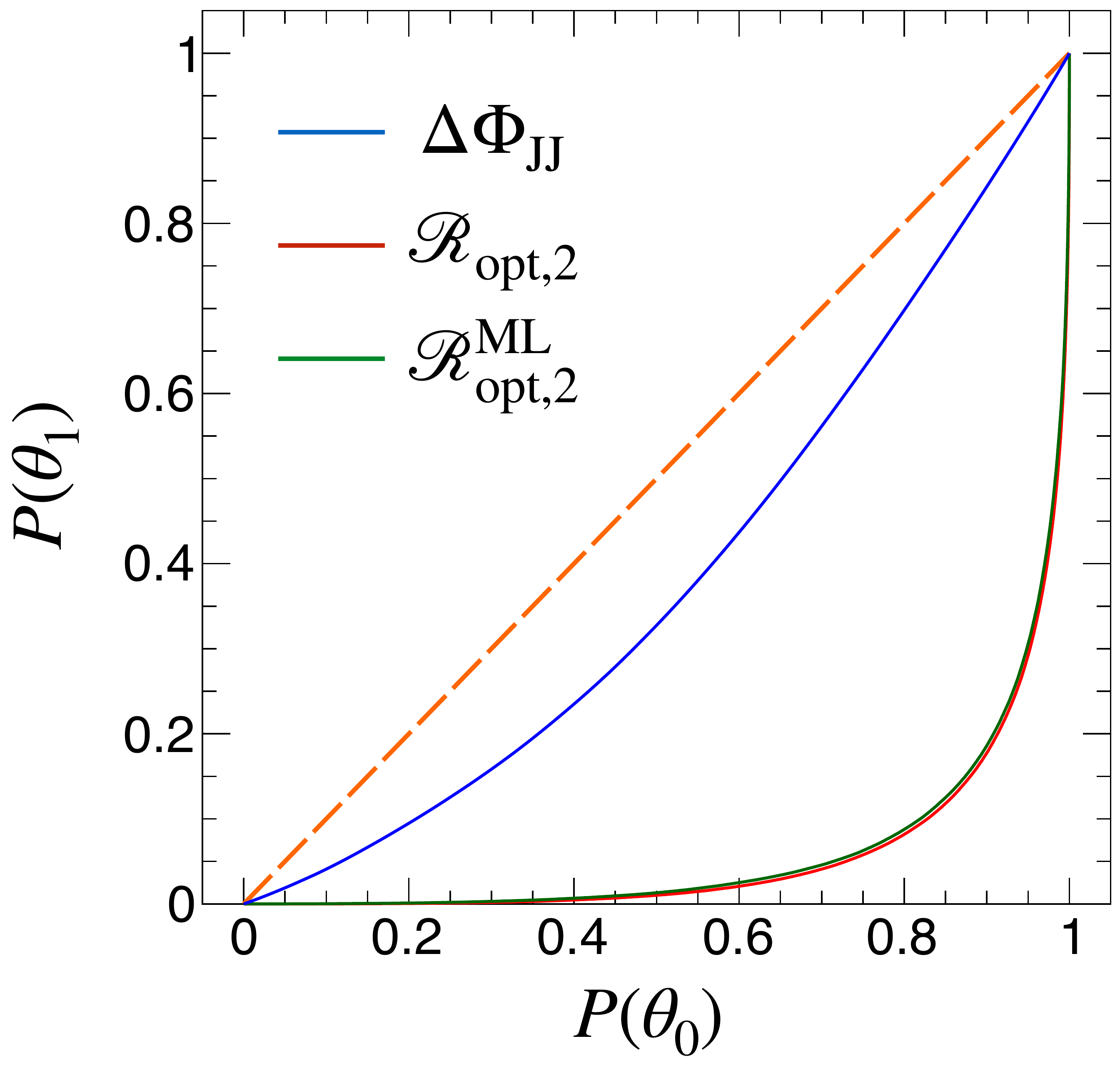}~~~~
    \includegraphics[width=0.32\linewidth]{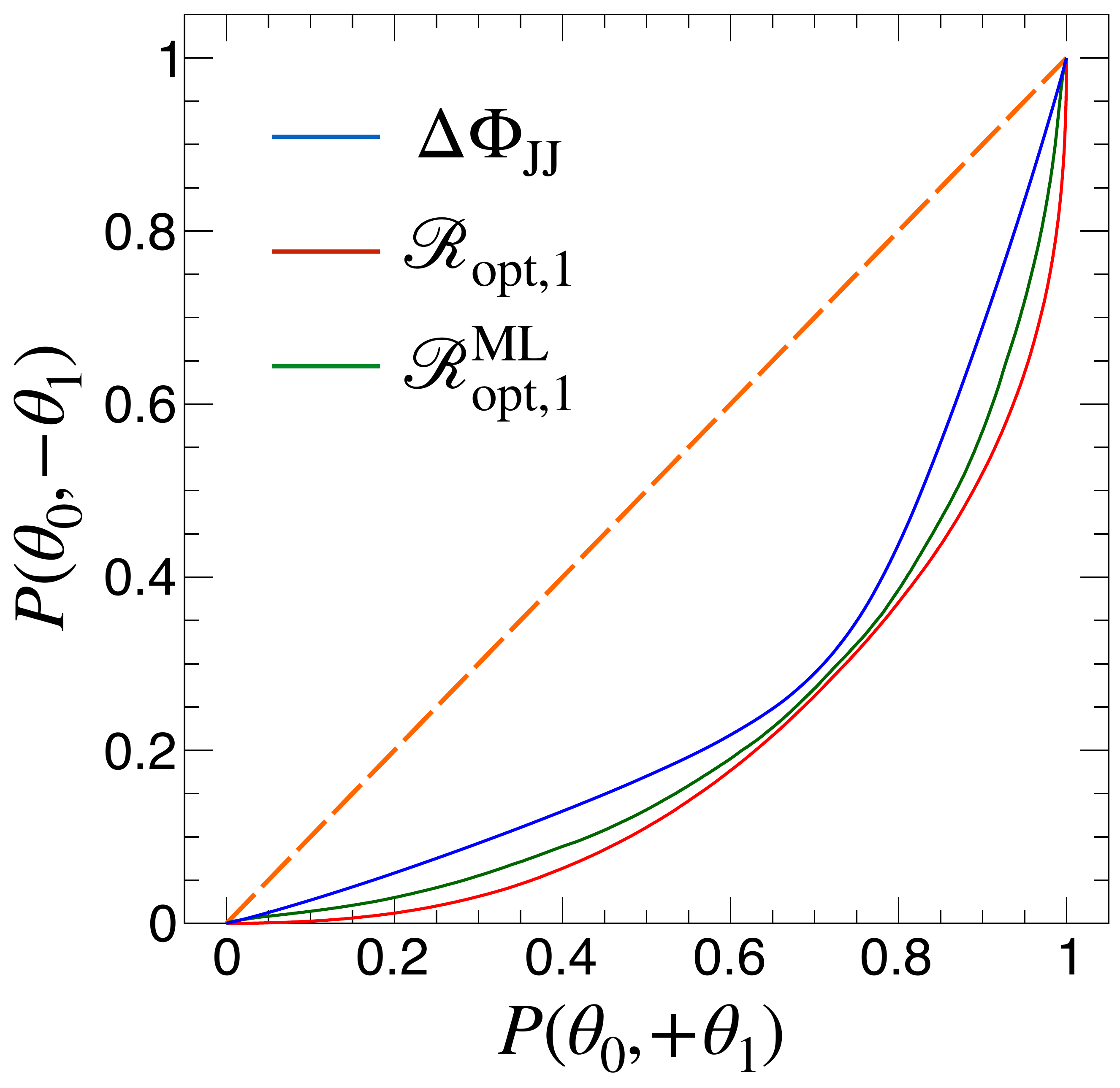}
    \caption{
    The ROC curves showing the separation power between the processes driven by a $CP$-odd coupling $\theta_1$
    and the SM $\theta_0$ (left) and between the 50\% mixture of the two processes with the positive and negative
    sign of interference (right). Three observables in the on-shell Higgs boson production in vector boson fusion 
     are tested: $\Delta\Phi_\mathrm{JJ}$ (blue), optimized with matrix element (red) and machine learning (green). 
    The study is inspired by Ref.~\cite{Gritsan:2020pib}. 
     }
    \label{fig:roc}
  \end{center}
\end{figure}

Examples of the  ${\cal R}_\mathrm{opt,2}$ and ${\cal R}_\mathrm{opt,1}$ observables for the Higgs 
boson production in the $ZZ$ and $WW$ fusion can be found in Fig.~\ref{fig:obs}~\cite{Gritsan:2020pib},
and compared to the EFT-sensitive observable in this process $\Delta\Phi_\mathrm{JJ}$.
Among the three operators, $C_{H\widetilde{W}B}$, $C_{H\widetilde{W}}$, and $C_{H\widetilde{B}}$,
there are two weak directions, corresponding to the $H\gamma\gamma$ and $HZ\gamma$ couplings, 
which are eliminated in the mass eigenstate basis. 
The separation power can be quantified with Receiver Operating Characteristic (ROC)
curves, shown in Fig.~\ref{fig:roc}.
Performance of the ${\cal R}_\mathrm{opt,2}$ observable is validated with two samples
corresponding to the $\vec\theta=(\theta_0,0)$ and $(0,\theta_1)$ models. 
For a given selection ${\cal R}_\mathrm{opt,2}>$ threshold, a point with probabilities to
select events from either one model $P(\theta_0)$ or the other $P(\theta_1)$ samples the ROC curve. 
In a similar manner, 
performance of the ${\cal R}_\mathrm{opt,1}$ observable is validated with two samples
corresponding to the $\vec\theta=(\theta_0,+\theta_1)$ and $(\theta_0,-\theta_1)$ models. 
The two models differ only by the sign of interference, and therefore the interference component
is emphasized in this test. 
Observables optimized with matrix element calculation exhibit the best performance
in isolating the operator of interest. 

\subsubsection{Observables optimized with machine learning}
\label{sect:obs-ml}

While the matrix-element calculations guarantee optimal performance from first principles, there are practical 
limitations to their applications. The most critical limitations are the transfer functions, which are difficult and
time-consuming to model. Parton shower and detector effects may confuse and distort the input to matrix elements
to such a degree that calculations become impractical. Machine-learning techniques may come to the rescue in 
such a case. Training of machine-learning algorithms is still based on MC samples utilizing the same
matrix elements that would be used for optimal discriminants discussed in Section~\ref{sect:obs-me}.
However, these MC samples reflect the parton shower and detector effects, 
and therefore allow construction of optimized observables which incorporate these effects. 

There are two important aspects in this training: which observables should enter the learning process and 
which samples should be used. 
Matrix-element approach gives answers to both and provides insight into the process of constructing the 
optimized observables with machine learning. 
First, the input observables should provide full information $\vec{x}_\mathrm{reco}^\mathrm{\,full}$,
which could be simply the four-vectors of all particles involved, like in the matrix elements, or better 
derived physics quantities which are equivalent to those. 
Second, there are two types of optimal observables, as pointed in Eqs.~(\ref{eq:optimized2}) 
and~(\ref{eq:optimized1}). 
The first observable corresponds to the classic problem of differentiating between two models, and 
a machine-learning algorithm is trained on two samples corresponding to models $\theta_0$ 
and $\theta_1$. The resulting observable is equivalent to Eq.~(\ref{eq:optimized2}).
Training of the second discriminant is less obvious, as it is expected to isolate the interference component. 
A discriminant trained to differentiate the two models with maximal quantum-mechanical mixing 
$(\theta_0,+\theta_1)$ and $(\theta_0,-\theta_1)$ becomes a machine-learning equivalent 
to that in Eq.~(\ref{eq:optimized1})~\cite{Gritsan:2020pib}. 

Performance of the two optimized observables ${\cal R}_\mathrm{opt,2}^\mathrm{ML}$ and 
${\cal R}_\mathrm{opt,1}^\mathrm{ML}$ obtained with machine learning (Boosted Decision Tree in this case) 
is illustrated in Fig.~\ref{fig:roc} for the same example discussed in Section~\ref{sect:obs-me}.
Since only acceptance effects were introduced in MC simulation for illustration purpose, 
and those cancel in the ratios in Eqs.~(\ref{eq:optimized2}) and~(\ref{eq:optimized1}),
the matrix-element and machine-learning optimized observables exhibit the same performance. 
There is a small degradation in performance of the ${\cal R}_\mathrm{opt,1}^\mathrm{ML}$,
which is attributed to the more challenging task of training in this case and should be recovered 
in the limit of perfect training.

This Section provides a straightforward path to observables optimized with machine learning.
A more elaborate approach to inference with machine learning is discussed in Section~\ref{sect:ml}.

\subsection{Approaches to experimental measurements}
\label{sect:meas}

The ideal situation  in Eq.~(\ref{eq:P}) would be 
$p(\vec{x}_\mathrm{reco}|\vec{x}_\mathrm{truth}) = \delta(\vec{x}_\mathrm{reco}-\vec{x}_\mathrm{truth})$,
which means that the truth-level quantities could be reconstructed in experiment without any loss
of information and $\vec{x}_\mathrm{reco}^\mathrm{\,full}=\vec{x}_\mathrm{truth}$.
This never happens in practice, and experimental techniques are needed to analyze the data. 
In experiment, the ${\cal P}_\mathrm{obs}(\vec{x}_\mathrm{reco})$ distribution is observed, 
and it is matched to the predicted distribution in order to obtain the confidence intervals of $\vec{\theta}$.
There are two conceptually different approaches taken in particle physics 
to use Eq.~(\ref{eq:P}) for the measurements:
\begin{itemize}
\item In the single-step approach, the observed ${\cal P}_\mathrm{obs}(\vec{x}_\mathrm{reco})$
distribution is matched directly to the reconstruction-level prediction ${\cal P}(\vec{x}_\mathrm{reco}|\vec{\theta})$
to obtain constraints on $\vec{\theta}$. 
\item In the two-step approach, first the ${\cal P}_\mathrm{obs}(\vec{x}_\mathrm{reco})$ distribution is unfolded and 
the best approximation to the truth-level ${\cal P}_\mathrm{obs}(\vec{x}^{\,\prime}_\mathrm{truth})$ distribution is reported,
where $\vec{x}^{\,\prime}_\mathrm{truth}$ is a reduced set of quantities describing the truth-level process. 
In the second step, the observed unfolded ${\cal P}_\mathrm{obs}(\vec{x}^{\,\prime}_\mathrm{truth})$ distribution 
is matched to the truth-level prediction ${\cal P}(\vec{x}^{\,\prime}_\mathrm{truth}|\vec{\theta})$ 
to obtain constraints on $\vec{\theta}$.
\end{itemize}
In the above, $\vec{x}^{\,\prime}_\mathrm{truth}$ is a subset of 
$\vec{x}_\mathrm{truth}=(\vec{x}^{\,\prime}_\mathrm{truth},\vec{x}^{\,\prime\prime}_\mathrm{truth})$ in Eq.~(\ref{eq:P}) which 
matches the $\vec{x}_\mathrm{reco}$ as close as possible.  It does not have to be complete and 
may represent just one quantity, e.g. the invariant mass of a system.
The parton-level prediction for a reduced set of quantities can be expressed as
\begin{equation}
{\cal P}(\vec{x}^{\,\prime}_\mathrm{truth}|\vec{\theta}) = 
\int \mathrm{d}\vec{x}^{\,\prime\prime}_\mathrm{truth} {\cal P}(\vec{x}_\mathrm{truth}|\vec{\theta}) \,,
\label{eq:partialP} 
\end{equation}
where integration is performed over the degrees of freedom $\vec{x}^{\,\prime\prime}_\mathrm{truth}$ not used in differential distributions. 

The two approaches have their pros and cons and both have been widely utilized in particle physics. 
The single-step approach is conceptually straightforward and unbiased, if modeling 
of the effects in Eq.~(\ref{eq:P}) is done correctly, e.g. with MC simulation. 
It may also be the most optimal approach, if the choice of observables $\vec{x}_\mathrm{reco}$ is optimal
and contains the full information as close to $\vec{x}_\mathrm{truth}$ as possible. 
However, this approach suffers from two main drawbacks, which are complementary 
to the two-step approach discussed below. 
First of all, re-analysis of the data requires the full chain of experimental processing,
starting from a modification of ${\cal P}(\vec{x}_\mathrm{truth}|\vec{\theta})$ in Eq.~(\ref{eq:P})
and MC simulation of experimental apparatus with the new model, e.g. after changing the list
of EFT operators targeted in analysis as described by $\vec{\theta}$. For this reason, 
theoretical collaborations cannot easily reuse the results of such an analysis to adopt
a new model. Second, this approach is rather complex because the EFT modeling should be done 
with the full detector simulation, and leads to a much more involved process than the analysis of 
unfolded distributions, which requires EFT modeling at parton level only. 

The two-step approach is attractive for two main reasons: First of all, it allows wide dissemination and 
preservation of the data in the form of differential truth-level distribution 
${\cal P}_\mathrm{obs}(\vec{x}^{\,\prime}_\mathrm{truth})$
for later re-analysis, e.g. by theoretical collaborations. Second, this approach 
decouples the experimental effects in the first step from analysis of EFT effects in the second step,
which greatly simplifies the EFT part of analysis. These two features make this approach especially 
attractive for analysis of differential distributions outside of the experimental collaborations. 
At the same time, this approach suffers from two main drawbacks, which potentially may lead to 
a non-optimal and/or biased analysis. 
The choice of $\vec{x}^{\,\prime}_\mathrm{truth}$ is usually limited to one or few observables, 
resulting in significant loss of information compared to the complete set $\vec{x}_\mathrm{truth}$. 
The unfolding from ${\cal P}_\mathrm{obs}(\vec{x}_\mathrm{reco})$ to  
${\cal P}_\mathrm{obs}(\vec{x}^{\,\prime}_\mathrm{truth})$ depends on $\vec{\theta}$,
but these parameters are not known in advance. The SM assumption is usually made in such unfolding,
which may lead to bias. With multiple processes, background subtraction for a given process is also usually 
done assuming SM parameters in the other processes, which may also lead to bias. 

While measurements in the single-step approach can enter the global EFT fit directly by building 
the joint likelihood of multiple measurements within experimental collaborations, reporting these
measurements outside of such collaborations may represent a challenge. The usual practice of 
reporting multi-gaussian approximation may not be sufficient for proper combination with other 
measurements. The proper treatment requires that the experiments release the full likelihood associated to their 
measurement, which represents technical challenges, but possible in principle. 
Alternatively, the intermediate reco-level parameterization of a single-step approach could be
reported, which would effectively lead to a two-step approach, but full information would be retained.
In such a case, the observed 
${\cal P}_\mathrm{obs}({x}^\prime_\mathrm{reco})$ distribution is reported by an experiment for a given observable 
${x}^\prime_\mathrm{reco}$, along with the expected distributions ${\cal P}_\mathrm{reco}({x}^\prime_\mathrm{reco}|\theta_i)$ 
for a variety of models of interest~$\theta_i$, and including background processes. This approach shares pros and 
cons of the above two main approaches. This approach eliminates the unfolding procedure with its own complications 
and allows easy interpretation of public differential distributions. However, this approach still limits the re-interpretation 
to only those models $\theta_i$ which have been pre-computed with the tools used by the experiments.
This approach also limits available information, as typically only 1D differential distribution of an observable 
${x}^\prime_\mathrm{reco}$ is reported. However, the latter limitation could be overcome if several optimized 
observables are utilized simultaneously. 

The dependence of the unfolding procedure on the EFT parameters can be illustrated with the following. 
Eliminating $\vec{x}^{\,\prime\prime}$ from Eq.~(\ref{eq:P}) leads to an equivalent expression
\begin{equation}
{\cal P}(\vec{x}_\mathrm{reco}|\vec{\theta}) = 
\int \mathrm{d}\vec{x}^{\,\prime}_\mathrm{truth} ~p^\prime(\vec{x}_\mathrm{reco}|\vec{x}^{\,\prime}_\mathrm{truth};\vec{\theta}\,) 
{\cal P}(\vec{x}^{\,\prime}_\mathrm{truth}|\vec{\theta}) \,,
\label{eq:unfoldingP} 
\end{equation}
with a very important difference in comparison to Eq.~(\ref{eq:P}): the modified transfer function \\
$p^\prime(\vec{x}_\mathrm{reco}|\vec{x}^{\,\prime}_\mathrm{truth};\vec{\theta}\,)$ must, in general, 
depend on the model parameters $\vec{\theta}$. This dependence becomes pronounced if 
detector effects $p(\vec{x}_\mathrm{reco}|\vec{x}_\mathrm{truth})$  are not uniform and distributions 
${\cal P}(\vec{x}_\mathrm{truth}|\vec{\theta})$ vary with $\vec{\theta}$ over the degrees of 
freedom~$\vec{x}^{\,\prime\prime}_\mathrm{truth}$. 
A curious reader can confirm this conclusion by deriving the expression for 
$p^\prime(\vec{x}_\mathrm{reco}|\vec{x}^{\,\prime}_\mathrm{truth};\vec{\theta}\,)$, 
which is a non-trivial function of $\vec{\theta}$ in the case of non-trivial detector effects. 

The reverse transformation of Eq.~(\ref{eq:unfoldingP}) 
becomes the unfolding procedure, which is discussed in Section~\ref{sect:unfold}. 
The dependence of unfolding on the model parameters $\vec{\theta}$ creates challenges in the EFT interpretation. 
This is usually sidestepped by assuming SM parameters $\vec{\theta}=\vec{\theta}_0$ in the transformation
\begin{equation}
{\cal P}_\mathrm{obs}(\vec{x}_\mathrm{reco}) \simeq
\int \mathrm{d}\vec{x}^{\,\prime}_\mathrm{truth} ~p^\prime(\vec{x}_\mathrm{reco}|\vec{x}^{\,\prime}_\mathrm{truth};\vec{\theta}_0\,) 
{\cal P}_\mathrm{obs}(\vec{x}^{\,\prime}_\mathrm{truth}) \,,
\label{eq:unfoldingPobs} 
\end{equation}
where the level of approximation in Eq.~(\ref{eq:unfoldingPobs}) with the assumption $\vec{\theta}=\vec{\theta}_0$
needs to be reported for the range of parameters~$\vec{\theta}$ under consideration. 
In practice, modeling of Eq.~(\ref{eq:unfoldingPobs}) is performed with MC simulation of the SM processes,
and the level of approximation needs to be tested with alternative simulations, including detector effects, 
for a range of $\vec{\theta}$. Ideally, experimental collaborations should also report prescription for correcting 
the bias for certain popular models with parameters $\vec{\theta}$.
These corrections are particularly important to evaluate for modification of EFT parameters because they can
lead to dramatic detector effects, e.g. very different response due to substitution of $Z$ and $\gamma^*$ 
particles in the propagators.

The choice between the single-step vs. the two-step approaches to experimental measurements
is driven by the tradeoff between the various pros and cons and by the available resources, data, and tools. 
The single-step approach is the most optimal and can use the full knowledge of detector information, 
and therefore it is most suitable for analysis by the experimental collaborations. 
The two-step approach is easier for re-interpretation, even if this comes at the cost of some 
loss of information and potential bias, and therefore it is most suitable for analysis by the 
theoretical collaborations without access to detector information. 

In the end, experimental measurements are experimentally delivered quantitative results, which are typically cross 
sections or related quantities, and can be further used in global fits for EFT parameters $\vec{\theta}$. 
(Applications of experimental measurements to global EFT fits are discussed in Area 4 of the group effort.)
These experimental measurements are obtained from analysis of observables $\vec{x}_\mathrm{reco}$ 
in a limited set of processes. 
We group experimental measurements sensitive to EFT effects in several categories, which progress from
simple to more involved:
\begin{itemize}
\item single-process cross section;
\item single observable differential distribution affected by a single or multiple processes;
\item multi-observable differential distribution or multiple single-observable differential distributions with correlations;
\item binned sub-process cross sections, such as STXS (Section~\ref{sect:stxs}) in Higgs boson physics;
\item dedicated EFT measurements, such as amplitude analysis with cross sections per EFT operator;
\item dedicated EFT operator extraction by experiments.
\end{itemize}		
The first four types of measurements can be generically called differential and fall under the
two-step approach. The last two types of measurements can be generically called dedicated 
and fall under the single-step approach. 
All of the above types of measurements can be either analyzed stand-alone or enter combination
with other measurements in the global fits.

\subsection{Statistical methods for experimental measurements}
\label{sect:stat}

We discuss further ingredients of experimental measurements, such as unfolding techniques
for calculation of differential distributions ${\cal P}_\mathrm{obs}(\vec{x}^{\,\prime}_\mathrm{truth})$, 
and likelihood-based and likelihood-free inference techniques.

\subsubsection{Unfolding}
\label{sect:unfold}

Unfolding or \emph{signal deconvolution} is a broad term that refers to techniques that seek to translate data from one form, 
which contains noise or distortions, into another, which does not. In the context of High Energy Physics, this usually refers 
to correcting experimental data for detector acceptance, efficiency, and resolution effects. The reconstructed data (colloquially 
referred to as the \emph{reco level}) is corrected for the effects of the detector to the \emph{truth level}. 
This is equivalent to inverting the transformation in Eq.~(\ref{eq:unfoldingPobs}) 
to convert the observed distribution of events ${\cal P}_\mathrm{obs}(\vec{x}_\mathrm{reco})$
into the distribution at truth level ${\cal P}_\mathrm{obs}(\vec{x}^{\,\prime}_\mathrm{truth})$.
As discussed at the beginning of Section~\ref{sect:area3_observables}, the \emph{truth level} can refer 
to any set of observables that are defined based on information in the MC event record, 
the \emph{particle level} or \emph{parton level}. 

The results that are unfolded are almost always binned data presented in histograms. 
Therefore, unfolding is, at its essence, a matrix inversion problem. 
All unfolding techniques rely on construction a matrix that maps truth events in bin $i$ to reco 
events in bin $j$ which is usually called a \emph{response matrix} or a \emph{migration matrix}. 
A general representation of unfolding is given by:
\begin{equation}
\label{eq:unfolding}
\frac{\mbox{d}\sigma}{\mbox{d}X_{i}} = \frac{1}{\mathcal{L}\cdot\mathcal{B}\cdot\Delta X_i \cdot \epsilon_{i}^{\mbox{\scriptsize{ext}}}} \cdot \sum_{j} R_{ij}^{-1} \cdot \epsilon^{\mbox{ \scriptsize{fid}}}_{j} \cdot (N_j^{\mbox{\scriptsize{obs}}} - N_j^{\mbox{\scriptsize{bkg}}})\mbox{ ,} 
\end{equation}
where $X$ is the observable to be unfolded, $\mathcal{L}$ is the integrated luminosity of the data, $\mathcal{B}$ 
is the branching ratio of the channel being studied, $\Delta X_i$ is the width of the bin in observable $X$, $N_j^{\mbox{\scriptsize{obs}}}$
 is the number of observed events in data in bin $j$, $N_j^{\mbox{\scriptsize{bkg}}}$ is the number of expected background events 
 in bin $j$. $R_{ij}$ is the response matrix, $\epsilon_{i}^{\mbox{\scriptsize{ext}}}$ is a bin-by-bin correction factor that accounts 
 for the limited phase-space of the event selection (sometimes called an \emph{extrapolation} correction, and 
 $\epsilon^{\mbox{ \scriptsize{fid}}}_{j}$ is a bin-by-bin correction factor that accounts for reco events that are 
 not present in the truth selection (sometimes called \emph{non-fiducial} backgrounds). Most unfolding techniques 
 work by inverting $R_{ij}$ and applying that inverted matrix to the observed data to remove detector effects. 
 Matrix inversion is a well-studied problem and, in the context of unfolding, the main challenge is to control 
 harmonic fluctuations in the statistical uncertainties that are inherent to the process of inverting the response matrix. 
 The key difference between different unfolding approaches are usually how these statistical uncertainties are controlled 
 (a process often called \emph{regularisation}). For example, in Iterative Bayesian Unfolding \cite{DAgostini:1994fjx}, 
 these fluctuations are regularised using an iterative technique based on Bayes theorem, whereas in Singular Value 
 Decomposition \cite{Hocker:1995kb} the technique for which it is named is used to control the statistical fluctuations. 

\subsubsection*{An aside on notation}
Experimental collaborations are not consistent in the notation used in unfolded results, both within and without, 
and great care must be taken to understand exactly how each result has formulated the unfolding problem. 
Equation~\ref{eq:unfolding} is a fair representation of the most factorised form of the general unfolding problem. 
However, some results may concatenate various aspects. For example, the matrix $R_{ij}$ may sometimes have 
the effects of $\epsilon_{i}^{\mbox{\scriptsize{ext}}}$ and $\epsilon^{\mbox{ \scriptsize{fid}}}_{j}$ folded in, or the 
denominator in the first fraction can be combined into a single factor. Experimental results also use the terms 
\emph{response} and \emph{migration} matrix interchangeably and inconsistently so the specific definition 
in each result must be taken into account.

\subsubsection*{Using unfolded results in EFT fits}
All unfolded results are, by definition, biased. The techniques that are used to tackle the matrix inversion problem, 
as well as to control for other experimental effects, invariably introduces some dependence on the model used to 
construct them, as discussed in reference to Eqs.~(\ref{eq:unfoldingP}) and~(\ref{eq:unfoldingPobs}).
However, it is not true to say that this bias is always large or prohibits the use of results in EFT fits. 
It is more correct to say that unfolded results have a \emph{region of validity} centred on the SM but which extends 
beyond it. How far that model extends, and in particular if it extends sufficiently far to encompass shifts in various 
Wilson Coefficients, depends on a number of things. The most important components to consider when using 
unfolded results in EFT fits are $R_{ij}$, $\epsilon_{i}^{\mbox{\scriptsize{ext}}}$, $\epsilon^{\mbox{ \scriptsize{fid}}}_{j}$, 
and $N_j^{\mbox{\scriptsize{bkg}}}$. $R_{ij}$ in its simplest form (i.e. without the $\epsilon$ factors folded in) 
is usually the same regardless of the model used to build it as it is only probing the detector smearing effects. 
Matrices that are highly diagonal and symmetric about the diagonal, are likely to have a much broader region 
of validity than, say, a matrix that is diagonal but off-centre and which has significant one-sided smearing. 
$\epsilon_{i}^{\mbox{\scriptsize{ext}}}$ is an extrapolation out of the phase space of the measurement and 
into some other phase-space, usually that of the total cross-section. It is entirely dependent on the model that 
constructs it. However, it can easily be undone post-hoc and redone with EFT effects included, and is therefore 
not a huge problem when considering EFT fits. $\epsilon^{\mbox{ \scriptsize{fid}}}_{j}$ cannot be redone post-hoc 
and is also somewhat dependent on the model used to construct it. In cases where this is flat and close to one, 
the bias that it causes is negligible (especially for normalised differential cross-sections). However, significant 
shapes in this curve is usually an indication that the region of validity of the result is small. Finally, 
$N_j^{\mbox{\scriptsize{bkg}}}$ is only problematic for EFT fits if the background is non-negligible 
and could be effected by the Wilson coefficients under study. Such effects can usually be avoided 
by only selecting results with high signal purity, such as Drell-Yan, top quark, and dijet results, 
though this is not an exhaustive list.

\subsubsection*{Recommendations for unfolded results}
The safest results to use in EFT fits have the following properties:
\begin{itemize}
	\item They have a very high signal to background ratio and, ideally, what backgrounds 
	to penetrate the signal region are not sensitive to the Wilson coefficients being probed.
	\item They have highly diagonal response matrices and/or response matrices which are highly 
	symmetric about the diagonal which makes the detector smearing effects marginal and well-controlled.
	\item They have flat $\epsilon^{\mbox{ \scriptsize{fid}}}_{j}$ curves. This is especially relevant for normalised 
	differential cross-sections, as it means the shape of the unfolded data is decoupled from the normalisation 
	and therefore safe to use in an EFT fit where the shape of the EFT effects is very different from the SM.
	\item They are measured in a tight fiducial region and at particle level (which, by extension, 
	means that $\epsilon_{i}^{\mbox{\scriptsize{ext}}}$ should be close to unity and flat.)
\end{itemize}
There are relatively few experimental results that satisfy all of these criteria (though, unfolded angular observables 
are usually ideal) and omitting one or more of these recommendations is reasonable as long as some considerations 
for the effect of the bias on any fits is considered in more detail.

\subsubsection{Simplified template cross sections}
\label{sect:stxs}

\begin{figure}[t]
  \begin{center}
    \captionsetup{justification=centerlast}
    \includegraphics[width=0.57\linewidth]{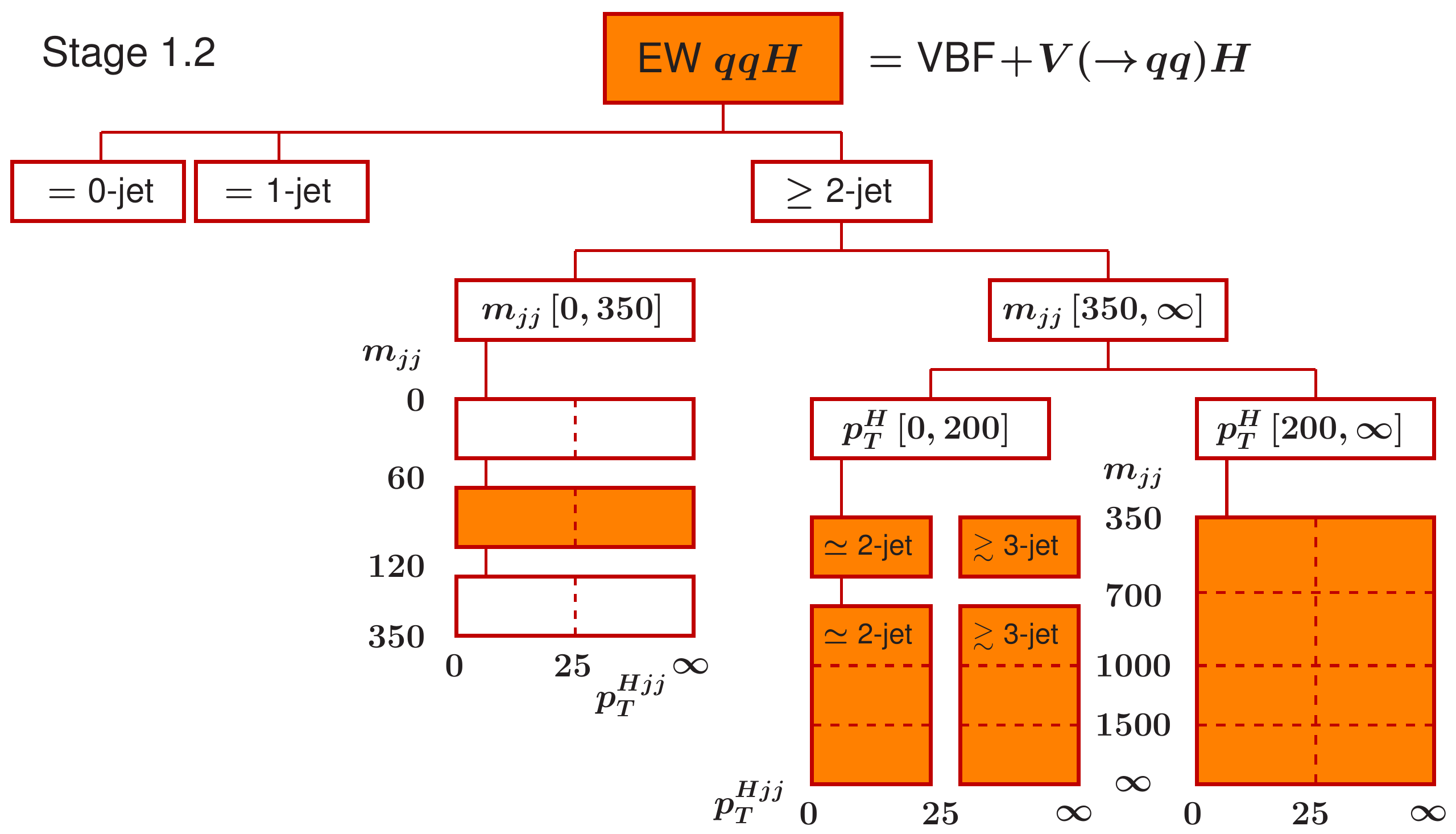}
    \caption{
Schematic representation of STXS 1.2 bins for the electroweak production of the Higgs boson in association with two jets~\cite{lhcxwg}.}
    \label{fig:stxs}
  \end{center}
\end{figure}

In the analysis of the Higgs boson data on LHC, a framework of the simplified template cross sections (STXS)~\cite{deFlorian:2016spz}
has been developed, which is designed for the measurement of the single-$H$ cross sections in mutually exclusive 
regions of the phase space $\vec{x}^{\,\prime}_\mathrm{truth}$ (which includes parton shower in this case). 
The observable quantities are the bins of kinematic templates for each Higgs boson production process. 
An example of such a template is shown in Fig.~\ref{fig:stxs} for the electroweak production of the Higgs boson 
in association with two jets, where ten bins are defined based on the quantities characterizing the process. 
These bins have been chosen to provide sensitivity to higher-dimension operators, which typically exhibit
enhancement at higher transverse momenta. 

This framework falls under the two-step measurement approach discussed in Section~\ref{sect:meas} and 
shares its pros and cons. It resembles the differential distributions with the unfolding procedure discussed in
Section~\ref{sect:unfold}, but has its unique features, and for this reason we discuss it in a separate section. 
This framework takes advantage of the fact that characterization of the Higgs boson production is independent from 
its decay. Therefore, the same quantities in STXS are measured with multiple Higgs boson decay channels. 
At the same time, the production processes are separated during the data analysis and results are
presented for each production process independently, which is different from the typical Higgs boson
differential distributions. These features are unique to Higgs boson physics and 
for this reason this framework has not found application in other areas so far. 
The main advantage of this approach is wide dissemination of results for (re)interpretation in a uniform way. 
The main disadvantage is its limited scope and potential dependence of cross section interpretation 
on the parameters of the model, as discussed in reference to Eqs.~(\ref{eq:unfoldingP}) and~(\ref{eq:unfoldingPobs}). 

\subsubsection{Template likelihood fit}
\label{sect:template}

The end goal of most analyses on LHC is to construct the likelihood ${\cal L}$ and maximize it to obtain 
the best-fit values of $\vec{\theta}$ and set their confidence intervals. 
This approach can be taken either in the single-step approach, when EFT-sensitive parameters $\vec{\theta}$ are measured directly, 
or in the two-step approach, when intermediate cross sections $\vec{\theta}$ are measured, e.g. in the STXS approach. 
There are various techniques to construct or approximate the likelihood, including the matrix element method and
machine learning, discussed below, but the most straightforward and widely used approach is the so-called template
approach. 

\begin{figure}[t]
  \begin{center}
    \captionsetup{justification=centerlast}
    \includegraphics[width=0.39\linewidth]{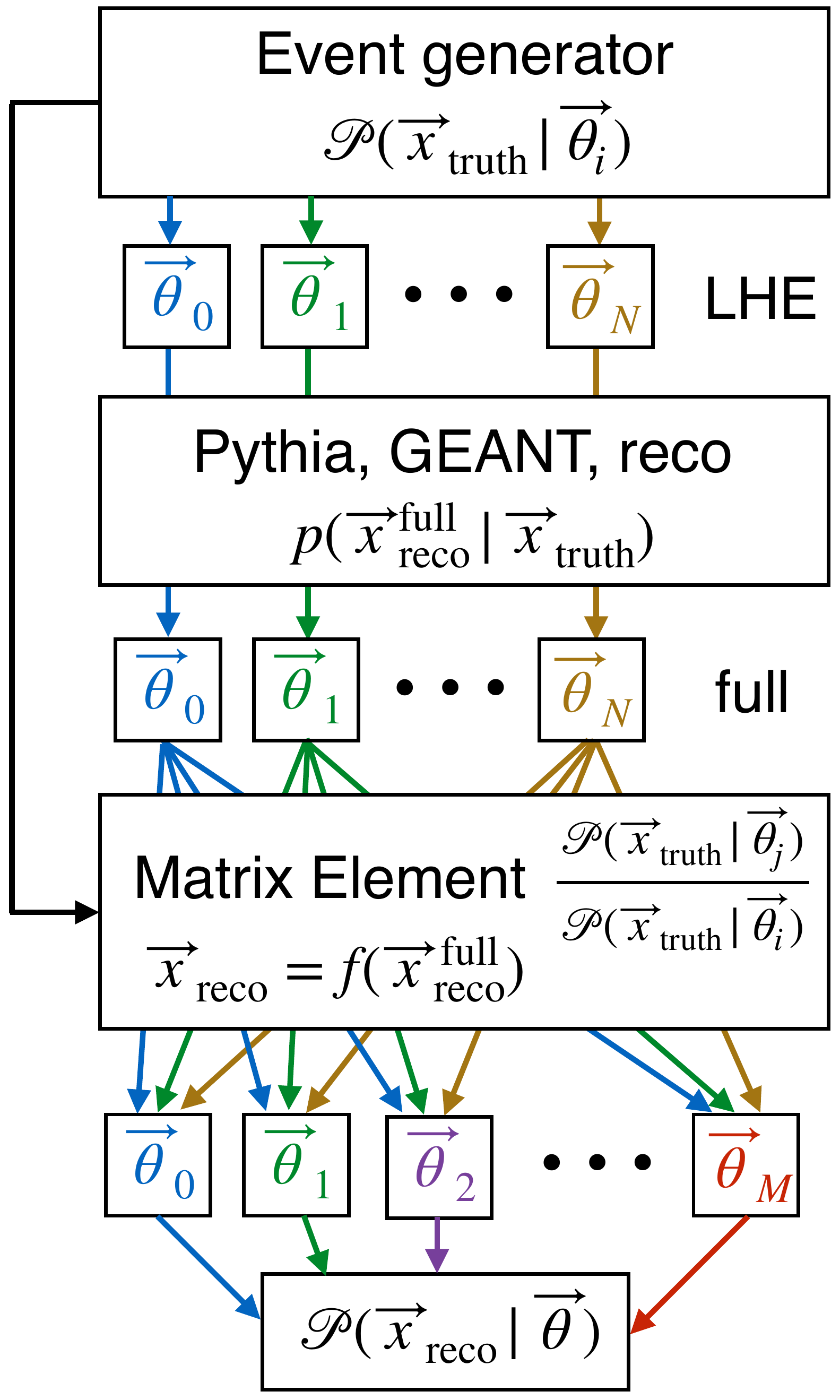}
    \caption{
The chain of MC modeling of probability density ${\cal P}(\vec{x}_\mathrm{reco}|\vec{\theta})$
as a function reconstructed observables $\vec{x}_\mathrm{reco}$ and EFT parameters $\vec{\theta}$
following Eq.~(\ref{eq:P}). 
An event generator produces LHE files for $N$ hypotheses $\vec\theta_i$, which are processed 
through full MC simulation, including parton shower and detector effects, and further re-weighted
into $M$ hypotheses necessary for analytical morphing of template parameterization.
The diagram is inspired by Ref.~\cite{CMS:2021nnc}.
     }
    \label{fig:templates}
  \end{center}
\end{figure}

In the template approach, the pdf ${\cal P}(\vec{x}_\mathrm{reco}|\vec{\theta})$ is a histogram, or template, 
of observables $\vec{x}_\mathrm{reco}$ binned in one or several dimensions. 
The dependence on parameters $\vec{\theta}$ in the special case of EFT couplings
is shown in Eq.~(\ref{eq:probreco}). 
This allows analytical morphing of template parameterization by generating a discrete number of 
$M=(K+2)!/(2K!)$ models corresponding to each term in Eq.~(\ref{eq:probreco}).
Construction of such templates is illustrated in Fig.~\ref{fig:templates}, where the chain of MC modeling of 
the pdf follows Eq.~(\ref{eq:P}). 
An event generator produces LHE files~\cite{Alwall:2006yp} for $N$ hypotheses $\vec\theta_i$, 
which are processed through full MC simulation, including parton shower and detector effects. It is important
to pick a diverse enough set of $N$ hypotheses in such a way that all corners of phase space are sampled well. 
However, this set does not need to be complete, as matrix element re-weighting of MC samples allows 
generating any number of models $M$. This last step of re-weighting is a lot less time consuming than 
the full MC simulation. Therefore, in order to minimize statistical uncertainties at minimal cost, any model
can be re-weighting to any other model. In the end, constraints could be obtained on $\theta_i/\theta_0$
and the overall cross section of the process. 

\subsubsection{Matrix element method}
\label{sect:mem}

The matrix element method (MEM) employs Eq.~(\ref{eq:P}) directly to construct a likelihood function. 
It is important to draw distinction between MEM, discussed in this section, and the observables created
with the help of matrix-element calculations, discussed in Section~\ref{sect:obs-me}. In the latter case, 
matrix elements are used to calculate observables $\vec{x}_\mathrm{reco}$ following the general approach 
of Eq.~(\ref{eq:xreco}). These observables can later be used in different measurements, from 
differential distributions to machine-learning inference. The MEM approach, on the other hand, is the
direct way to perform a measurement through construction of a likelihood with the help of transfer functions 
$p(\vec{x}_\mathrm{reco}|\vec{x}_\mathrm{truth})$. 

The MEM approach is particularly suited to EFT measurements for two main reasons. 
First of all, such an approach may be built close to optimal because nearly 
the full information $\vec{x}_\mathrm{reco}^\mathrm{\,full}$ can be employed. Second, the
large number of EFT parameters $\vec{\theta}$ can be easily incorporated in this approach because those 
appear in the readily available matrix element calculation ${\cal P}(\vec{x}_\mathrm{truth}|\vec{\theta})$. 
In other words, once the transfer functions are incorporated, there is no limitation on the matrix element 
calculations, other than availability of the matrix elements themselves. Once a MEM analysis is built for the 
SM process, it can easily be extended to EFT by plugging in the right matrix elements. 

At the same time, the MEM approach has its own limitations. The first challenge is to build the transfer functions 
$p(\vec{x}_\mathrm{reco}|\vec{x}_\mathrm{truth})$, which is never possible to do from first principle and certain 
approximations have to be employed. This becomes particularly challenging in the case of high multiplicity of 
particles generated in parton shower and considered in analysis. Correct particles could escape detection,
wrong particles could be picked, and permutations of particles could create confusion. Therefore, the MEM
approach is most successful when full reconstruction of the process with little confusion is possible. 
The second challenge is computational, as numerical technics for calculations and integration often have
to be employed and would lead to significant CPU requirements. The MEM approach is most successful when 
a large degree of analytical calculations could be maintained. The third challenge is availability of the matrix
elements for all processes considered. In particular, not all background processes could be described 
analytically, such as instrumental background. One may approximate such backgrounds either with 
empirical functions or templates of observables. However, the former is not necessarily possible and 
the latter eliminates some of the main advantages of the MEM approach.

The first proposal of MEM in application to a non-trivial environment of hadron collisions with missing
particles was made in Ref.~\cite{Kondo:1988yd}.
There have been a number of successful applications of the MEM approach to the Higgs, top, electroweak, 
and $b$-quark flavor measurements at LHC and other experiments. At the same time, the MEM approach 
has not become the main approach in performing the measurements on LHC due to its challenges.

\subsubsection{Inference with machine learning}
\label{sect:ml}

The heart of the challenge for inferring the EFT parameters $\theta$ from data is that the likelihood function 
defined by Eq.~(\ref{eq:P}) involves an intractable integral. As expressed in Eq.~(\ref{eq:P}) even the transfer 
function $p(\vec{x}_\mathrm{reco}|\vec{x}_\mathrm{truth})$ is intractable as it would involve an enormous 
integral over the truth-level quantities that describe the parton shower, hadronization, and the electromagnetic 
and hadronic showers that occur when final state particles interact with the detector. 
We introduce $z$ to denote the full MC truth record that includes the hard scattering as well as 
the other steps in the full simulation chain, and use it interchangeably with $\vec{x}_\mathrm{truth}$.
In the language of statistics and machine learning, 
our simulation chain is a generative model and all the random variables in the MC 
simulation that we refer to as MC truth are referred to as latent variables. Simulators like this that can be 
used to generate synthetic data with MC, but which do not admit a tractable likelihood are referred to as 
implicit statistical models~\cite{Diggle1984MonteCM}. 

The task of performing statistical inference when the data generating process does not have a tractable likelihood 
is known as \emph{simulation-based} or \emph{likelihood-free inference}~\cite{Cranmer:2019eaq}. This case is not 
at all unique to particle physics. The formulation of this problem in a common, abstract language has led to statisticians, 
computer scientists, and domain scientists from various fields developing powerful methods for simulation-based 
inference together. 

The particle physics community has been coping with problems like this for decades. The strategy is to use histograms 
or kernel density estimation to approximate the intractable integral from samples of synthetic data generated with 
MC. Since it is not practical to histogram the high dimensional data $\vec{x}_\mathrm{reco}$, 
we identify observables (summary statistics) that carry as much information about the parameters $\vec\theta$ as possible.  
This is the well known template likelihood fit strategy, discussed in Section~\ref{sect:template}.
Alternatively, the matrix element method, discussed in Section~\ref{sect:mem}, seeks to approximate 
${\cal P}(\vec{x}_\mathrm{reco}|\vec{\theta})$ through numerical integration, but must rely on a simplification of the 
transfer function $p(\vec{x}_\mathrm{reco}|\vec{x}_\mathrm{truth})$ that does not involve the details of the parton 
shower and detector interaction. However, machine learning opens up new ways to approximate the likelihood function 
that does not require sacraficiing power with the introduction of summary statistics nor does it require a simplification 
of the transfer functions that one would obtain with the full parton shower and detector simulation and reconstruction. 

\begin{figure}[t]
    \centering
    \includegraphics[width=\textwidth]{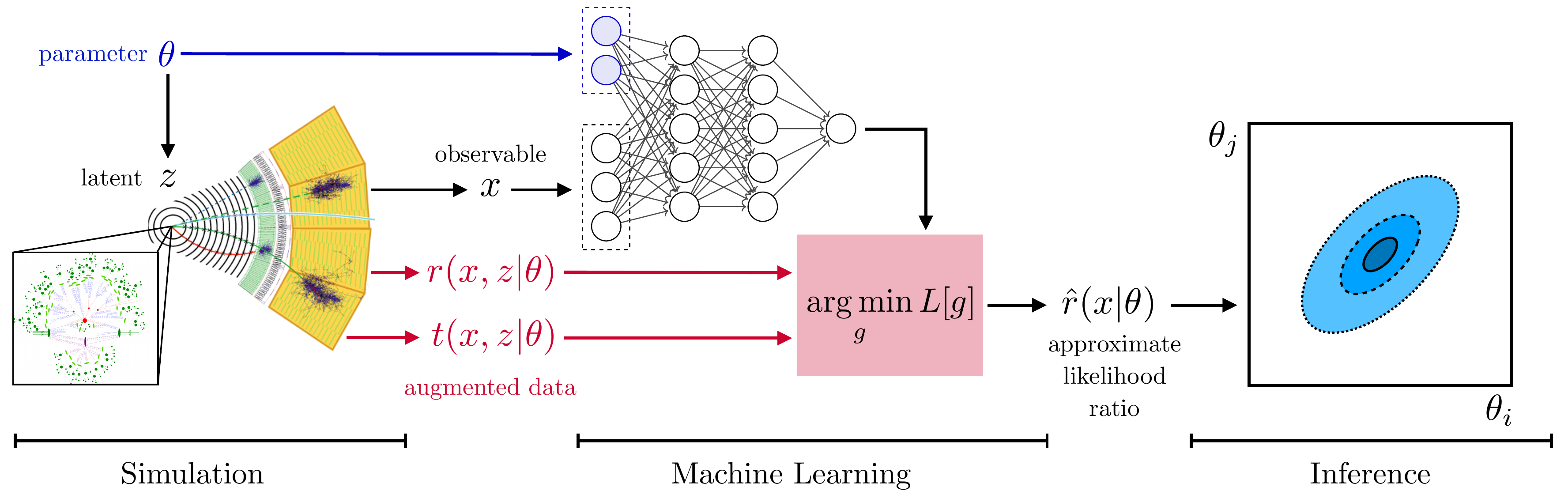}
    \caption{Illustration of simulation-based inference techniques that  train a neural network surrogate for the likelihood 
or likelihood ratio function. Figure taken from Ref.~\cite{Brehmer:2018kdj}. Here $x$ corresponds to~$\vec{x}_\mathrm{reco}$.
and $z$ denotes the full MC truth record.
}
    \label{fig:rascal_explainer}
\end{figure}

A schematic of this approach is shown in Fig.~\ref{fig:rascal_explainer}. A parametrized neural 
network~\cite{Cranmer:2015bka,Baldi:2016fzo} is used to learn the likelihood ratio 
$r(\vec{x}_\mathrm{reco} | \theta) = {\cal P}(\vec{x}_\mathrm{reco}|\vec{\theta})/{\cal P}(\vec{x}_\mathrm{reco}|\vec{\theta}_0)$, 
where ${\cal P}(\vec{x}_\mathrm{reco}|\vec{\theta}_0)$ is some reference distribution such as the SM or 
a potentially unphysical mixture of different EFT points. The choice of this reference distribution simply leads to 
an offset in the log-likelihood and does not affect the maximum likelihood estimate or the resulting contours. 
The reason that the reference distribution is used is that it allows us to reframe the task of approximating the 
likelihood from one of density estimation to classification. We can train a parametrized neural network on the 
supervised learning task of binary classification between data generated from ${\cal P}(\vec{x}_\mathrm{reco}|\vec{\theta}_0)$ 
and data generated from ${\cal P}(\vec{x}_\mathrm{reco}|\vec{\theta})$. Instead of using a different neural network for 
each EFT parameter point, a single parametrized network learns to interpolate. 

The most basic version of this approach was first laid out in Ref.~\cite{Cranmer:2015bka}, but generating 
MC training samples that cover the full EFT space is prohibitive. Furthermore, in the region of the 
EFT space we are most interested in, the deviation from the SM prediction is small. 
This means that in the two samples the classifier is trying to distinguish two distributions that are nearly 
the same, which leads to a poor approximation of the likelihood ratio. Both of these challenges can be 
overcome~\cite{Brehmer:2018kdj,Brehmer:2018eca}. The first point is that the analytical morphing allows 
us not only to morph template histograms but also to continuously reweight individual events~\cite{ATLAS:morphing}. 
Together with a resampling procedure, we can produce training data that uniformly covers the EFT parameter 
space from a finite number of basis samples. 
References~\cite{Brehmer:2018kdj,Brehmer:2018eca,Chen:2020mev} 
also describe how the analytical morphing can be used to encode the $\vec{\theta}$ dependence directly 
into morphing-aware neural network architectures. 
However, these tricks do not address the problem that the deviation from the SM is small. 
This is addressed by a trick known as \textit{mining gold}~\cite{Brehmer:2018hga,Stoye:2018ovl}. 
The insight here is that while ${\cal P}(\vec{x}_\mathrm{reco}|\vec{\theta})$ involves an intractable integral, 
the \textit{joint likelihood ratio} 
$r(\vec{x}_\mathrm{reco}, z|\vec{\theta}) = 
p(\vec{x}_\mathrm{reco}, z|\vec{\theta})/p(\vec{x}_\mathrm{reco}, z|\vec{\theta}_0)$ 
and the \textit{joint score} 
$t(\vec{x}_\mathrm{reco}, z|\vec{\theta}) = 
\nabla_{\vec{\theta}} \log p(\vec{x}_\mathrm{reco}, z)$ 
are tractable because the terms that depend on the transfer functions drop out when the 
$\vec{\theta}$-dependence is isolated to the hard scattering. Moreover, both of these quantities can be 
computed directly from the analytic morphing equations. These can then be used to augment the training 
data and lead to dramatically more efficient training. The joint score in particular is of interest because it is 
a local quantity and directly characterizes how much more or less likely it would be for a region of phase 
space to be populated if the EFT parameters were modified from their SM values. 
By regressing on these quantities it is possible to recover the likelihood ratio or the score (having marginalized 
out the latent variables associated to MC truth). It turns out that the score function provides locally 
sufficient statistics, coincides with the concept of optimal observables when detector effects are negligible, 
as introduced in Section~\ref{sect:obs-me}, and generalizes them to settings where detector effects must 
be taken into account, as shown in Section~\ref{sect:obs-ml}.
The software package \texttt{MadMiner} \cite{Brehmer:2019xox,brehmer_johann_2021_5654822} 
provides a reference implementation of these techniques.

\clearpage

\section{Operators and measurements for global SMEFT fits}
\label{sect:area3_operators}
One of the goals of this activity area on Experimental Measurements and Observables is to survey which 
experimental channels are sensitive to which EFT operators. The aim is to establish a map between operators
and experimental measurements. 
As a first step we want to determine which operators are relevant and as a second step to 
examine the amount of information that each process can provide for a given operator, given the accuracy of any
given experimental process. The second step can be 
achieved by considering some appropriate metric such as the Fisher information. 
 
\begin{figure}[htbp]
  \begin{center}
    \includegraphics[width=0.5\linewidth]{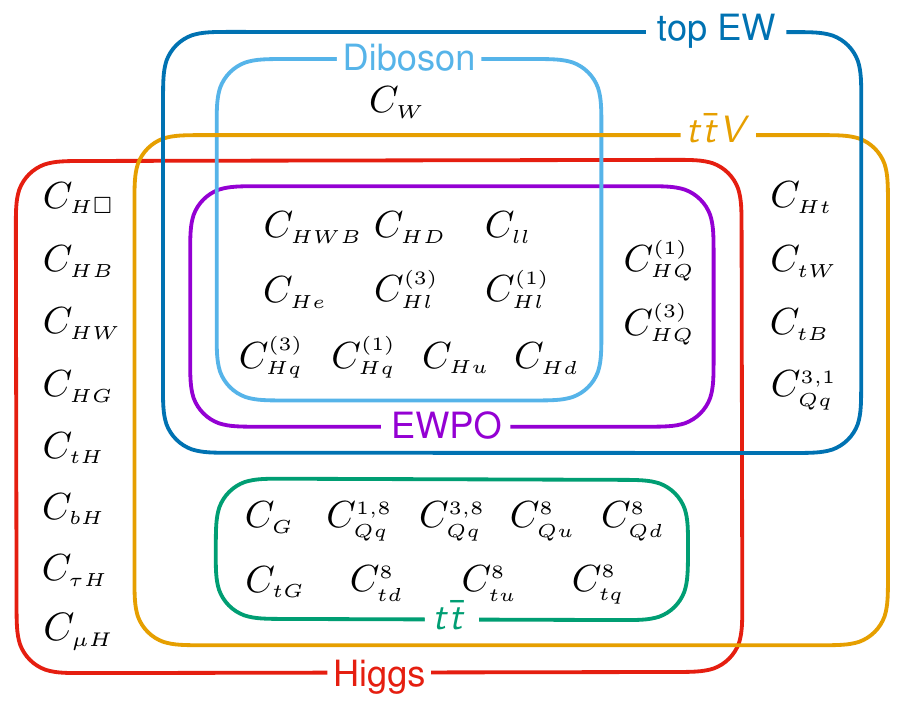}
    \caption{\small 
     \label{venn} Schematic representation of the datasets and their overlapping dependences on the 34 Wilson coefficients included in the analysis of Ref.~\cite{Ellis:2020unq}}
    \label{fig:schematic}
  \end{center}
\end{figure}

A valuable source of information in this endeavour are the global fits which exist in the literature. 
The set of operators relevant for each fit is determined by the processes included in the fit as well as the choice of 
 flavour assumption which determines the relevant degrees of freedom. As an example, in Fig.~\ref{fig:schematic}
 we show a schematic representation of the datasets and their overlapping dependences on the 34 Wilson coefficients 
 included in the global Higgs, top, diboson and EWPO analysis of \verb|fitmaker|~\cite{Ellis:2020unq}. 

A major part of global interpretations of HEP data in the SMEFT framework is establishing the connection between operators and experimental measurements and eventually the generation of predictions for the dependence of each measurement on the Wilson Coefficients of interest. The first step involves starting from a particular flavour assumption and then creating a list of operators entering a given process at some perturbative order. The second step is more quantitative and typically involves a significant undertaking, requiring large scale numerical calculations that take into account details of the experimental analyses. The map from operators to measurements contains a great deal of information in itself, indicating both how the data may collectively constrain each degree of freedom and the origin of potential correlations among them, post-fit. 

Such results only make up half of the required ingredients of a global fit, the other half being the experimental data and covariance matrix. Nevertheless, it is worthwhile to present and dissect them in a systematic way for a number of reasons. From a fundamental perspective, we may understand the potential impacts and interplay that the data imposes on the allowed parameter space of the theory, which can guide our expectations before doing any elaborate fits and also help us to understand the results of such an exercise. Moreover, it serves as a useful validation exercise, where we can verify the consistency of the predictions both internally and against other existing calculations. This is especially important given the fact that such a large scale generation of predictions is inevitably prone to human error. Finally, making the results public allows them to be re-used in future interpretations, avoiding a duplication of effort and therefore hastening progress towards maximal indirect sensitivity to new physics from precision measurements.

In this note, we establish a map between operators and experimental measurements and quantify the sensitivity of the Higgs, top, and Electroweak measurements that went into the recently published global analyses of \verb|fitmaker|~\cite{Ellis:2020unq} and SMEFiT~\cite{Ethier:2021bye} collaborations.  Whilst the two collaborations follow similar approaches, some differences exist and these will be mentioned in the note where relevant. 


\subsection{Calculational framework}

In this note we work in the Warsaw basis of SMEFT dimension-6 operators ~\cite{Grzadkowski:2010es}.
The operator coefficients are normalised as
\begin{equation}
    \mathcal{L}_{\text{Dim-6}} = \sum_{i} \frac{C_i}{\Lambda^2} \mathcal{O}_i \, .
\end{equation}
%

%
%
Before discussing the operators entering each analysis we briefly 
mention the corresponding flavour assumptions as employed by the
two fitting collaborations. SMEFiT~\cite{Ethier:2021bye} follows  the strategy presented
in~\cite{Hartland:2019bjb,AguilarSaavedra:2018nen}, namely
minimal flavour violation (MFV) hypothesis~\cite{DAmbrosio:2002vsn} in the
quark sector as the baseline scenario. 
 \verb|fitmaker|~\cite{Ellis:2020unq} considers both a flavour universal 
and an MFV scenario. We discuss in detail the operators considered in each analysis below. 

\subsubsection{Operator basis in SMEFiT 2021 analysis}
The SMEFiT analysis adopts a  $U(2)_q\times U(2)_u \times U(3)_d$ symmetry
where the Yukawa couplings are nonzero only for the top quark,
consistent with the {\tt SMEFT@NLO} model~\cite{Degrande:2020evl}.
Whilst strictly not consistent with the flavour assumption, the bottom and
charm quark Yukawa operators are included in the fit, to account for the current LHC
sensitivity to these parameters, while all other Yukawas are set to zero.
A universal $(U(1)_\ell \times
U(1)_e)^3$ symmetry is adopted in the lepton sector,
which sets  all the lepton masses as well as their Yukawa couplings to
zero. Again, a non-zero $\tau$ Yukawa operator is allowed.

\begin{table}[htbp] 
  \begin{center}
    \renewcommand{\arraystretch}{1.30}
        \begin{tabular}{cccc}
          \toprule
  Class  &  $N_{\rm dof}$ &  Independent DOFs  & DoF in EWPOs\\
          \midrule
          \multirow{4}{*}{four-quark}    &  \multirow{5}{*}{14}
          & $c_{Qq}^{1,8}$, $c_{Qq}^{1,1}$, $c_{Qq}^{3,8}$,   \\
          \multirow{4}{*}{(two-light-two-heavy)}   &
          &  $c_{Qq}^{3,1}$,  $c_{tq}^{8}$,  $c_{tq}^{1}$,  \\
             &    & $c_{tu}^{8}$, $c_{tu}^{1}$, $c_{Qu}^{8}$,\\
            &    & $c_{Qu}^{1}$, $c_{td}^{8}$, $c_{td}^{1}$,   \\
          &    &  $c_{Qd}^{8}$, $c_{Qd}^{1}$   \\
          \midrule
                    \multirow{1}{*}{four-quark}      &  \multirow{2}{*}{5}
                    & $c_{QQ}^1$, $c_{QQ}^8$, $c_{Qt}^1$, &   \\
          \multirow{1}{*}{(four-heavy)}      &   & $c_{Qt}^8$, $c_{tt}^1$ &  \\
\midrule
                    \multirow{1}{*}{four-lepton}      &  \multirow{1}{*}{1}
                    &   &  $c_{\ell\ell}$ \\
	  \midrule
      \multirow{4}{*}{two-fermion}     &  \multirow{5}{*}{23} &  $c_{t\varphi}$, $c_{tG}$,   $c_{b\varphi}$,    &  
                     $c_{\varphi \ell_1}^{(1)}$, $c_{\varphi \ell_1}^{(3)}$, $c_{\varphi \ell_2}^{(1)}$ \\
       \multirow{4}{*}{(+ bosonic fields)}      &    & $c_{c\varphi}$, $c_{\tau\varphi}$,   $c_{tW}$,   &
       $c_{\varphi \ell_2}^{(3)}$, $c_{\varphi \ell_3}^{(1)}$, $c_{\varphi \ell_3}^{(3)}$,\\
            &    &  $c_{tZ}$,  $c_{\varphi Q}^{(3)}$, $c_{\varphi Q}^{(-)}$,     &
            $c_{\varphi e}$, $c_{\varphi \mu}$, $c_{\varphi \tau}$,  \\
       &    &  $c_{\varphi t}$     & $c_{\varphi q}^{(3)}$, $c_{\varphi q}^{(-)}$, \\
       &    &       &  $c_{\varphi u i}$,
       $c_{\varphi d i}$ \\
       \midrule
      \multirow{2}{*}{Purely bosonic}     &  \multirow{2}{*}{7} &
      $c_{\varphi G}$, $c_{\varphi B}$, $c_{\varphi W}$,   &  $c_{\varphi W B}$, $c_{\varphi D}$   \\
               &   &  $c_{\varphi d}$,  $c_{WWW}$   &  \\
            \midrule
          Total  & 50 (36 independent)   & 34   & 16 (2 independent)   \\
         \bottomrule
  \end{tabular}
  \caption{\small \label{tab:operatorbasis} Summary of the degrees of freedom
  considered in the present work.
   We categorize these DoFs into five disjoint classes: four-quark (two-light-two-heavy),
   four-quark (four-heavy), four-lepton, two-fermion, and purely bosonic DoFs.
   The 16 DoFs displayed in the last columns are subject to 14 constraints from the EWPOs,
   leaving only 2 independent combinations to be constrained by the fit.
 }
  \end{center}
\end{table}


\begin{table}[t] 
  \begin{center}
    \renewcommand{\arraystretch}{1.30}
        \begin{tabular}{lll}
          \toprule
          Operator $\qquad$ & Coefficient $\qquad\qquad\qquad$ & Definition \\
        \midrule
        $\Op{\varphi G}$ & $c_{\varphi G}$  & $\left(\pdp\right)G^{\mu\nu}_{\sss A}\,
        G_{\mu\nu}^{\sss A}$  \\ \hline
        $\Op{\varphi B}$ & $c_{\varphi B}$ & $\left(\pdp\right)B^{\mu\nu}\,B_{\mu\nu}$\\ \hline
        $\Op{\varphi W}$ &$c_{\varphi W}$ & $\left(\pdp\right)W^{\mu\nu}_{\sss I}\,
        W_{\mu\nu}^{\sss I}$ \\ \hline
        $\Op{\varphi W B}$ &$c_{\varphi W B}$ & $(\varphi^\dagger \tau_{\sss I}\varphi)\,B^{\mu\nu}W_{\mu\nu}^{\sss I}\,$ \\ \hline
        $\Op{\varphi d}$ & $c_{\varphi d}$ & $\partial_\mu(\pdp)\partial^\mu(\pdp)$ \\ \hline
        $\Op{\varphi D}$ & $c_{\varphi D}$ & $(\varphi^\dagger D^\mu\varphi)^\dagger(\varphi^\dagger D_\mu\varphi)$ \\ \hline
         $\mathcal{O}_{W}$&   $c_{WWW}$ & $\epsilon_{IJK}W_{\mu\nu}^I W^{J,\nu\rho} W^{K,\mu}_\rho$ \\
       \bottomrule
        \end{tabular}
        \caption{Purely bosonic dimension-six operators that
          modify the production and decay of Higgs bosons and
          the interactions of the electroweak gauge bosons.
          For each operator, we indicate its definition in terms of the SM
          fields,
          and the notational conventions that will be used
          both for the operator and for the Wilson coefficient.
          The operators $O_{\varphi WB}$ and $O_{\varphi D}$
          are severely
          constrained by the EWPOs together with several of
          the two-fermion operators from Table~\ref{tab:oper_ferm_bos}.
           \label{tab:oper_bos}}
\end{center}
\end{table}


With this considerations, one ends up with 50 dimension-six EFT degrees of freedom
to be constrained by experimental data.
These are summarised in Table~\ref{tab:operatorbasis},
categorised into five disjoint classes, from top to bottom: four-quark (two-light-two-heavy), four-quark (four-heavy), four-lepton, two-fermion, and purely bosonic DoFs. Flavour indices are labelled by $i,j,k$ and $l$; left-handed
quark and lepton fermion SU(2)$_L$ doublets are denoted by $q_i$, $\ell_i$;
the right-handed quark singlets by $u_i$, $d_i$, while
the right-handed lepton singlets are denoted by $e$, $\mu$, $\tau$ without using
flavor index.
We  use $Q$ and $t$ to denote the left-handed top-bottom doublet and the right-handed
top singlet, and the Higgs doublet is denoted by $\varphi$.
Of these 50 EFT coefficients, 36 are independent fit parameters
while 14 are indirectly constrained by the LEP EWPOs, following
the procedure described in~\cite{Ethier:2021bye}. 

When presenting results for operator sensitivity,
the SMEFiT analysis selects the purely bosonic operators $c_{\varphi W B}$ and $c_{\varphi D}$,
for illustration purposes.

Table~\ref{tab:oper_bos} provides the definition of the 
purely bosonic dimension-six operators considered in the analysis, which
modify the production and decay of Higgs bosons as well as the interactions
of the electroweak gauge bosons.
In addition to the information provided by the input dataset,
the  operators $O_{\varphi WB}$ and $O_{\varphi D}$
are also severely
constrained by the EWPOs.
The triple gauge operator $O_{W}$ generates a TGC coupling modification which is purely transversal
and is hence constrained only by diboson data.

Table~\ref{tab:oper_ferm_bos} collects, using the same format
as in Table~\ref{tab:oper_bos}, the relevant Warsaw-basis operators
that contain two fermion fields, either quarks or leptons,
plus a single four-lepton operator.
From top to bottom, the table lists the two-fermion operators involving 3rd generation quarks,
those involving 1st and 2nd generation quarks, and
operators containing two leptonic fields (of any generation).
In this list the four-lepton operator $\mathcal{O}_{\ell\ell}$ is also included.
The operators that involve a top-quark field, either $Q$ (left-handed doublet) or $t$
(right-handed singlet),
are crucial for the interpretation of LHC top-quark measurements
and all of them involve at least one Higgs-boson field, which
introduces an interplay between the top and Higgs sectors of the SMEFT.
We point out that
most of the operator coefficients defined in Table~\ref{tab:oper_ferm_bos} correspond
directly to degrees of freedom used in
the fit, except for three of them, which are
indicated with a (*) in the second column,
for which 
three additional degrees of freedom are defined from the linear
combinations  following ~\cite{AguilarSaavedra:2018nen}, see~\cite{Ethier:2021ydt}.

\begin{table}[htbp]
  \begin{center}
    \renewcommand{\arraystretch}{1.3}
    \begin{small}
    \begin{tabular}{lll}
      \toprule
      Operator $\qquad$ & Coefficient & Definition \\
                \midrule \midrule
		&3rd generation quarks&\\
                \midrule \midrule
    $\Op{\varphi Q}^{(1)}$ & $c_{\varphi Q}^{(1)}$~(*) & $i\big(\varphi^\dagger\lra{D}_\mu\,\varphi\big)
 \big(\bar{Q}\,\gamma^\mu\,Q\big)$ \\\hline
    $\Op{\varphi Q}^{(3)}$ & $c_{\varphi Q}^{(3)}$  & $i\big(\varphi^\dagger\lra{D}_\mu\,\tau_{\sss I}\varphi\big)
 \big(\bar{Q}\,\gamma^\mu\,\tau^{\sss I}Q\big)$ \\ \hline
    $\Op{\varphi t}$ & $c_{\varphi t}$& $i\big(\varphi^\dagger\,\lra{D}_\mu\,\,\varphi\big)
 \big(\bar{t}\,\gamma^\mu\,t\big)$ \\ \hline
      $\Op{tW}$ & $c_{tW}$ & $i\big(\bar{Q}\tau^{\mu\nu}\,\tau_{\sss I}\,t\big)\,
 \tilde{\varphi}\,W^I_{\mu\nu}
 + \text{h.c.}$ \\  \hline
 $\Op{tB}$ & $c_{tB}$~(*) &
 $i\big(\bar{Q}\tau^{\mu\nu}\,t\big)
 \,\tilde{\varphi}\,B_{\mu\nu}
 + \text{h.c.}$ \\\hline
    $\Op{t G}$ & $c_{tG}$ & $ig{\sss S}\,\big(\bar{Q}\tau^{\mu\nu}\,T_{\sss A}\,t\big)\,
 \tilde{\varphi}\,G^A_{\mu\nu}
 + \text{h.c.}$ \\  \hline
    $\Op{t \varphi}$ & $c_{t\varphi}$ & $\left(\pdp\right)
 \bar{Q}\,t\,\tilde{\varphi} + \text{h.c.}$  \\\hline
    $\Op{b \varphi}$ & $c_{b\varphi}$ & $\left(\pdp\right)
 \bar{Q}\,b\,\varphi + \text{h.c.}$ \\
                \midrule \midrule
		&1st, 2nd generation quarks&\\
                \midrule \midrule
    $\Op{\varphi q}^{(1)}$ & $c_{\varphi q}^{(1)}$~(*) & $\sum\limits_{\sss i=1,2} i\big(\varphi^\dagger\lra{D}_\mu\,\varphi\big)
 \big(\bar{q}_i\,\gamma^\mu\,q_i\big)$ \\\hline
    $\Op{\varphi q}^{(3)}$ & $c_{\varphi q}^{(3)}$ & $\sum\limits_{\sss i=1,2} i\big(\varphi^\dagger\lra{D}_\mu\,\tau_{\sss I}\varphi\big)
 \big(\bar{q}_i\,\gamma^\mu\,\tau^{\sss I}q_i\big)$ \\  \hline
  ${\Op{\varphi u i}}$ &
      ${{c_{\varphi u i}}}$ & $\sum\limits_{\sss i=1,2,3} i\big(\varphi^\dagger\,\lra{D}_\mu\,\,\varphi\big)
 \big(\bar{u}_i\,\gamma^\mu\,u_i\big)$\\ \hline
 ${\Op{\varphi d i}}$ &
     ${{c_{\varphi d i}}}$ & $\sum\limits_{\sss i=1,2,3} i\big(\varphi^\dagger\,\lra{D}_\mu\,\,\varphi\big)
 \big(\bar{d}_i\,\gamma^\mu\,d_i\big)$\\ \hline
    $\Op{c \varphi}$ & $c_{c \varphi}$ & $\left(\pdp\right)
 \bar{q}_2\,c\,\tilde\varphi + \text{h.c.}$ \\
                \midrule \midrule
		&two-leptons&\\
                \midrule \midrule
    $\Op{\varphi \ell_i}^{(1)}$ & $c_{\varphi \ell_i}^{(1)}$ & $ i\big(\varphi^\dagger\lra{D}_\mu\,\varphi\big)
   \big(\bar{\ell}_i\,\gamma^\mu\,\ell_i\big)$ \\\hline 
    $\Op{\varphi \ell_i}^{(3)}$ & $c_{\varphi \ell_i}^{(3)}$ & $ i\big(\varphi^\dagger\lra{D}_\mu\,\tau_{\sss I}\varphi\big)
 \big(\bar{\ell}_i\,\gamma^\mu\,\tau^{\sss I}\ell_i\big)$ \\  \hline
    $\Op{\varphi e}$ & $c_{\varphi e}$ & $ i\big(\varphi^\dagger\lra{D}_\mu\,\varphi\big)
 \big(\bar{e}\,\gamma^\mu\,e\big)$ \\\hline
    $\Op{\varphi \mu}$ & $c_{\varphi \mu}$ & $ i\big(\varphi^\dagger\lra{D}_\mu\,\varphi\big)
 \big(\bar{\mu}\,\gamma^\mu\,\mu\big)$ \\  \hline
    $\Op{\varphi \tau}$ & $c_{\varphi \tau}$ & $i\big(\varphi^\dagger\lra{D}_\mu\,\varphi\big)
 \big(\bar{\tau}\,\gamma^\mu\,\tau\big)$ \\  \hline
    $\Op{\tau \varphi}$ & $c_{\tau \varphi}$ & $\left(\pdp\right)
 \bar{\ell_3}\,\tau\,{\varphi} + \text{h.c.}$ \\
                \midrule \midrule
		&four-lepton &\\
                \midrule \midrule
 $\Op{\ell\ell}$ & $c_{\ell\ell}$ & $\left(\bar \ell_1\gamma_\mu \ell_2\right) \left(\bar \ell_2\gamma^\mu \ell_1\right)$ \\
 \hline
  \bottomrule
\end{tabular}
\end{small}
\caption{Same as Table~\ref{tab:oper_bos}
  for the operators containing two fermion fields, either
  quarks or leptons, as well as the four-lepton operator $\Op{\ell\ell}$.
  The flavor index $i$ runs from 1 to 3.
  The coefficients indicated with (*) in the second column do not correspond to physical degrees of freedom
  in the fit, but are rather replaced by  $c_{\varphi q_i}^{(-)}$, $c_{\varphi Q_i}^{(-)}$, and
  $c_{tZ}$.
\label{tab:oper_ferm_bos}}
\end{center}
\end{table}


\clearpage

Finally, SMEFiT considers four-quark operators which involve the top quark 
fields and thus modify the production of top quarks at hadron colliders.
These  can be classified into
 operators composed by four heavy quark fields (top and/or bottom quarks) and 
 operators composed by two light and two heavy quark fields.
 The EFT coefficients  are
constructed in terms of suitable linear combinations of the four fermion
coefficients in the Warsaw basis, 
\begin{align}
	\qq{1}{qq}{ijkl}
	&= (\bar q_i \gamma^\mu q_j)(\bar q_k\gamma_\mu q_l)
	 \nonumber
	,\\
	\qq{3}{qq}{ijkl}
	&= (\bar q_i \gamma^\mu \tau^I q_j)(\bar q_k\gamma_\mu \tau^I q_l)
 \nonumber
	,\\
	\qq{1}{qu}{ijkl}
	&= (\bar q_i \gamma^\mu q_j)(\bar u_k\gamma_\mu u_l)
         \nonumber
	,\\
	\qq{8}{qu}{ijkl}
	&= (\bar q_i \gamma^\mu T^A q_j)(\bar u_k\gamma_\mu T^A u_l)
         \nonumber
	,\\
	\qq{1}{qd}{ijkl}
	&= (\bar q_i \gamma^\mu q_j)(\bar d_k\gamma_\mu d_l)
         \nonumber
	,\\
	\qq{8}{qd}{ijkl}
	&= (\bar q_i \gamma^\mu T^A q_j)(\bar d_k\gamma_\mu T^A d_l)
        \label{eq:FourQuarkOp} 
	,\\
	\qq{}{uu}{ijkl}
	&=(\bar u_i\gamma^\mu u_j)(\bar u_k\gamma_\mu u_l)
         \nonumber
	,\\
	\qq{1}{ud}{ijkl}
	&=(\bar u_i\gamma^\mu u_j)(\bar d_k\gamma_\mu d_l)
         \nonumber
	,\\
	\qq{8}{ud}{ijkl}
	&=(\bar u_i\gamma^\mu T^A u_j)(\bar d_k\gamma_\mu T^A d_l)
         \nonumber \, ,
\end{align}
In Table~\ref{eq:summaryOperatorsTop} we provide the definition of all four-fermion EFT
degrees of freedom
that enter the fit
in terms of the coefficients of Warsaw basis operators of Eq.~(\ref{eq:FourQuarkOp}).
Recall that within our flavour assumptions, the coefficients associated to different values of
the generation indices $i$ ($i=1,2$) or $j$ ($j=1,2,3$) will be the same.

\begin{table}[htbp] 
  \begin{center}
    \renewcommand{\arraystretch}{1.20}
        \begin{tabular}{ll}
          \toprule
          DoF $\qquad$ &  Definition (in  Warsaw basis notation) \\
          \midrule
      $c_{QQ}^1$    &   $2\ccc{1}{qq}{3333}-\frac{2}{3}\ccc{3}{qq}{3333}$ \\ \hline
    $c_{QQ}^8$       &         $8\ccc{3}{qq}{3333}$\\  \hline
 $c_{Qt}^1$         &         $\ccc{1}{qu}{3333}$\\   \hline
 $c_{Qt}^8$         &         $\ccc{8}{qu}{3333}$\\   \hline
  $c_{tt}^1$         &     $\ccc{}{uu}{3333}$  \\    \hline
            \midrule      
  $c_{Qq}^{1,8}$       &  	 $\ccc{1}{qq}{i33i}+3\ccc{3}{qq}{i33i}$     \\   \hline
  $c_{Qq}^{1,1}$         &   $\ccc{1}{qq}{ii33}+\frac{1}{6}\ccc{1}{qq}{i33i}+\frac{1}{2}\ccc{3}{qq}{i33i} $   \\    \hline
   $c_{Qq}^{3,8}$         &   $\ccc{1}{qq}{i33i}-\ccc{3}{qq}{i33i} $   \\   \hline
  $c_{Qq}^{3,1}$          & 	$\ccc{3}{qq}{ii33}+\frac{1}{6}(\ccc{1}{qq}{i33i}-\ccc{3}{qq}{i33i}) $   \\     \hline
   $c_{tq}^{8}$         &  $ \ccc{8}{qu}{ii33}   $ \\    \hline
   $c_{tq}^{1}$       &   $  \ccc{1}{qu}{ii33} $\\    \hline
   $c_{tu}^{8}$      &   $2\ccc{}{uu}{i33i}$  \\     \hline
    $c_{tu}^{1}$        &   $ \ccc{}{uu}{ii33} +\frac{1}{3} \ccc{}{uu}{i33i} $ \\   \hline
    $c_{Qu}^{8}$         &  $  \ccc{8}{qu}{33ii}$\\     \hline
    $c_{Qu}^{1}$     &  $  \ccc{1}{qu}{33ii}$  \\     \hline
    $c_{td}^{8}$        &   $\ccc{8}{ud}{33jj}$ \\    \hline
    $c_{td}^{1}$          &  $ \ccc{1}{ud}{33jj}$ \\     \hline
    $c_{Qd}^{8}$        &   $ \ccc{8}{qd}{33jj}$ \\     \hline
    $c_{Qd}^{1}$         &   $ \ccc{1}{qd}{33jj}$\\
         \bottomrule
  \end{tabular}
  \caption{\small Definition of the four-fermion degrees of freedom that enter into
    the fit in terms of the coefficients of Warsaw basis operators of Eq.~(\ref{eq:FourQuarkOp}).
    These DoFs are classified into four-heavy (upper) and two-light-two-heavy
    (bottom part) operators. The flavor index $i$ is either 1 or 2, 
    and $j$ is either 1, 2 or 3: with our flavor assumptions,  these coefficients will be the same
    regardless of the specific values that $i$ and $j$ take.
\label{eq:summaryOperatorsTop}}
  \end{center}
\end{table}


\subsubsection{Operators in the fitmaker analysis}

There are 20 operators relevant for Higgs, diboson and electroweak measurements in the flavour universal scenario. In the notation of Ref.~\cite{Grzadkowski:2010es}, these are
\begin{align}
    &\mathcal{O}_{H W B}, \mathcal{O}_{H D}, \mathcal{O}_{u}, \mathcal{O}_{H l}^{(3)}, \mathcal{O}_{H l}^{(1)}, \mathcal{O}_{H e}, \mathcal{O}_{H q}^{(3)}, \mathcal{O}_{H q}^{(1)}, \mathcal{O}_{H d}, \mathcal{O}_{H u}, \\
    &\mathcal{O}_{H \square}, \mathcal{O}_{H G}, \mathcal{O}_{H W}, \mathcal{O}_{H B}, \mathcal{O}_{W}, \mathcal{O}_{G},
    \mathcal{O}_{\tau H}, \mathcal{O}_{\mu H}, \mathcal{O}_{b H}, \mathcal{O}_{t H} \,. 
\end{align}
We note here the slightly different notation used as $H$ denotes the Higgs doublet for which Tables~\ref{tab:operatorbasis} and~\ref{tab:oper_ferm_bos} use a $\varphi$. Otherwise, the above operator choice corresponds to a subset of the SMEFiT operators with the following differences:
\begin{itemize}
    \item $\mathcal{O}_{H\Box}$ is equivalent to $\mathcal{O}_{\phi d}$ in SMEFiT up to a minus sign from integration-by-parts
    \item $C_W$ and the SMEFiT $c_{WWW}$ differ by a conventional minus sign
    \item \verb|fitmaker| additionally includes the muon Yukawa operator, $\mathcal{O}_{\mu H}$
\end{itemize}
As mentioned, for the purposes of this note we focus on the Higgs, diboson and electroweak measurements, and relevant operators. We refer the reader to Ref.~\cite{Ellis:2020unq} for the global analysis, which also includes top physics measurements.

\subsection{Experimental measurements}

\subsubsection{Input experimental data in SMEFiT 2021 analysis}
We discuss next the experimental data used
in~\cite{Ethier:2021bye}.
In Table~\ref{eq:table_dataset_overview} we summarise the number of data points in the baseline dataset
for each of the data categories and processes considered in this analysis, as well
as the total per category and the overall total.
SMEFiT include 150, 97, and 70 cross-sections from top-quark production, Higgs boson production
and decay, and diboson production
cross-sections from LEP and the LHC respectively in the baseline dataset,
for a total of 317 cross-section measurements.
For all processes, we consider only parton-level measurements, since the theoretical EFT
interpretation and simulation of particle-level measurements is more challenging.

\begin{table}[htbp]
  \centering
  \small
   \renewcommand{\arraystretch}{1.30}
  \begin{tabular}{c|c|c}
 Category   & Processes    &  $n_{\rm dat}$     \\
    \toprule
    \multirow{6}{*}{Top quark production}   &  $t\bar{t}$ (inclusive)   &  94  \\
    &  $t\bar{t}Z$, $t\bar{t}W$    & 14 \\
    &   single top (inclusive)   & 27 \\
    &  $tZ, tW$   &  9\\
    &  $t\bar{t}t\bar{t}$, $t\bar{t}b\bar{b}$    & 6 \\
    &  {\bf Total}    & {\bf 150 }  \\
    \midrule
    \multirow{3.3}{*}{Higgs production} & Run I signal strengths  &22   \\
    \multirow{3.1}{*}{and decay} & Run II  signal strengths  & 40  \\
    & Run II, differential distributions \& STXS  & 35  \\
    &  {\bf Total}    & {\bf 97}  \\
    \midrule
    \multirow{3}{*}{Diboson production} & LEP-2 &40   \\
     & LHC & 30  \\
    &  {\bf Total}    & {\bf 70}  \\
    \bottomrule
   Baseline dataset     & {\bf Total}      & {\bf 317}  \\
\bottomrule
  \end{tabular}
  \caption{\small The number of data points $n_{\rm dat}$ in our baseline dataset
    for each of the categories of processes considered here.
 \label{eq:table_dataset_overview}
}
\end{table}


Concerning top-quark production measurements,
we consider four
different categories: inclusive top-quark pair production, top-quark pair
production in association with vector bosons or heavy quarks, inclusive single
top-quark production, and single top-quark production in association with
vector bosons.
Top-quark pair production in association
with a Higgs boson is considered part of the Higgs processes.
The bulk of the top quark measurements corresponds to
inclusive top-quark pair production, with measurements
of single and double differential distributions in the dilepton
and lepton+jets final states from ATLAS and CMS
~\cite{Aad:2015mbv,Khachatryan:2015oqa,Sirunyan:2017azo,Aaboud:2016hsq,Khachatryan:2016fky,Khachatryan:2016mnb,Sirunyan:2018wem,Sirunyan:2017mzl,Aaboud:2016iot,Aad:2019ntk,Sirunyan:2018ucr}.
In general, the $m_{t\bar{t}}$ invariant mass distributions are found
to be the most constraining observables.
We also include top-quark pair charge asymmetry measurements:
the ATLAS and CMS combined dataset at 8~TeV~\cite{Sirunyan:2017lvd},
and the ATLAS dataset at 13~TeV~\cite{ATLAS:2019czt}.

For associated top-quark pair production together with gauge bosons
and heavy quarks, we consider
the ATLAS and CMS measurements of the total cross-sections for $t\bar{t}t\bar{t}$
and $b\bar{b}b\bar{b}$ production~\cite{Sirunyan:2017snr,Sirunyan:2017roi,Sirunyan:2019wxt,Aad:2020klt,Aaboud:2018eki,Sirunyan:2019jud},
in the ATLAS and CMS measurements of inclusive $tW$ and $tZ$ production at
8~TeV and 13~TeV~\cite{Khachatryan:2015sha,Sirunyan:2017uzs,Aad:2015eua,
  Aaboud:2016xve,Aaboud:2019njj,CMS:2019too}.
In addition to inclusive measurements, we also consider the measurement
of the $p_T^Z$ differential distribution in $t\bar{t}Z$ production from CMS.
The $t\bar{t}V$ measurements are
especially useful to constrain EFT effects that modify the electroweak
couplings of the top-quark.
We note that the  $t\bar{t}t\bar{t}$
and $b\bar{b}b\bar{b}$  measurements can only be meaningfully described
within a EFT analysis that accounts for the quadratic $\mathcal{O}\lp \Lambda^{-4}\rp$
corrections.

In the case of inclusive single top-quark production
both in the $t$-channel and in the $s$-channel,
we account for total cross-sections and rapidity distributions
from ATLAS and CMS~\cite{Khachatryan:2014iya,CMS-PAS-TOP-14-004,Aaboud:2017pdi,Aad:2015upn,Khachatryan:2016ewo,Aaboud:2016ymp,CMS:2016xnv,Sirunyan:2016cdg,Sirunyan:2019hqb}.
We also consider the associated single top-quark production with weak bosons,
with the $tW$ and $tZ$ measurements from ATLAS and CMS
at 8 and 13 TeV~\cite{Aad:2015eto,Chatrchyan:2014tua,Aaboud:2016lpj,Sirunyan:2018lcp,Sirunyan:2017nbr,Aaboud:2017ylb,Aad:2020zhd,Aad:2015eto,Aad:2020wog,Sirunyan:2018zgs}.

Concerning  Higgs boson production and decay measurements,
we consider  inclusive fiducial cross-section measurements (signal
strengths) as well as  differential
distributions and STXS measurements.
For the LHC Run I, we take into account the inclusive
measurements of  Higgs boson production and decay rates from the ATLAS and CMS
combination based on the full 7 and 8~TeV datasets~\cite{Khachatryan:2016vau}.
For the LHC Run II, we consider the ATLAS measurement of signal strengths
corresponding to an integrated luminosity of
$\mathcal{L}=80$~fb$^{-1}$~\cite{Aad:2019mbh}, and the CMS measurement
corresponding to an integrated luminosity of
$\mathcal{L}=35.9$~fb$^{-1}$~\cite{Sirunyan:2018koj}.
As in the case of the Run I signal strengths, we keep into account
correlations between the various production and final state combinations.

In the case of differential measurements, we consider the ATLAS and CMS differential distributions
in the Higgs boson kinematic variables obtained from the combination of the
$h\to \gamma\gamma$, $h\to ZZ$, and (in the CMS case)
$h \to b\bar{b}$ final states at 13~TeV based
on $\mathcal{L}=36$~fb$^{-1}$~\cite{Aaboud:2018ezd,Sirunyan:2018sgc}.
Specifically, we consider the differential distributions in the Higgs boson
transverse momentum $p_T^h$.
We also include the ATLAS measurement of the associated production of Higgs
bosons, $Vh$, in the $h\to b\bar{b}$ final state at
13~TeV~\cite{Aaboud:2019nan}.
Then we also include selected differential measurements presented in the ATLAS
Run II Higgs combination paper~\cite{Aad:2019mbh}.
Specifically, we include the measurements of Higgs production
in gluon fusion, $gg \to h$, in different bins of
$p_T^h$ and in the number of jets in the event.
Furthermore, we consider the differential STXS Higgs boson production measurements
presented by CMS at 13~TeV based on an integrated luminosity of
$\mathcal{L}=77.4$~fb$^{-1}$ and corresponding to the final state
$\gamma\gamma$~\cite{CMS:1900lgv}.
In all differential measurements, whenever available the
information on the experimental correlated systematic uncertainties is included.

Finally we consider diboson production cross-sections measured by LEP and the LHC.
To begin with, we consider the LEP-2
legacy measurements of $WW$ production~\cite{Schael:2013ita}.
Specifically, we include the cross-sections differential
in $\cos\theta_W$ in four different bins in the center of
mass energy, from $\sqrt{s}=182$ GeV up to $\sqrt{s}=206$ GeV.
Concerning the LHC datasets, we include measurements of the differential
distributions for $W^{\pm}Z$ production at 13~TeV from
ATLAS~\cite{ATLAS-CONF-2018-034} and CMS~\cite{Sirunyan:2019bez} based on a
luminosity of $\mathcal{L}=36.1$ fb$^{-1}$. In both cases, the two gauge bosons
are reconstructed by means of the fully leptonic final state.
Our baseline choice will be to include the $m_{T}^{WZ}$
distribution for the ATLAS measurement, which extends up to transverse masses of $m_{T}^{WZ}=600$ GeV,
while for  the corresponding CMS measurement, the normalised
differential distributions in $p_T^Z$ is chosen.
In addition, we also consider
the differential distributions for $WW$ production from ATLAS at 13~TeV
based on a luminosity of $\mathcal{L}=36.1$ fb$^{-1}$~\cite{Aaboud:2019nkz}.
We note that diboson measurements are the only ones providing direct information
on the triple gauge operator $c_{WWW}$.

To finalise this discussion of the experimental data, we mention that
the {\sc\small SMEFiT} analysis framework was used in~\cite{Ethier:2021ydt}
for the EFT interpretation of vector-boson scattering (VBS) measurements
from the LHC Run II, together with diboson data.
The constraints on the electroweak EFT operators provided by VBS
were found to be compatible with those derived in the global fit.

\subsubsection{Input experimental data in fitmaker analysis}

The \verb|fitmaker| code contains a database of 342 measurements. We will consider here a subset pertaining to Higgs, diboson and electroweak precision data. \verb|fitmaker| is publicly available such that the maps produced here can be extended to include Top data and different assumptions on the flavour structure.

The \verb|fitmaker| dataset for the flavour universal scenario fit consists of the electroweak pseudo-observables reported by LEP/SLD, together with the $W$ mass measurements at ATLAS and Tevatron, a range of diboson measurements from LEP and ATLAS, as well as a large number of Higgs measurements, both inclusive and differential including STXS bins. It is constructed to avoid potential statistical overlap as much as possible, while favouring experimental combinations that provide access to correlation information. For example, only one type of Higgs measurement is included per experiment, i.e., an STXS combination for ATLAS and a signal strength combination for CMS, both of which publish correlation matrices. The datasets are further discussed below when the linear dependence of operators and measurements is presented. For more details, we refer the reader to Ref.~\cite{Ellis:2020unq}. This study precedes the latest CDF $W$-mass measurement, and we refer interested readers to our recent dedicated study of the impact of this measurement in Ref~\cite{Bagnaschi:2022whn}.

\subsection{Mapping between data and EFT coefficients.}
In this section we discuss which operators enter which measurements following two different approaches. First we present the linear dependence of various measurements on the Wilson coefficients as presented in the \verb|fitmaker| analysis. We then present the Fisher information table which determines the impact of various operators on different measurements taking into account the experimental precision. 
\subsubsection{Linear dependences of measurements on operators (Fitmaker)}

\begin{figure}[t]
    \centering
    \includegraphics[width=1.0\textwidth]{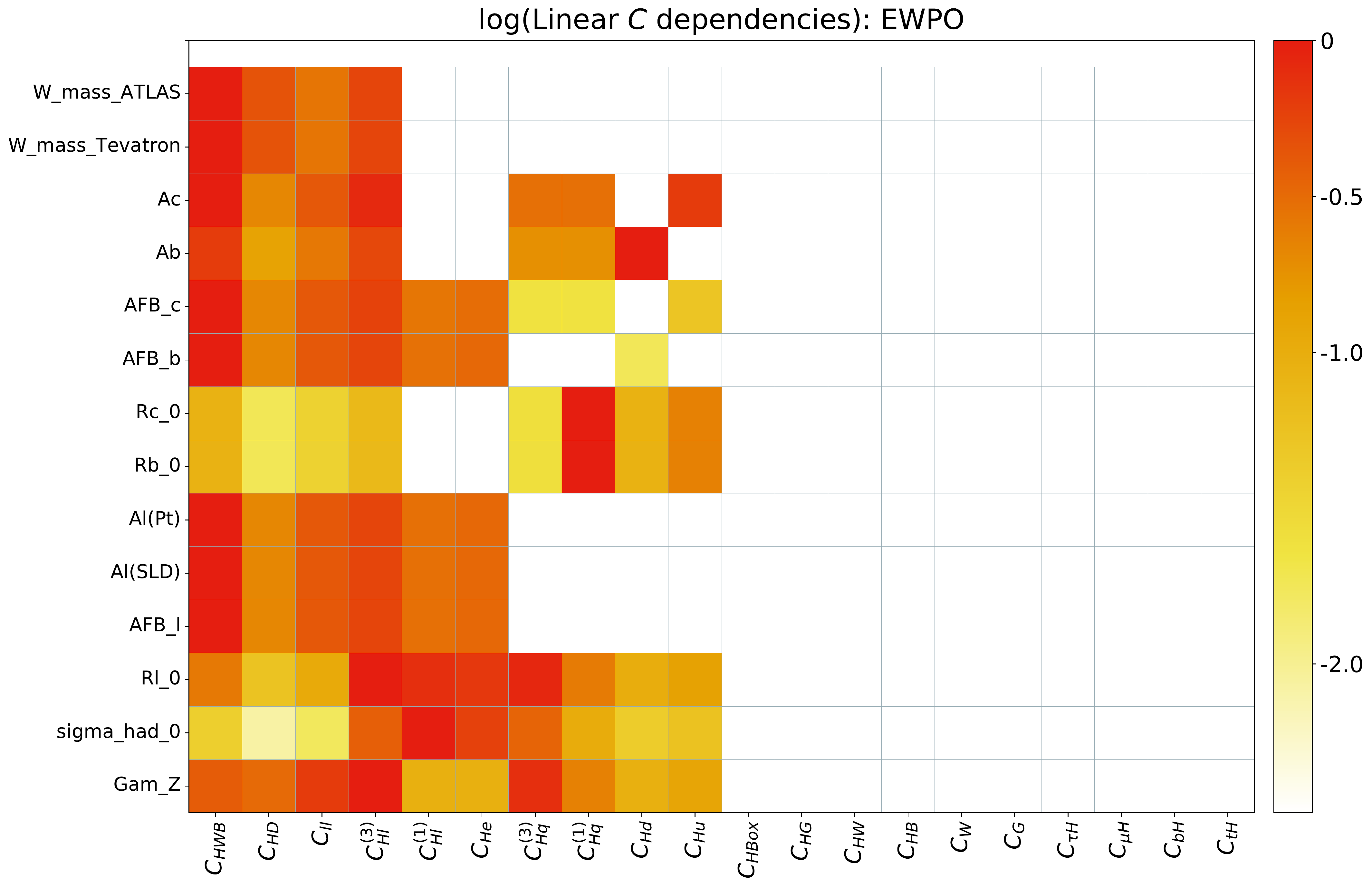}
    \caption{Logarithm of normalised linear dependences for electroweak measurements. The entries are normalised by dividing each one by the largest operator dependence of a given measurement, $a_{\mathrm{max}}^X$, such that the colour map depicts $\log( a_i^X /a_{\mathrm{max}}^X )$.}
    \label{fig:EWPO}
\end{figure}

In the \verb|fitmaker| analysis the linear dependence, $a_i^X$, on the Wilson coefficients, $C_i$, is computed using MadGraph5 with the SMEFTsim UFO model for a measured quantity $X$ relative to the SM value $X_{SM}$,  
\begin{equation}
\mu_{X} \equiv \frac{X}{X_{S M}}=1+\sum_{i} a_{i}^{X} \frac{C_{i}}{\Lambda^{2}}+\mathcal{O}\left(\frac{1}{\Lambda^{4}}\right) \, .
\end{equation}
In other words,   $a_i^X$ indicates the linear EFT cross-section for process $X$ and operator $\mathcal{O}_i$ once the corresponding Wilson coefficient $C_i$ has been factored out.
This is done at tree-level for all measurements except for the Higgs coupling to gluons which is calculated at one loop using SMEFT@NLO.  Wherever relevant, measurements are of unfolded observables, such that the dependences are computed at parton-level in the unfolded phase space, not including any showering or detector effects. This means that using this set of dependences neglects any acceptance differences between the SM and the SMEFT predictions,  which, in some cases, have been shown to be relevant (See, \emph{e.g.}, the impact of $\mathcal{O}_{HWB}$ on the measurement of $h\to 4\ell$ presented in Ref.~\cite{ATLAS:2020rej}).  See Ref.~\cite{Ellis:2020unq} for more details on our extraction of the linear dependences. 
After normalising the $a_i$ linear dependences of $\mu_{X}$ on the Wilson coefficients by dividing by the largest absolute value of the $a_i$ for a given measurement $X$, the resulting normalised dependence is colour coded on a log scale in Fig.~\ref{fig:EWPO}. Each row corresponds to a measurement while each column corresponds to a coefficient. The darkest red colours represent the strongest linear dependencies while the uncoloured white blocks indicate no dependence at linear order. We see that the right half of the grid is uncoloured since they correspond mostly to operators that can only be constrained in Higgs physics or diboson measurements. We can also read off that the operator coefficients most relevant for a large number of measurements is $C_{HWB}$, though the operators $C_{Hu}$ and $C_{Hd}$ have the largest dependencies in the $A_c$ and $A_b$ asymmetries respectively. For the $Z$ decay width, $\Gamma_Z$, it is the $Z$ coupling to fermions modified by the operators corresponding to $C_{Hl}^{(3)}$ and $C_{Hq}^{(3)}$ that are the most important contributions to this measurement. It is well known that, of the 10 Warsaw basis operators that impact these measurements, only 8 linear combinations are actually constrained by a fit to this data alone~\cite{Falkowski:2014tna}.

\begin{figure}[t]
    \centering
    \includegraphics[width=1.0\textwidth]{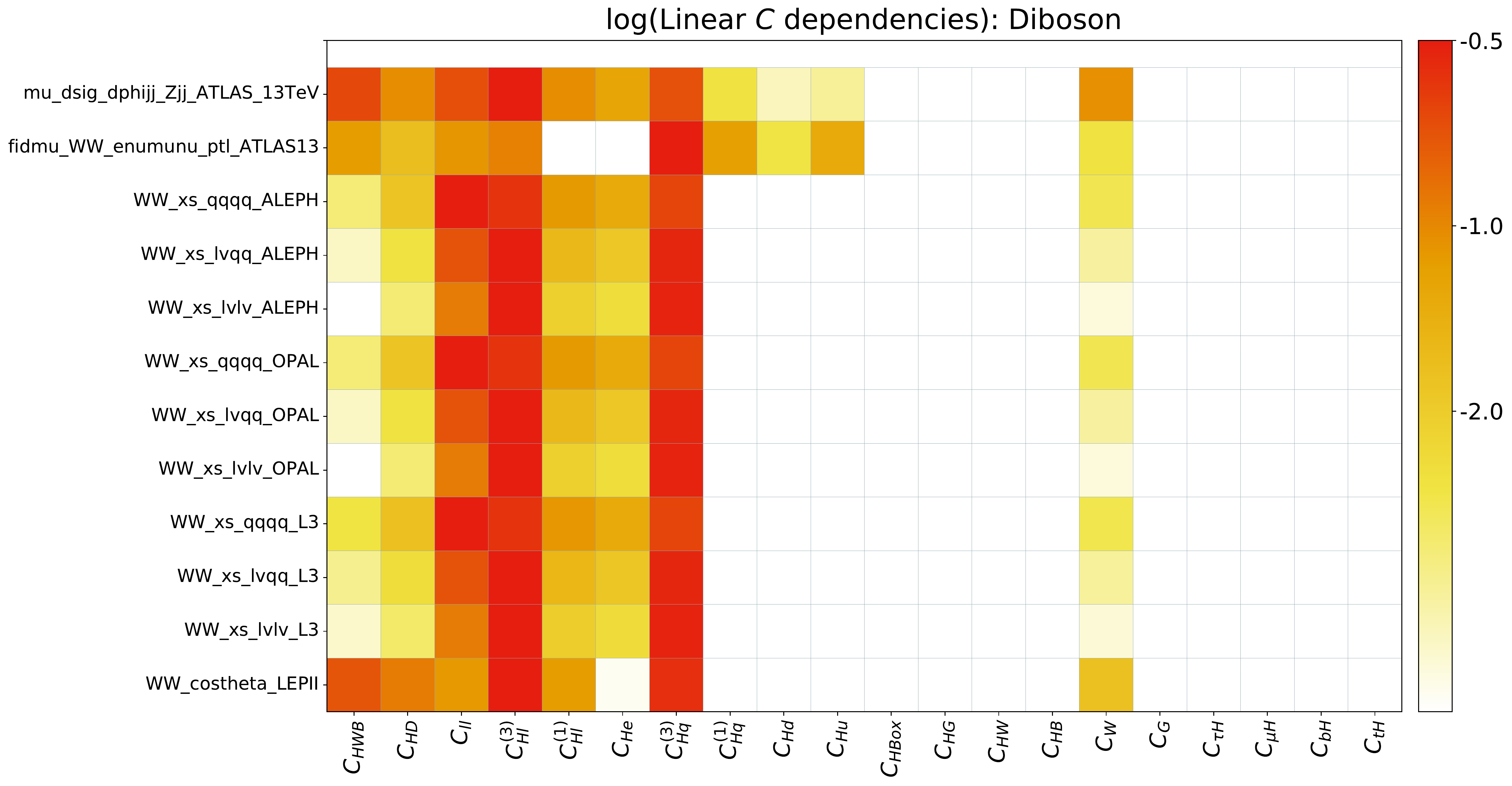}
    \caption{Logarithm of normalised linear dependences for diboson measurements, as in Fig.~\ref{fig:EWPO}.}
    \label{fig:diboson}
\end{figure}

\begin{figure}[t]
    \centering
    \includegraphics[width=1.0\textwidth]{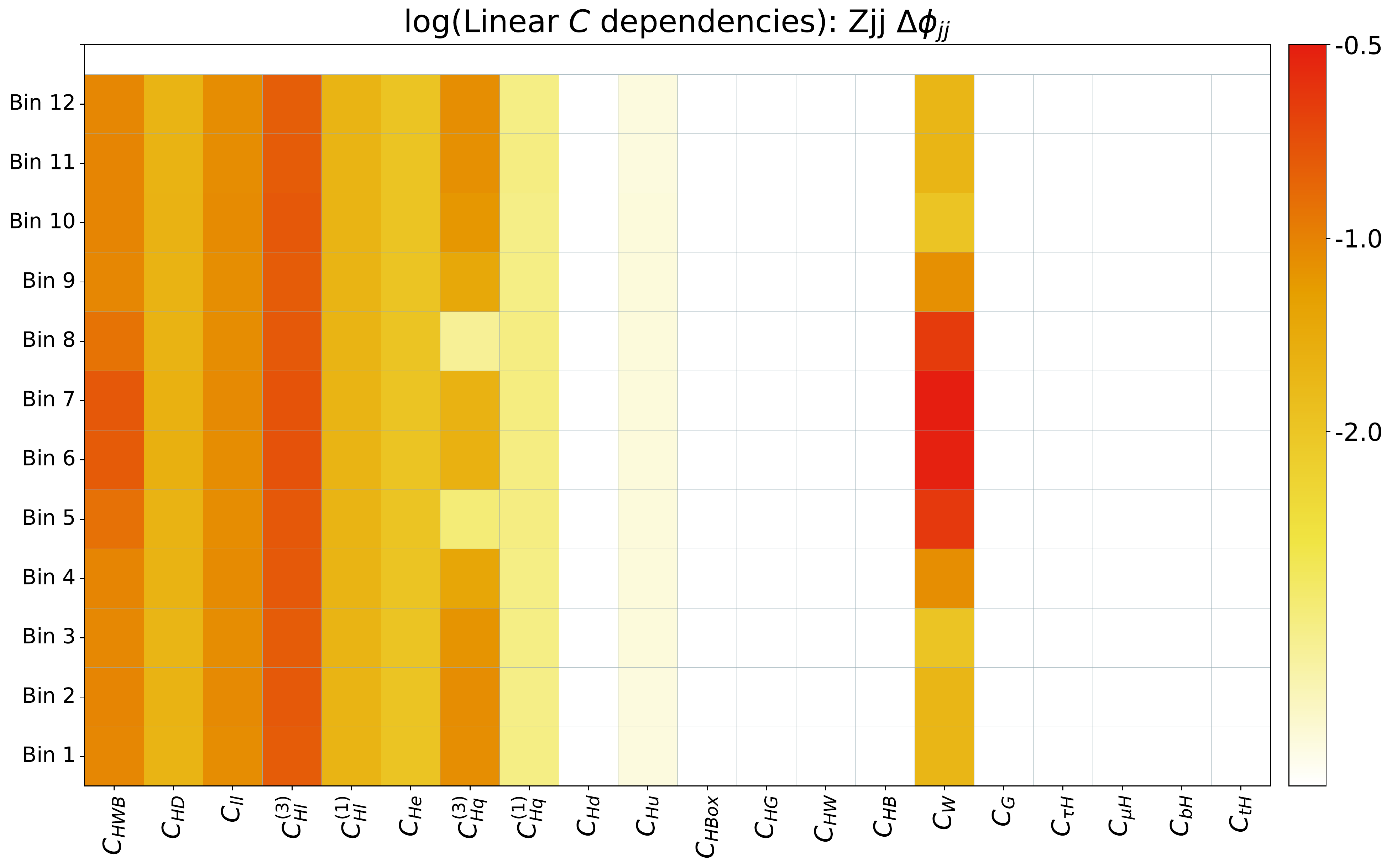}
    \caption{Logarithm of normalised linear dependences for each $\Delta\phi_{jj}$ bin from $-\pi$ to $\pi$ in the differential $Zjj$ measurement. The normalisation here is with respect to the strongest linear dependence across all bins of the measurement.}
    \label{fig:Zjj}
\end{figure}

Similarly, Fig.~\ref{fig:diboson} shows the mapping of linear dependences for a variety of diboson measurements in the same normalisation and colour scheme. These are differential distribution measurements that include either several kinematic bins or cross section measurements are different collider energy; we show here only the linear dependence in the last bin for simplicity. Just as in the $Z$ decay width case, $C_{Hl}^{(3)}$ and $C_{Hq}^{(3)}$ are the largest contributions to these measurements. In this case, however, this is expected from the fact that these operators directly modify the interactions of the $W$ boson, while the others enter via $Z$-coupling modifications and input parameter shifts. Crucially, these measurements probe different linear combinations of the 10 operators constrained by the $Z/W$ pole data. Moreover, we also see that the operator coefficient $C_W$ can now be constrained, in particular by the new $Zjj$ data that was found by ATLAS to eliminate blind directions and be particularly sensitive to linear contributions from this operator in a global fit. Since the operator $\mathcal{O}_W$ involves three gauge field strengths, it can only be constrained by diboson data and not by electroweak or Higgs processes. In Fig.~\ref{fig:Zjj} we show the linear dependences for all the bins of this differential measurement, where the linear dependencies are now normalised to 1 with respect to the strongest $a_i$ in all the bins. The non-trivial shape induced by $C_W$ is evident. This is in contrast to other, differential measurements such as $W/Z$ boson $p_T$'s that suffer from a suppressed interference between the SM and $C_W$ that leads to relatively weaker sensitivity at linear level.

\begin{figure}[t]
    \centering
    \includegraphics[width=1.0\textwidth]{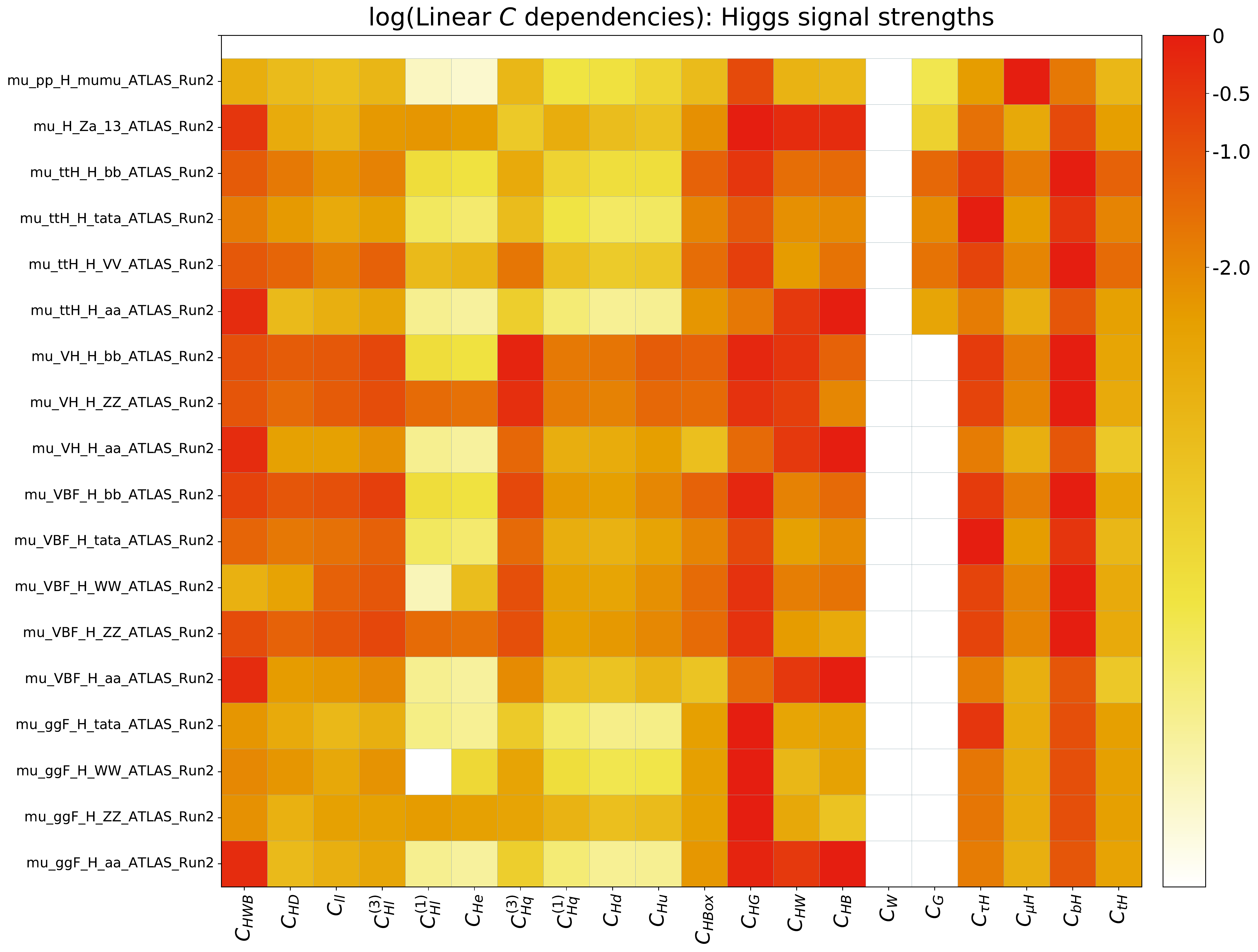}
    \caption{Logarithm of normalised linear dependences for Higgs signal strength measurements, as in Fig.~\ref{fig:EWPO}. Dependences include effects in both production and decay.}
    \label{fig:Higgs}
\end{figure}

\begin{figure}[t]
    \centering
    \includegraphics[width=1.0\textwidth]{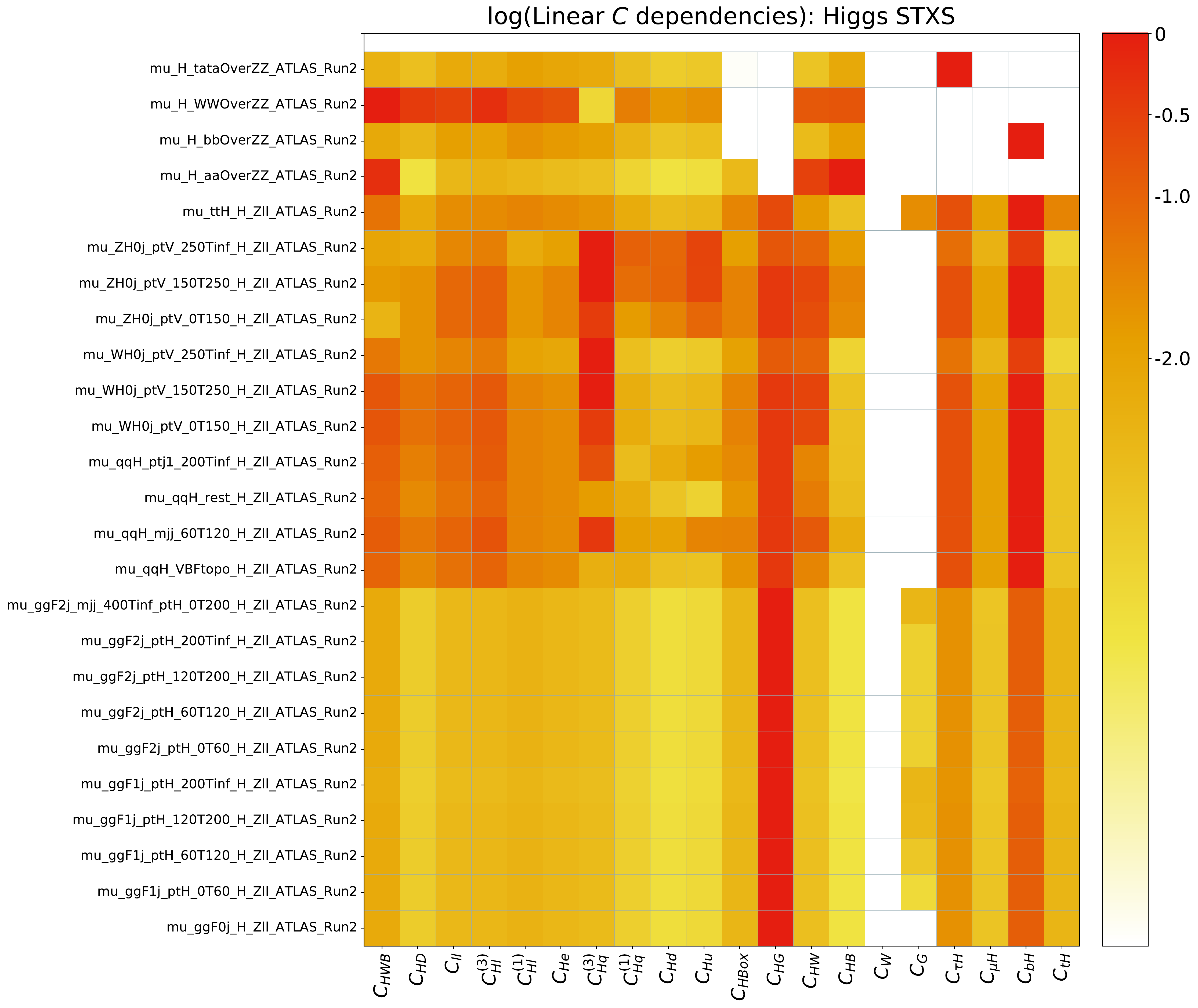}
    \caption{Logarithm of normalised linear dependences for Higgs STXS measurements, as in Fig.~\ref{fig:EWPO}. Although the figure is labelled ``ATLAS'', since our analysis makes use of those specific measurements, STXS definitions, and therefore SMEFT dependences, are universal. Dependences include effects in both production and the nominal, $h\to 4\ell$ decay, except the first four. These quantify the dependences of the ratio of the other Higgs decay branching fractions to the nominal one.}
    \label{fig:STXS}
\end{figure}

Finally in Figs.~\ref{fig:Higgs} and \ref{fig:STXS} are the maps for Higgs signal strengths and stage 1.1 Simplified Template Cross-section (STXS) bins, respectively. The operators that can only be constrained by Higgs physics are now populated. As expected, the gluon fusion channels depend on $C_{HG}$ the strongest while the one-loop calculation using SMEFT@NLO~\cite{Degrande:2020evl} allows this to be properly calculated. This is also crucial to account for the effects of the triple gluon field strength operator coefficient, $C_G$ as well as the top quark chromomagnetic operator $C_{tG}$, which we considered in a generalised flavour assumption employed when combining with top quark data. The EW Higgs production modes provide additional handles on the current operators probed by the $W/Z$ pole and diboson data. In general, the new kinematic granularity provided by the STXS bins breaks degeneracies among many operators present in the signal strengths. 

\clearpage

\subsubsection{Fisher information matrix analysis (SMEFiT)}
A clean mapping between the input experimental measurements
is provided by the tools underlying information geometry~\cite{Brehmer:2017lrt}, specifically
by means of the Fisher information matrix~\cite{Ethier:2021bye}. defined as
\be
\label{eq:FisherDef}
I_{ij}\lp {\boldsymbol c} \rp = -{\rm E}\lc \frac{\partial^2 \ln f \lp {\boldsymbol \sigma}_{\rm exp}|
{\boldsymbol c} \rp}{\partial c_i \partial c_j} \rc \, , \qquad i,j=1,\ldots,n_{\rm op} \, ,
\ee
 Here $E$ indicates the expectation value, $f$ is the functional dependence of the cross-section with the EFT coefficients $\boldsymbol{c}$
and ${\boldsymbol \sigma}_{\rm exp}$ is the array of central experimental
cross-section measurements.
The  functional dependence $f \lp {\boldsymbol \sigma}_{\rm exp}|
{\boldsymbol c} \rp$ corresponds to the full likelihood function,
which is most cases is provided as a multi-Gaussian distribution.
It is important to emphasize however that the absolute size of the entries of the Fisher matrix does not
contain physical information: one is always allowed to redefine the overall normalisation
of an operator such that $c_i\sigma^{(\rm eft)}_{m,i} = c_i'\sigma'^{(\rm eft)}_{m,i}$, with
$c_i' = B_i c_i$ and $\sigma^{(\rm eft)}_{m,i} =\sigma'^{(\rm eft)}_{m,i}/B_i$
with $B_i$ being arbitrary constants.
For a given operator the relative value of $I_{ii}$ between two groups of processes is independent
of this choice of normalisation and
thus conveys meaningful information.
For this reason, in the following we present results for the Fisher information matrix normalised
such that 
the sum of the diagonal entries associated to a given EFT coefficient adds up to a fixed reference value
which is taken to be 100.

\begin{figure}[htbp]
  \begin{center}
    \includegraphics[width=0.85\linewidth]{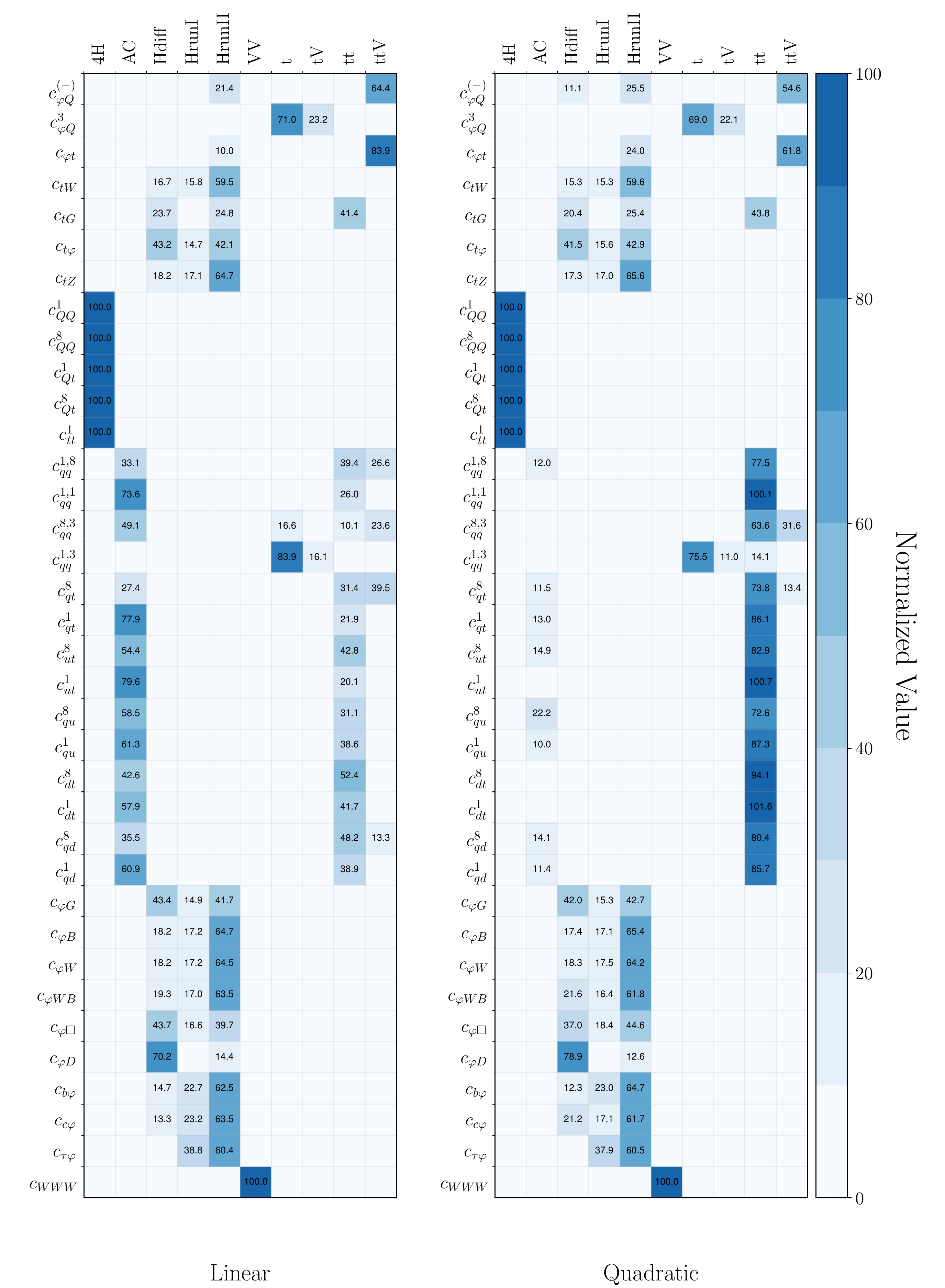}
    \caption{\small The values of the diagonal entries of the Fisher
      information matrix evaluated for the dataset of the global linear
      (left) and quadratic (right panel) SMEFiT analysis
      The normalisation here is such that the sum of the entries associated to each EFT
      coefficient adds up to 100.
      For entries in the heat map larger than 10, we also indicate the corresponding
      numerical values.}
     \label{fig:FisherMatrix} 
  \end{center}
\end{figure}

Fig.~\ref{fig:FisherMatrix} displays the values of the diagonal entries of the Fisher
information matrix both at the linear and at the quadratic level.
We note that at linear order in the EFT  expansion the dependence on the coefficients cancels out and the Fisher information
matrix is strictly fit-independent. At the quadratic level one
has some dependence on the Wilson coeffictions and
the Fisher information needs to be computed in an iterative manner.
One can identify, for each EFT coefficient, which datasets
provide the dominant constraints.
For instance, one observes that the two-light-two-heavy operators are overwhelmingly constrained
by inclusive top quark pair production data, except for $c_{Qq}^{3,1}$ for
which single top is the most important set of processes.
At the linear level, the information on the two-light-two-heavy coefficients provided
by the differential distributions and by the charge asymmetry $A_C$ data is comparable,
while the latter is less important in the quadratic fits.
In the case of the two-fermion operators, the leading constraints typically arise from Higgs data, in particular
from the Run II signal strengths measurements, and then to a lesser extent from the Run I data
and the Run II differential distributions.
Two  exceptions are $c_{\varphi t}$, which at the linear level (but not at the quadratic one)
is dominated by $t\bar{t}V$, and the chromo-magnetic operator $c_{tG}$, for which inclusive
$t\bar{t}$ production is most important.
Also for the purely bosonic operators the Higgs data provides most of the information,
except for $c_{WWW}$, as expected since this operator is only accessible in
diboson processes.
One observes that the $\mathcal{O}\lp \Lambda^{-4}\rp$ corrections induce in most
cases a moderate change in the Fisher information
matrix, but in others they can significantly alter
the balance between processes.
  
By fine-graining the calculation of the Fisher information matrix,
one can gain insight about not only which group of processeses dominates
the constraints on a given EFT coefficient, but also within a given group
of processes which is the specific dataset that dominates.
To illustrate this, \cref{tab:FisherMatrix_AC,tab:FisherMatrix_tt8,tab:FisherMatrix_tt13} display
a similar comparison as that shown for Fig.~\ref{fig:FisherMatrix} now in a fine-grained version
and restricted to the inclusive top quark pair production datasets.
Among various interesting observations,
we see that in the quadratic EFT fit the constraints on the 2-light-2-heavy
operators are dominated by the CMS measurements at 13 TeV in the dilepton and lepton+jets
channels based on the 2016 dataset.
This finding is not completely surprising, since these two datasets are the only
ones based on 36 fb$^{-1}$ of luminosity, yet it is reassuring that we can identify
this dominance {\it a priori}, without needing to redo the actual fit.


\begin{table}
\scriptsize
\centering
\begin{tabular}{|c|c|c|c|}
\hline
 \multicolumn{2}{|c|}{} & \multicolumn{2}{c|}{Processes} \\ \hline
 Class & Coefficient & \href{https://arxiv.org/abs/1709.05327}{${\rm ATLAS\_CMS\_tt\_AC\_8TeV}$} & \href{https://cds.cern.ch/record/2682109}{${\rm ATLAS\_tt\_AC\_13TeV}$} \\ \hline
\multirow{10}{*}{2FB}
 & $c_{\varphi Q}^{(-)}$ & { \color{black} 0.000}({\color{black} 0.000}) & { \color{black} 0.000}({\color{black} 0.000})\\ \cline{2-4} 
 & $c_{\varphi Q}^{3}$ & { \color{black} 0.000}({\color{black} 0.000}) & { \color{black} 0.000}({\color{black} 0.000})\\ \cline{2-4} 
 & $c_{\varphi t}$ & { \color{black} 0.000}({\color{black} 0.000}) & { \color{black} 0.000}({\color{black} 0.000})\\ \cline{2-4} 
 & $c_{tW}$ & { \color{black} 0.000}({\color{black} 0.000}) & { \color{black} 0.000}({\color{black} 0.000})\\ \cline{2-4} 
 & $c_{tG}$ & { \color{black} 0.000}({\color{black} -0.000}) & { \color{black} 0.000}({\color{black} 0.002})\\ \cline{2-4} 
 & $c_{t \varphi}$ & { \color{black} 0.000}({\color{black} 0.000}) & { \color{black} 0.000}({\color{black} 0.000})\\ \cline{2-4} 
 & $c_{tZ}$ & { \color{black} 0.000}({\color{black} 0.000}) & { \color{black} 0.000}({\color{black} 0.000})\\ \cline{2-4} 
 & $c_{b \varphi}$ & { \color{black} 0.000}({\color{black} 0.000}) & { \color{black} 0.000}({\color{black} 0.000})\\ \cline{2-4} 
 & $c_{c \varphi}$ & { \color{black} 0.000}({\color{black} 0.000}) & { \color{black} 0.000}({\color{black} 0.000})\\ \cline{2-4} 
 & $c_{\tau \varphi}$ & { \color{black} 0.000}({\color{black} 0.000}) & { \color{black} 0.000}({\color{black} 0.000})\\ \hline
\multirow{14}{*}{2L2H}
 & $c_{qq}^{1,8}$ & { \color{black} 3.656}({\color{black} 1.068}) & { \color{blue} 29.418}({\color{blue} 10.975})\\ \cline{2-4} 
 & $c_{qq}^{1,1}$ & { \color{black} 6.186}({\color{black} -1.948}) & { \color{blue} 67.425}({\color{black} -0.434})\\ \cline{2-4} 
 & $c_{qq}^{8,3}$ & { \color{black} 5.417}({\color{black} -0.492}) & { \color{blue} 43.636}({\color{black} 2.502})\\ \cline{2-4} 
 & $c_{qq}^{1,3}$ & { \color{black} 0.005}({\color{black} -0.386}) & { \color{black} 0.063}({\color{black} -0.447})\\ \cline{2-4} 
 & $c_{qt}^{8}$ & { \color{black} 3.607}({\color{black} 1.915}) & { \color{blue} 23.810}({\color{black} 9.620})\\ \cline{2-4} 
 & $c_{qt}^{1}$ & { \color{black} 6.626}({\color{black} 2.269}) & { \color{blue} 71.322}({\color{blue} 10.718})\\ \cline{2-4} 
 & $c_{ut}^{8}$ & { \color{black} 5.775}({\color{black} 1.058}) & { \color{blue} 48.661}({\color{blue} 13.818})\\ \cline{2-4} 
 & $c_{ut}^{1}$ & { \color{black} 6.527}({\color{black} -2.794}) & { \color{blue} 73.039}({\color{black} -2.238})\\ \cline{2-4} 
 & $c_{qu}^{8}$ & { \color{black} 7.163}({\color{black} 3.653}) & { \color{blue} 51.370}({\color{blue} 18.553})\\ \cline{2-4} 
 & $c_{qu}^{1}$ & { \color{black} 4.991}({\color{black} 1.881}) & { \color{blue} 56.302}({\color{black} 8.139})\\ \cline{2-4} 
 & $c_{dt}^{8}$ & { \color{black} 4.623}({\color{black} -0.268}) & { \color{blue} 38.021}({\color{black} 4.517})\\ \cline{2-4} 
 & $c_{dt}^{1}$ & { \color{black} 5.293}({\color{black} -1.517}) & { \color{blue} 52.588}({\color{black} -2.034})\\ \cline{2-4} 
 & $c_{qd}^{8}$ & { \color{black} 4.445}({\color{black} 2.603}) & { \color{blue} 31.046}({\color{blue} 11.473})\\ \cline{2-4} 
 & $c_{qd}^{1}$ & { \color{black} 5.504}({\color{black} 2.325}) & { \color{blue} 55.378}({\color{black} 9.071})\\ \hline
\multirow{5}{*}{4H}
 & $c_{QQ}^{1}$ & { \color{black} 0.000}({\color{black} 0.000}) & { \color{black} 0.000}({\color{black} 0.000})\\ \cline{2-4} 
 & $c_{QQ}^{8}$ & { \color{black} 0.000}({\color{black} 0.000}) & { \color{black} 0.000}({\color{black} 0.000})\\ \cline{2-4} 
 & $c_{Qt}^{1}$ & { \color{black} 0.000}({\color{black} 0.000}) & { \color{black} 0.000}({\color{black} 0.000})\\ \cline{2-4} 
 & $c_{Qt}^{8}$ & { \color{black} 0.000}({\color{black} 0.000}) & { \color{black} 0.000}({\color{black} 0.000})\\ \cline{2-4} 
 & $c_{tt}^{1}$ & { \color{black} 0.000}({\color{black} 0.000}) & { \color{black} 0.000}({\color{black} 0.000})\\ \hline
\multirow{7}{*}{B}
 & $c_{\varphi G}$ & { \color{black} 0.000}({\color{black} 0.000}) & { \color{black} 0.000}({\color{black} 0.000})\\ \cline{2-4} 
 & $c_{\varphi B}$ & { \color{black} 0.000}({\color{black} 0.000}) & { \color{black} 0.000}({\color{black} 0.000})\\ \cline{2-4} 
 & $c_{\varphi W}$ & { \color{black} 0.000}({\color{black} 0.000}) & { \color{black} 0.000}({\color{black} 0.000})\\ \cline{2-4} 
 & $c_{\varphi WB}$ & { \color{black} 0.000}({\color{black} 0.000}) & { \color{black} 0.000}({\color{black} 0.000})\\ \cline{2-4} 
 & $c_{\varphi \Box}$ & { \color{black} 0.000}({\color{black} 0.000}) & { \color{black} 0.000}({\color{black} 0.000})\\ \cline{2-4} 
 & $c_{\varphi D}$ & { \color{black} 0.000}({\color{black} 0.000}) & { \color{black} 0.000}({\color{black} 0.000})\\ \cline{2-4} 
 & $c_{WWW}$ & { \color{black} 0.000}({\color{black} 0.000}) & { \color{black} 0.000}({\color{black} 0.000})\\ \hline
\end{tabular}
\caption{\small Same as Fig.~\ref{fig:FisherMatrix} now in a fine-grained version
      and restricted to the datasets measuring the charge asymmetry in top quark pair production.
      The number outside (inside) brackets corresponds to the value of the Fisher
      information matrix in the linear (quadratic) fit.
      See text for more details.
    }
\label{tab:FisherMatrix_AC}
\end{table}
\begin{table}
\scriptsize
\centering
\begin{tabular}{|c|c|c|c|c|c|}
\hline
 \multicolumn{2}{|c|}{} & \multicolumn{4}{c|}{Processes} \\ \hline
 Class & Coefficient & \rot{\href{https://arxiv.org/abs/1511.04716}{${\rm ATLAS\_tt\_8TeV\_ljets\_Mtt}$}} & \rot{\href{https://arxiv.org/abs/1607.07281}{${\rm ATLAS\_tt\_8TeV\_dilep\_Mtt}$}} & \rot{\href{https://arxiv.org/abs/1505.04480}{${\rm CMS\_tt\_8TeV\_ljets\_Ytt}$}} & \rot{\href{https://arxiv.org/abs/1703.01630}{${\rm CMS\_tt2D\_8TeV\_dilep\_MttYtt}$}} \\ \hline
\multirow{10}{*}{2FB}
 & $c_{\varphi Q}^{(-)}$ & { \color{black} 0.000}({\color{black} 0.000}) & { \color{black} 0.000}({\color{black} 0.000}) & { \color{black} 0.000}({\color{black} 0.000}) & { \color{black} 0.000}({\color{black} 0.000})\\ \cline{2-6} 
 & $c_{\varphi Q}^{3}$ & { \color{black} 0.000}({\color{black} 0.000}) & { \color{black} 0.000}({\color{black} 0.000}) & { \color{black} 0.000}({\color{black} 0.000}) & { \color{black} 0.000}({\color{black} 0.000})\\ \cline{2-6} 
 & $c_{\varphi t}$ & { \color{black} 0.000}({\color{black} 0.000}) & { \color{black} 0.000}({\color{black} 0.000}) & { \color{black} 0.000}({\color{black} 0.000}) & { \color{black} 0.000}({\color{black} 0.000})\\ \cline{2-6} 
 & $c_{tW}$ & { \color{black} 0.000}({\color{black} 0.000}) & { \color{black} 0.000}({\color{black} 0.000}) & { \color{black} 0.000}({\color{black} 0.000}) & { \color{black} 0.000}({\color{black} 0.000})\\ \cline{2-6} 
 & $c_{tG}$ & { \color{black} 1.002}({\color{black} 0.976}) & { \color{black} 3.025}({\color{black} 3.120}) & { \color{blue} 17.981}({\color{blue} 18.770}) & { \color{blue} 10.349}({\color{blue} 10.159})\\ \cline{2-6} 
 & $c_{t \varphi}$ & { \color{black} 0.000}({\color{black} 0.000}) & { \color{black} 0.000}({\color{black} 0.000}) & { \color{black} 0.000}({\color{black} 0.000}) & { \color{black} 0.000}({\color{black} 0.000})\\ \cline{2-6} 
 & $c_{tZ}$ & { \color{black} 0.000}({\color{black} 0.000}) & { \color{black} 0.000}({\color{black} 0.000}) & { \color{black} 0.000}({\color{black} 0.000}) & { \color{black} 0.000}({\color{black} 0.000})\\ \cline{2-6} 
 & $c_{b \varphi}$ & { \color{black} 0.000}({\color{black} 0.000}) & { \color{black} 0.000}({\color{black} 0.000}) & { \color{black} 0.000}({\color{black} 0.000}) & { \color{black} 0.000}({\color{black} 0.000})\\ \cline{2-6} 
 & $c_{c \varphi}$ & { \color{black} 0.000}({\color{black} 0.000}) & { \color{black} 0.000}({\color{black} 0.000}) & { \color{black} 0.000}({\color{black} 0.000}) & { \color{black} 0.000}({\color{black} 0.000})\\ \cline{2-6} 
 & $c_{\tau \varphi}$ & { \color{black} 0.000}({\color{black} 0.000}) & { \color{black} 0.000}({\color{black} 0.000}) & { \color{black} 0.000}({\color{black} 0.000}) & { \color{black} 0.000}({\color{black} 0.000})\\ \hline
\multirow{14}{*}{2L2H}
 & $c_{qq}^{1,8}$ & { \color{black} 0.989}({\color{black} 0.116}) & { \color{black} 0.910}({\color{black} 0.499}) & { \color{black} 3.939}({\color{black} 2.715}) & { \color{black} 2.907}({\color{black} 0.308})\\ \cline{2-6} 
 & $c_{qq}^{1,1}$ & { \color{black} 0.746}({\color{black} -1.338}) & { \color{black} 0.809}({\color{black} 0.769}) & { \color{black} 0.594}({\color{black} 5.602}) & { \color{black} 2.204}({\color{black} -4.524})\\ \cline{2-6} 
 & $c_{qq}^{8,3}$ & { \color{black} 0.568}({\color{black} -0.514}) & { \color{black} 0.552}({\color{black} 0.335}) & { \color{black} 2.638}({\color{black} 2.356}) & { \color{black} 1.823}({\color{black} -1.572})\\ \cline{2-6} 
 & $c_{qq}^{1,3}$ & { \color{black} 0.000}({\color{black} -0.271}) & { \color{black} 0.000}({\color{black} 0.440}) & { \color{black} 0.002}({\color{black} 1.051}) & { \color{black} 0.005}({\color{black} -0.862})\\ \cline{2-6} 
 & $c_{qt}^{8}$ & { \color{black} 0.968}({\color{black} -0.029}) & { \color{black} 1.177}({\color{black} 0.662}) & { \color{black} 6.098}({\color{black} 4.034}) & { \color{black} 3.923}({\color{black} 0.290})\\ \cline{2-6} 
 & $c_{qt}^{1}$ & { \color{black} 1.008}({\color{black} -1.219}) & { \color{black} 0.840}({\color{black} 0.749}) & { \color{black} 0.584}({\color{black} 5.045}) & { \color{black} 0.722}({\color{black} -4.150})\\ \cline{2-6} 
 & $c_{ut}^{8}$ & { \color{black} 1.324}({\color{black} -0.321}) & { \color{black} 1.260}({\color{black} 0.987}) & { \color{black} 4.812}({\color{black} 5.256}) & { \color{black} 3.462}({\color{black} -0.637})\\ \cline{2-6} 
 & $c_{ut}^{1}$ & { \color{black} 0.466}({\color{black} -1.945}) & { \color{black} 0.594}({\color{black} 1.070}) & { \color{black} 0.815}({\color{black} 8.012}) & { \color{black} 1.001}({\color{black} -5.962})\\ \cline{2-6} 
 & $c_{qu}^{8}$ & { \color{black} 1.741}({\color{black} -0.569}) & { \color{black} 1.890}({\color{black} 1.267}) & { \color{blue} 10.610}({\color{black} 7.625}) & { \color{black} 6.911}({\color{black} -0.784})\\ \cline{2-6} 
 & $c_{qu}^{1}$ & { \color{black} 0.516}({\color{black} -1.112}) & { \color{black} 0.495}({\color{black} 0.704}) & { \color{black} 0.870}({\color{black} 4.351}) & { \color{black} 1.406}({\color{black} -3.318})\\ \cline{2-6} 
 & $c_{dt}^{8}$ & { \color{black} 1.722}({\color{black} -0.370}) & { \color{black} 1.774}({\color{black} 0.678}) & { \color{black} 7.789}({\color{black} 3.768}) & { \color{black} 5.872}({\color{black} -1.217})\\ \cline{2-6} 
 & $c_{dt}^{1}$ & { \color{black} 0.803}({\color{black} -1.097}) & { \color{black} 0.713}({\color{black} 0.770}) & { \color{black} 2.627}({\color{black} 4.482}) & { \color{black} 2.696}({\color{black} -4.220})\\ \cline{2-6} 
 & $c_{qd}^{8}$ & { \color{black} 2.681}({\color{black} -0.397}) & { \color{black} 3.058}({\color{black} 2.103}) & { \color{blue} 13.187}({\color{black} 7.253}) & { \color{black} 8.826}({\color{black} -1.971})\\ \cline{2-6} 
 & $c_{qd}^{1}$ & { \color{black} 0.610}({\color{black} -1.607}) & { \color{black} 0.692}({\color{black} 0.845}) & { \color{black} 1.493}({\color{black} 5.918}) & { \color{black} 2.854}({\color{black} -5.613})\\ \hline
\multirow{5}{*}{4H}
 & $c_{QQ}^{1}$ & { \color{black} 0.000}({\color{black} 0.000}) & { \color{black} 0.000}({\color{black} 0.000}) & { \color{black} 0.000}({\color{black} 0.000}) & { \color{black} 0.000}({\color{black} 0.000})\\ \cline{2-6} 
 & $c_{QQ}^{8}$ & { \color{black} 0.000}({\color{black} 0.000}) & { \color{black} 0.000}({\color{black} 0.000}) & { \color{black} 0.000}({\color{black} 0.000}) & { \color{black} 0.000}({\color{black} 0.000})\\ \cline{2-6} 
 & $c_{Qt}^{1}$ & { \color{black} 0.000}({\color{black} 0.000}) & { \color{black} 0.000}({\color{black} 0.000}) & { \color{black} 0.000}({\color{black} 0.000}) & { \color{black} 0.000}({\color{black} 0.000})\\ \cline{2-6} 
 & $c_{Qt}^{8}$ & { \color{black} 0.000}({\color{black} 0.000}) & { \color{black} 0.000}({\color{black} 0.000}) & { \color{black} 0.000}({\color{black} 0.000}) & { \color{black} 0.000}({\color{black} 0.000})\\ \cline{2-6} 
 & $c_{tt}^{1}$ & { \color{black} 0.000}({\color{black} 0.000}) & { \color{black} 0.000}({\color{black} 0.000}) & { \color{black} 0.000}({\color{black} 0.000}) & { \color{black} 0.000}({\color{black} 0.000})\\ \hline
\multirow{7}{*}{B}
 & $c_{\varphi G}$ & { \color{black} 0.000}({\color{black} 0.000}) & { \color{black} 0.000}({\color{black} 0.000}) & { \color{black} 0.000}({\color{black} 0.000}) & { \color{black} 0.000}({\color{black} 0.000})\\ \cline{2-6} 
 & $c_{\varphi B}$ & { \color{black} 0.000}({\color{black} 0.000}) & { \color{black} 0.000}({\color{black} 0.000}) & { \color{black} 0.000}({\color{black} 0.000}) & { \color{black} 0.000}({\color{black} 0.000})\\ \cline{2-6} 
 & $c_{\varphi W}$ & { \color{black} 0.000}({\color{black} 0.000}) & { \color{black} 0.000}({\color{black} 0.000}) & { \color{black} 0.000}({\color{black} 0.000}) & { \color{black} 0.000}({\color{black} 0.000})\\ \cline{2-6} 
 & $c_{\varphi WB}$ & { \color{black} 0.000}({\color{black} 0.000}) & { \color{black} 0.000}({\color{black} 0.000}) & { \color{black} 0.000}({\color{black} 0.000}) & { \color{black} 0.000}({\color{black} 0.000})\\ \cline{2-6} 
 & $c_{\varphi \Box}$ & { \color{black} 0.000}({\color{black} 0.000}) & { \color{black} 0.000}({\color{black} 0.000}) & { \color{black} 0.000}({\color{black} 0.000}) & { \color{black} 0.000}({\color{black} 0.000})\\ \cline{2-6} 
 & $c_{\varphi D}$ & { \color{black} 0.000}({\color{black} 0.000}) & { \color{black} 0.000}({\color{black} 0.000}) & { \color{black} 0.000}({\color{black} 0.000}) & { \color{black} 0.000}({\color{black} 0.000})\\ \cline{2-6} 
 & $c_{WWW}$ & { \color{black} 0.000}({\color{black} 0.000}) & { \color{black} 0.000}({\color{black} 0.000}) & { \color{black} 0.000}({\color{black} 0.000}) & { \color{black} 0.000}({\color{black} 0.000})\\ \hline
\end{tabular}
\caption{\small Same as Tab.~\ref{tab:FisherMatrix_AC} now for differential parton-level distributions in inclusive top quark pair production datasets at $\sqrt{s}=8\ TeV$.
}
\label{tab:FisherMatrix_tt8}
\end{table}
\begin{table}
\scriptsize
\centering
\begin{tabular}{|c|c|c|c|c|c|c|}
\hline
 \multicolumn{2}{|c|}{} & \multicolumn{5}{c|}{Processes} \\ \hline
 Class & Coefficient & \rot{\href{https://arxiv.org/abs/1610.04191}{${\rm CMS\_tt\_13TeV\_ljets\_2015\_Mtt}$}}& \rot{\href{https://arxiv.org/abs/1708.07638}{${\rm CMS\_tt\_13TeV\_dilep\_2015\_Mtt}$}} & \rot{\href{https://arxiv.org/abs/1803.08856}{${\rm CMS\_tt\_13TeV\_ljets\_2016\_Mtt}$}} & \rot{\href{https://arxiv.org/abs/1811.06625}{${\rm CMS\_tt\_13TeV\_dilep\_2016\_Mtt}$}} & \rot{\href{https://arxiv.org/abs/1908.07305}{${\rm ATLAS\_tt\_13TeV\_ljets\_2016\_Mtt}$}} \\ \hline
\multirow{10}{*}{2FB}
 & $c_{\varphi Q}^{(-)}$ & { \color{black} 0.000}({\color{black} 0.000}) & { \color{black} 0.000}({\color{black} 0.000}) & { \color{black} 0.000}({\color{black} 0.000}) & { \color{black} 0.000}({\color{black} 0.000}) & { \color{black} 0.000}({\color{black} 0.000})\\ \cline{2-7} 
 & $c_{\varphi Q}^{3}$ & { \color{black} 0.000}({\color{black} 0.000}) & { \color{black} 0.000}({\color{black} 0.000}) & { \color{black} 0.000}({\color{black} 0.000}) & { \color{black} 0.000}({\color{black} 0.000}) & { \color{black} 0.000}({\color{black} 0.000})\\ \cline{2-7} 
 & $c_{\varphi t}$ & { \color{black} 0.000}({\color{black} 0.000}) & { \color{black} 0.000}({\color{black} 0.000}) & { \color{black} 0.000}({\color{black} 0.000}) & { \color{black} 0.000}({\color{black} 0.000}) & { \color{black} 0.000}({\color{black} 0.000})\\ \cline{2-7} 
 & $c_{tW}$ & { \color{black} 0.000}({\color{black} 0.000}) & { \color{black} 0.000}({\color{black} 0.000}) & { \color{black} 0.000}({\color{black} 0.000}) & { \color{black} 0.000}({\color{black} 0.000}) & { \color{black} 0.000}({\color{black} 0.000})\\ \cline{2-7} 
 & $c_{tG}$ & { \color{black} 0.939}({\color{black} 1.078}) & { \color{black} 1.423}({\color{black} 1.579}) & { \color{black} 3.431}({\color{black} 4.254}) & { \color{black} 2.072}({\color{black} 2.486}) & { \color{black} 1.207}({\color{black} 1.349})\\ \cline{2-7} 
 & $c_{t \varphi}$ & { \color{black} 0.000}({\color{black} 0.000}) & { \color{black} 0.000}({\color{black} 0.000}) & { \color{black} 0.000}({\color{black} 0.000}) & { \color{black} 0.000}({\color{black} 0.000}) & { \color{black} 0.000}({\color{black} 0.000})\\ \cline{2-7} 
 & $c_{tZ}$ & { \color{black} 0.000}({\color{black} 0.000}) & { \color{black} 0.000}({\color{black} 0.000}) & { \color{black} 0.000}({\color{black} 0.000}) & { \color{black} 0.000}({\color{black} 0.000}) & { \color{black} 0.000}({\color{black} 0.000})\\ \cline{2-7} 
 & $c_{b \varphi}$ & { \color{black} 0.000}({\color{black} 0.000}) & { \color{black} 0.000}({\color{black} 0.000}) & { \color{black} 0.000}({\color{black} 0.000}) & { \color{black} 0.000}({\color{black} 0.000}) & { \color{black} 0.000}({\color{black} 0.000})\\ \cline{2-7} 
 & $c_{c \varphi}$ & { \color{black} 0.000}({\color{black} 0.000}) & { \color{black} 0.000}({\color{black} 0.000}) & { \color{black} 0.000}({\color{black} 0.000}) & { \color{black} 0.000}({\color{black} 0.000}) & { \color{black} 0.000}({\color{black} 0.000})\\ \cline{2-7} 
 & $c_{\tau \varphi}$ & { \color{black} 0.000}({\color{black} 0.000}) & { \color{black} 0.000}({\color{black} 0.000}) & { \color{black} 0.000}({\color{black} 0.000}) & { \color{black} 0.000}({\color{black} 0.000}) & { \color{black} 0.000}({\color{black} 0.000})\\ \hline
\multirow{14}{*}{2L2H}
 & $c_{qq}^{1,8}$ & { \color{black} 1.003}({\color{black} 1.973}) & { \color{black} 0.333}({\color{black} 0.560}) & { \color{black} 8.035}({\color{blue} 16.655}) & { \color{blue} 13.744}({\color{blue} 39.714}) & { \color{black} 7.522}({\color{blue} 14.983})\\ \cline{2-7} 
 & $c_{qq}^{1,1}$ & { \color{black} 1.441}({\color{black} 6.796}) & { \color{black} 0.227}({\color{black} 2.201}) & { \color{black} 9.698}({\color{blue} 52.347}) & { \color{black} 6.597}({\color{blue} 34.891}) & { \color{black} 3.678}({\color{black} 3.338})\\ \cline{2-7} 
 & $c_{qq}^{8,3}$ & { \color{black} 0.430}({\color{black} 2.986}) & { \color{black} 0.130}({\color{black} 0.874}) & { \color{black} 2.441}({\color{blue} 37.461}) & { \color{black} 0.937}({\color{blue} 18.134}) & { \color{black} 0.564}({\color{black} 3.590})\\ \cline{2-7} 
 & $c_{qq}^{1,3}$ & { \color{black} 0.001}({\color{black} 1.243}) & { \color{black} 0.000}({\color{black} 0.420}) & { \color{black} 0.007}({\color{black} 7.540}) & { \color{black} 0.003}({\color{black} 4.673}) & { \color{black} 0.002}({\color{black} -0.152})\\ \cline{2-7} 
 & $c_{qt}^{8}$ & { \color{black} 1.095}({\color{black} 3.119}) & { \color{black} 0.401}({\color{black} 0.779}) & { \color{blue} 12.938}({\color{blue} 43.148}) & { \color{black} 3.043}({\color{blue} 16.920}) & { \color{black} 1.733}({\color{black} 4.893})\\ \cline{2-7} 
 & $c_{qt}^{1}$ & { \color{black} 1.148}({\color{black} 5.566}) & { \color{black} 0.215}({\color{black} 1.966}) & { \color{blue} 10.525}({\color{blue} 36.109}) & { \color{black} 4.370}({\color{blue} 36.495}) & { \color{black} 2.456}({\color{black} 5.531})\\ \cline{2-7} 
 & $c_{ut}^{8}$ & { \color{black} 1.407}({\color{black} 5.182}) & { \color{black} 0.377}({\color{black} 1.296}) & { \color{blue} 17.865}({\color{blue} 40.618}) & { \color{black} 7.932}({\color{blue} 27.325}) & { \color{black} 4.398}({\color{black} 3.189})\\ \cline{2-7} 
 & $c_{ut}^{1}$ & { \color{black} 0.878}({\color{black} 8.309}) & { \color{black} 0.154}({\color{black} 3.035}) & { \color{black} 6.966}({\color{blue} 44.848}) & { \color{black} 5.931}({\color{blue} 40.126}) & { \color{black} 3.270}({\color{black} 3.230})\\ \cline{2-7} 
 & $c_{qu}^{8}$ & { \color{black} 1.060}({\color{black} 4.667}) & { \color{black} 0.670}({\color{black} 1.851}) & { \color{black} 4.893}({\color{blue} 43.891}) & { \color{black} 2.067}({\color{blue} 12.474}) & { \color{black} 1.293}({\color{black} 2.146})\\ \cline{2-7} 
 & $c_{qu}^{1}$ & { \color{black} 1.523}({\color{black} 3.869}) & { \color{black} 0.143}({\color{black} 1.609}) & { \color{blue} 20.696}({\color{blue} 48.683}) & { \color{black} 8.364}({\color{blue} 26.455}) & { \color{black} 4.580}({\color{black} 6.013})\\ \cline{2-7} 
 & $c_{dt}^{8}$ & { \color{black} 2.273}({\color{black} 4.198}) & { \color{black} 0.779}({\color{black} 1.207}) & { \color{blue} 17.028}({\color{blue} 44.740}) & { \color{black} 9.665}({\color{blue} 31.876}) & { \color{black} 5.477}({\color{black} 9.261})\\ \cline{2-7} 
 & $c_{dt}^{1}$ & { \color{black} 1.816}({\color{black} 6.135}) & { \color{black} 0.339}({\color{black} 1.948}) & { \color{blue} 16.044}({\color{blue} 53.957}) & { \color{blue} 10.761}({\color{blue} 35.154}) & { \color{black} 5.924}({\color{black} 4.508})\\ \cline{2-7} 
 & $c_{qd}^{8}$ & { \color{black} 1.601}({\color{black} 3.070}) & { \color{black} 1.207}({\color{black} 2.155}) & { \color{black} 6.384}({\color{blue} 32.430}) & { \color{black} 7.078}({\color{blue} 29.956}) & { \color{black} 4.157}({\color{black} 5.835})\\ \cline{2-7} 
 & $c_{qd}^{1}$ & { \color{black} 2.438}({\color{black} 6.856}) & { \color{black} 0.410}({\color{black} 2.536}) & { \color{blue} 18.526}({\color{blue} 37.766}) & { \color{black} 7.595}({\color{blue} 36.886}) & { \color{black} 4.297}({\color{black} 2.158})\\ \hline
\multirow{5}{*}{4H}
 & $c_{QQ}^{1}$ & { \color{black} 0.000}({\color{black} 0.000}) & { \color{black} 0.000}({\color{black} 0.000}) & { \color{black} 0.000}({\color{black} 0.000}) & { \color{black} 0.000}({\color{black} 0.000}) & { \color{black} 0.000}({\color{black} 0.000})\\ \cline{2-7} 
 & $c_{QQ}^{8}$ & { \color{black} 0.000}({\color{black} 0.000}) & { \color{black} 0.000}({\color{black} 0.000}) & { \color{black} 0.000}({\color{black} 0.000}) & { \color{black} 0.000}({\color{black} 0.000}) & { \color{black} 0.000}({\color{black} 0.000})\\ \cline{2-7} 
 & $c_{Qt}^{1}$ & { \color{black} 0.000}({\color{black} 0.000}) & { \color{black} 0.000}({\color{black} 0.000}) & { \color{black} 0.000}({\color{black} 0.000}) & { \color{black} 0.000}({\color{black} 0.000}) & { \color{black} 0.000}({\color{black} 0.000})\\ \cline{2-7} 
 & $c_{Qt}^{8}$ & { \color{black} 0.000}({\color{black} 0.000}) & { \color{black} 0.000}({\color{black} 0.000}) & { \color{black} 0.000}({\color{black} 0.000}) & { \color{black} 0.000}({\color{black} 0.000}) & { \color{black} 0.000}({\color{black} 0.000})\\ \cline{2-7} 
 & $c_{tt}^{1}$ & { \color{black} 0.000}({\color{black} 0.000}) & { \color{black} 0.000}({\color{black} 0.000}) & { \color{black} 0.000}({\color{black} 0.000}) & { \color{black} 0.000}({\color{black} 0.000}) & { \color{black} 0.000}({\color{black} 0.000})\\ \hline
\multirow{7}{*}{B}
 & $c_{\varphi G}$ & { \color{black} 0.000}({\color{black} 0.000}) & { \color{black} 0.000}({\color{black} 0.000}) & { \color{black} 0.000}({\color{black} 0.000}) & { \color{black} 0.000}({\color{black} 0.000}) & { \color{black} 0.000}({\color{black} 0.000})\\ \cline{2-7} 
 & $c_{\varphi B}$ & { \color{black} 0.000}({\color{black} 0.000}) & { \color{black} 0.000}({\color{black} 0.000}) & { \color{black} 0.000}({\color{black} 0.000}) & { \color{black} 0.000}({\color{black} 0.000}) & { \color{black} 0.000}({\color{black} 0.000})\\ \cline{2-7} 
 & $c_{\varphi W}$ & { \color{black} 0.000}({\color{black} 0.000}) & { \color{black} 0.000}({\color{black} 0.000}) & { \color{black} 0.000}({\color{black} 0.000}) & { \color{black} 0.000}({\color{black} 0.000}) & { \color{black} 0.000}({\color{black} 0.000})\\ \cline{2-7} 
 & $c_{\varphi WB}$ & { \color{black} 0.000}({\color{black} 0.000}) & { \color{black} 0.000}({\color{black} 0.000}) & { \color{black} 0.000}({\color{black} 0.000}) & { \color{black} 0.000}({\color{black} 0.000}) & { \color{black} 0.000}({\color{black} 0.000})\\ \cline{2-7} 
 & $c_{\varphi \Box}$ & { \color{black} 0.000}({\color{black} 0.000}) & { \color{black} 0.000}({\color{black} 0.000}) & { \color{black} 0.000}({\color{black} 0.000}) & { \color{black} 0.000}({\color{black} 0.000}) & { \color{black} 0.000}({\color{black} 0.000})\\ \cline{2-7} 
 & $c_{\varphi D}$ & { \color{black} 0.000}({\color{black} 0.000}) & { \color{black} 0.000}({\color{black} 0.000}) & { \color{black} 0.000}({\color{black} 0.000}) & { \color{black} 0.000}({\color{black} 0.000}) & { \color{black} 0.000}({\color{black} 0.000})\\ \cline{2-7} 
 & $c_{WWW}$ & { \color{black} 0.000}({\color{black} 0.000}) & { \color{black} 0.000}({\color{black} 0.000}) & { \color{black} 0.000}({\color{black} 0.000}) & { \color{black} 0.000}({\color{black} 0.000}) & { \color{black} 0.000}({\color{black} 0.000})\\ \hline
\end{tabular}
\caption{\small Same as Tab.~\ref{tab:FisherMatrix_AC} now for differential parton-level distributions in inclusive top quark pair production datasets at $\sqrt{s}=13\ TeV$.
}
\label{tab:FisherMatrix_tt13}
\end{table}


\clearpage

\subsection{Impact of measurements on individual constraints (Fitmaker)}

The linear dependence maps from measurements to Wilson coefficients provide important information to understand the result of a fit. However, the final constraint on a Wilson coefficient depends not only on how strongly a measurement varies with the corresponding coefficient but also on the sensitivity of those measurements. Some information in this direction is thus provide by the Fisher information tables shown in the previous section.  This can be further quantified by extracting individual constraints obtained by switching on operators one by one in the fit while setting all others to zero. 

\begin{figure}
    \centering
    \includegraphics[width=0.8\textwidth]{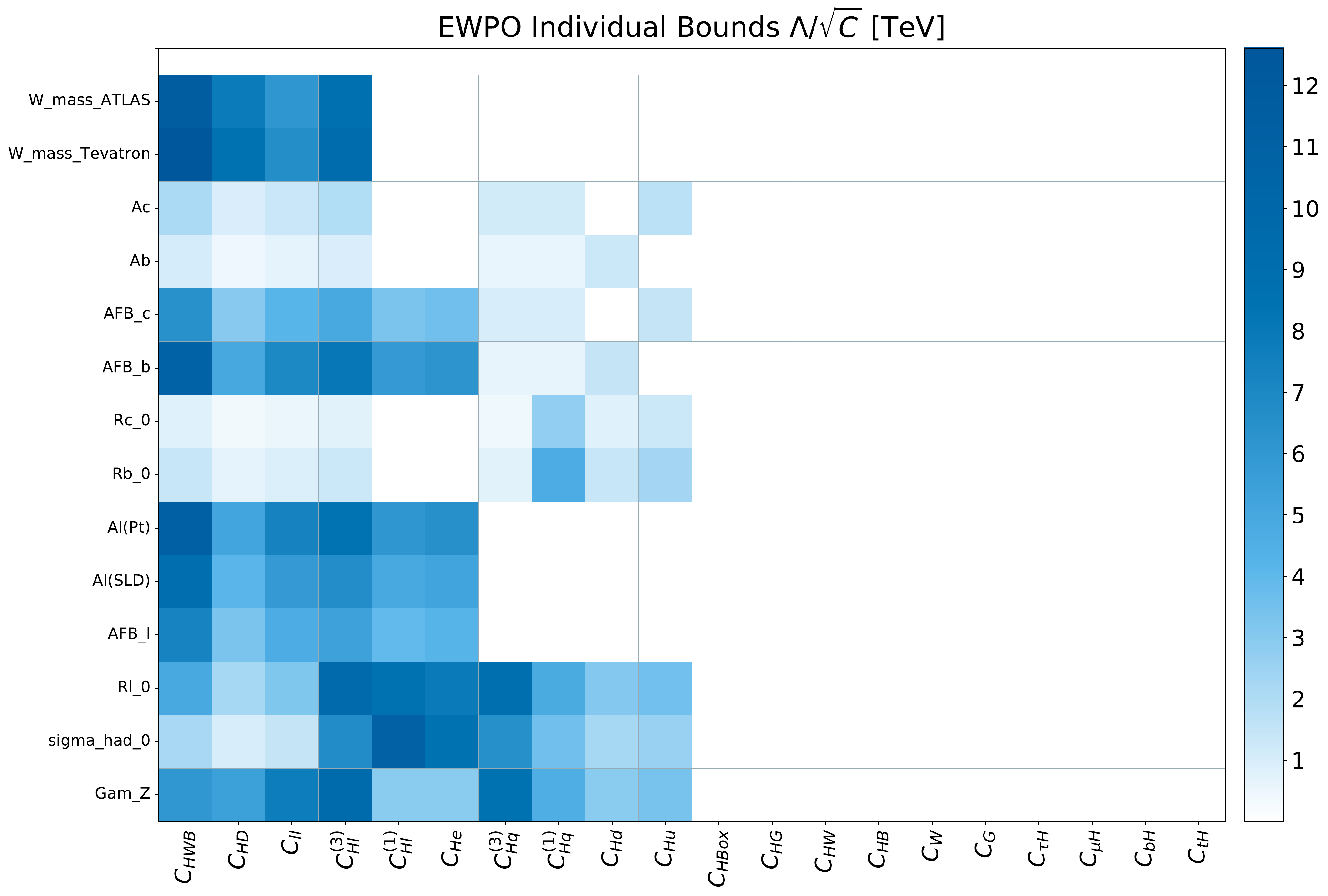}
    \caption{95\% CL individual limits from electroweak measurements. }
    \label{fig:EWPO_constraints}
\end{figure}

In Fig.~\ref{fig:EWPO_constraints} we show a table of the 95\% CL individual limits for electroweak measurements in each row and the corresponding constraint on a Wilson coefficient in each row. The limit is interpreted in terms of a scale $\Lambda/\sqrt{C_i}$ in units of TeV, colour coded in blue. We see that the bounds extend up to $\Lambda \sim 12$ TeV for $C_i \sim 1$, though this would be lower by a loop factor for loop-induced new physics and further lowered by a weak coupling. By picking out the darkest blue squares across a column, we can easily identify the measurement responsible for setting the strongest constraint on a particular operator coefficient. For example, $C_{Hl}^{(3)}$ and $C_{Hq}^{(3)}$ obtain some of their strongest constraints from $\Gamma_Z$, as expected from the fact that this measurement has a large linear dependence on them. 

\begin{figure}
    \centering
    \includegraphics[width=0.8\textwidth]{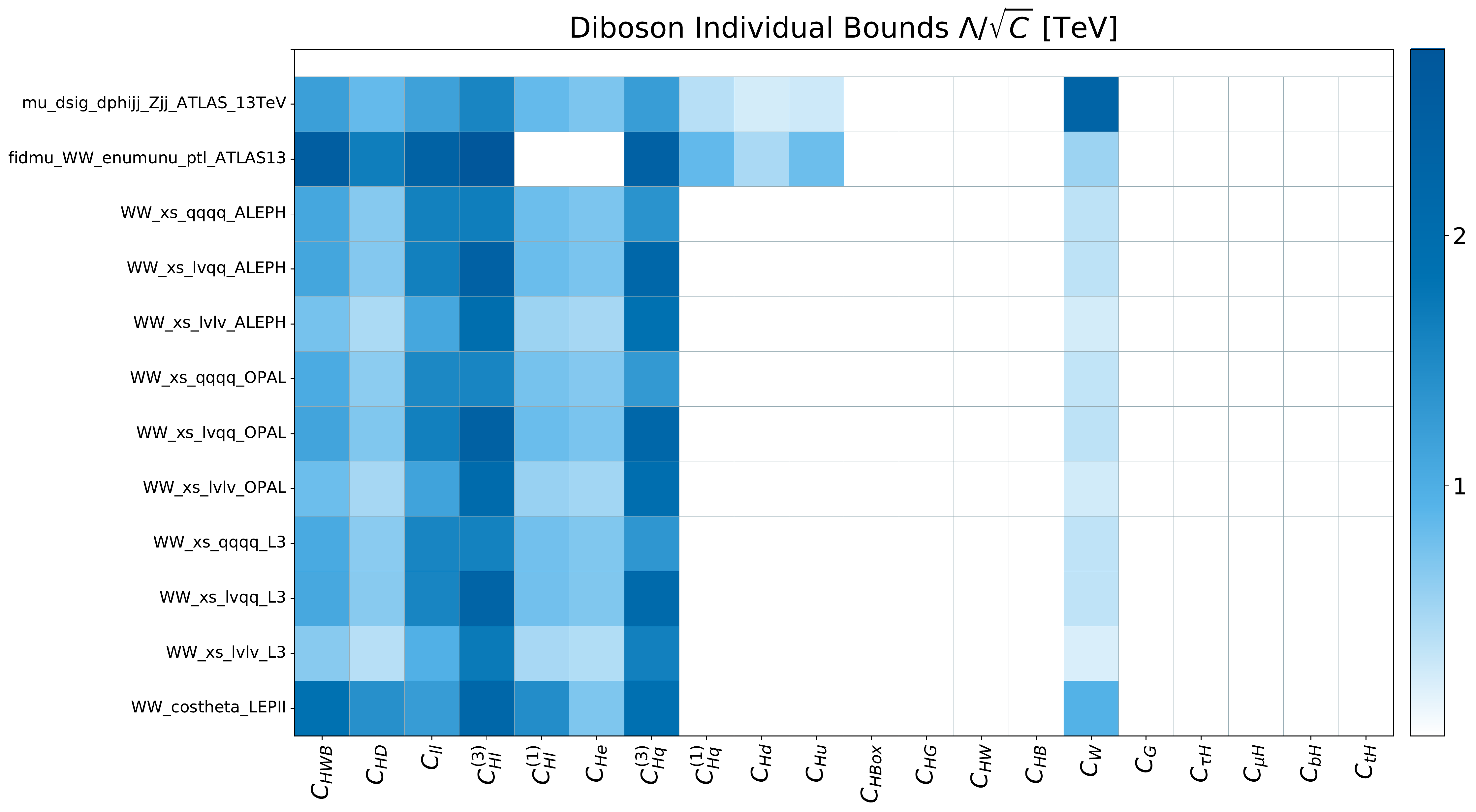}
    \caption{95\% CL individual limits from diboson measurements.}
    \label{fig:diboson_constraints}
\end{figure}

Fig.~\ref{fig:diboson_constraints} shows the individual bounds from diboson measurements. The scale sensitivity here is limited to about $2$ TeV, though this could be increased for new physics that is more strongly coupled. We see here the importance of the $Zjj$ measurement in constraining $C_W$. This provides the highest scale sensitivity, reflecting the linear dependence for $C_W$ in Fig.~\ref{fig:diboson} showing a larger relative linear dependence here than in $WW$ (note though that the $a_i$'s are normalised to 1 across each row). 

\begin{figure}
    \centering
    \includegraphics[width=0.8\textwidth]{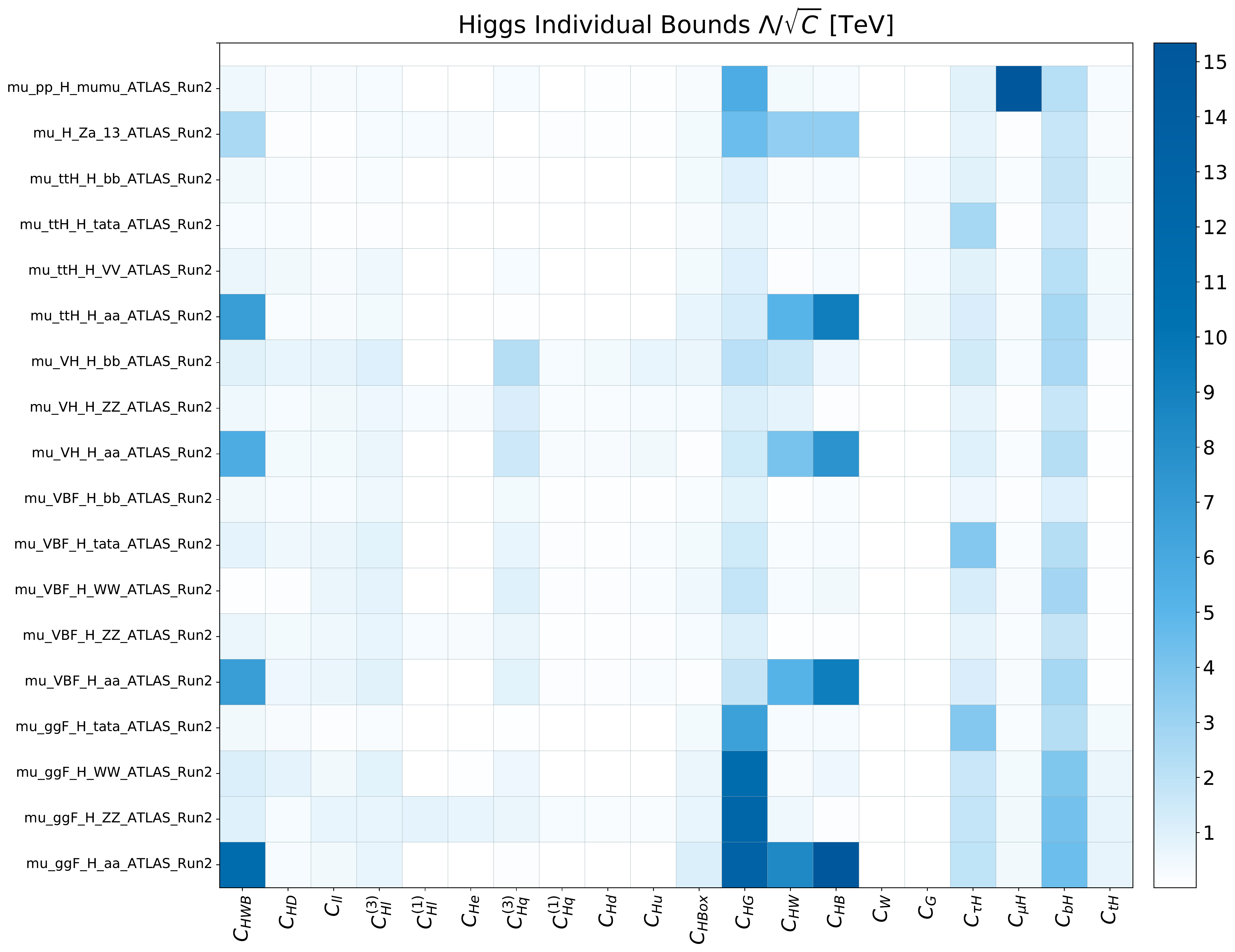}
    \caption{95\% CL individual limits from Higgs measurements.}
    \label{fig:Higgs_constraints}
\end{figure}

\begin{figure}
    \centering
    \includegraphics[width=0.8\textwidth]{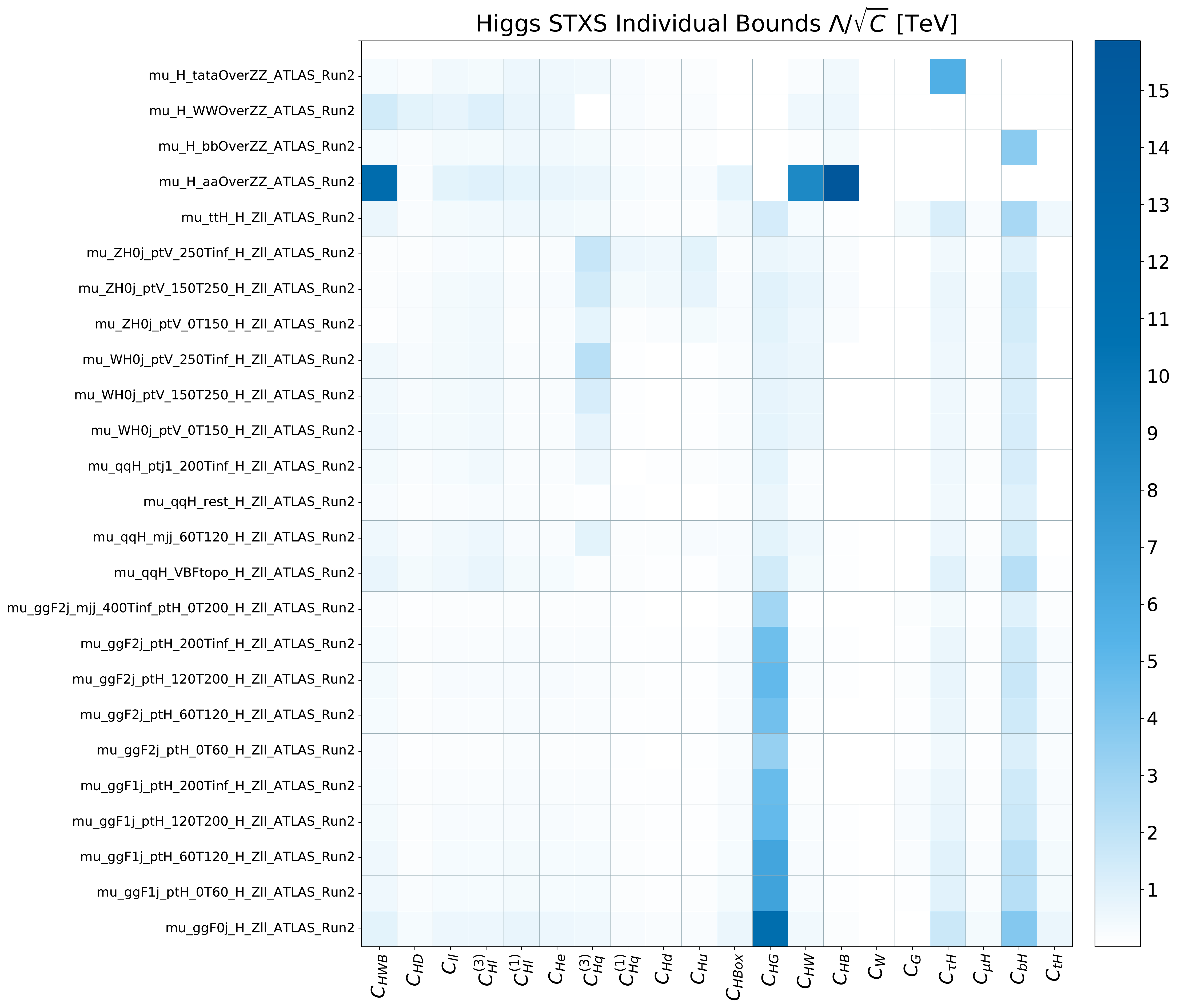}
    \caption{95\% CL individual limits from Higgs STXS measurements.}
    \label{fig:STXS_constraints}
\end{figure}

The Higgs signal strengths and STXS bounds in Figs.~\ref{fig:Higgs_constraints} and \ref{fig:STXS_constraints}, respectively, demonstrate a sensitivity to scales above 10 TeV (again, this would be lowered for weakly coupled and loop-induced new physics, while the $C_{\mu H}$ operator should be lowered by a tiny muon Yukawa factor due to this particular operator's normalisation). However, we see that the operator coefficients in the first half of the table columns are sensitive to much lower scales than the sensitivity coming from electroweak precision measurements, with the exception of $C_{HWB}$. For the operators most relevant for the Higgs in the second half of the table, the STXS constraints are also less sensitive than those coming from Higgs signal strengths. The STXS measurements are more important for marginalised limits where all operator coefficients are allowed to vary simultaneously.

\clearpage
\subsection{Global sensitivity from dataset variations (SMEFiT)}

\begin{figure}[t]
  \begin{center}
    \includegraphics[width=0.8\linewidth]{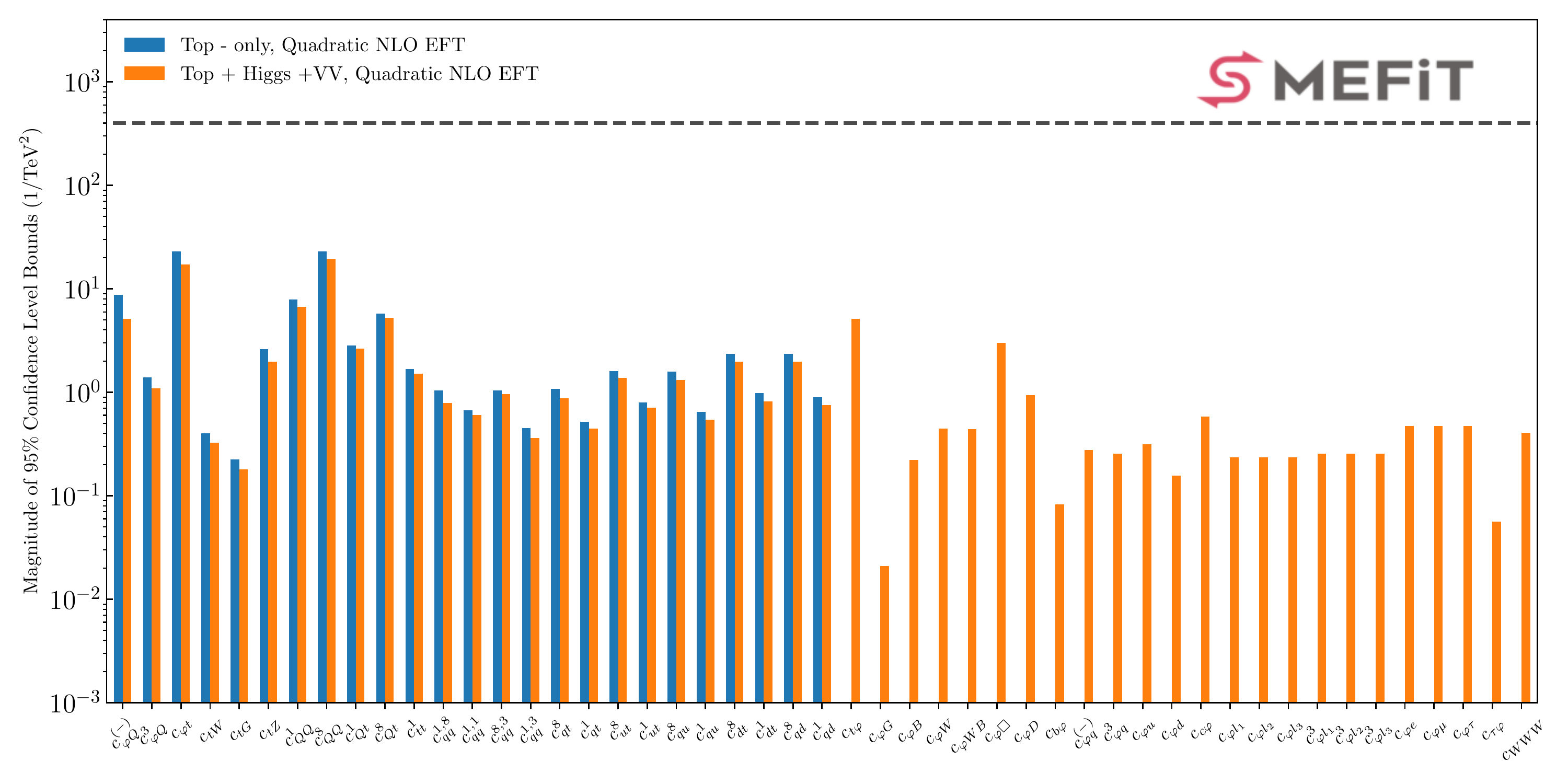}
    \includegraphics[width=0.8\linewidth]{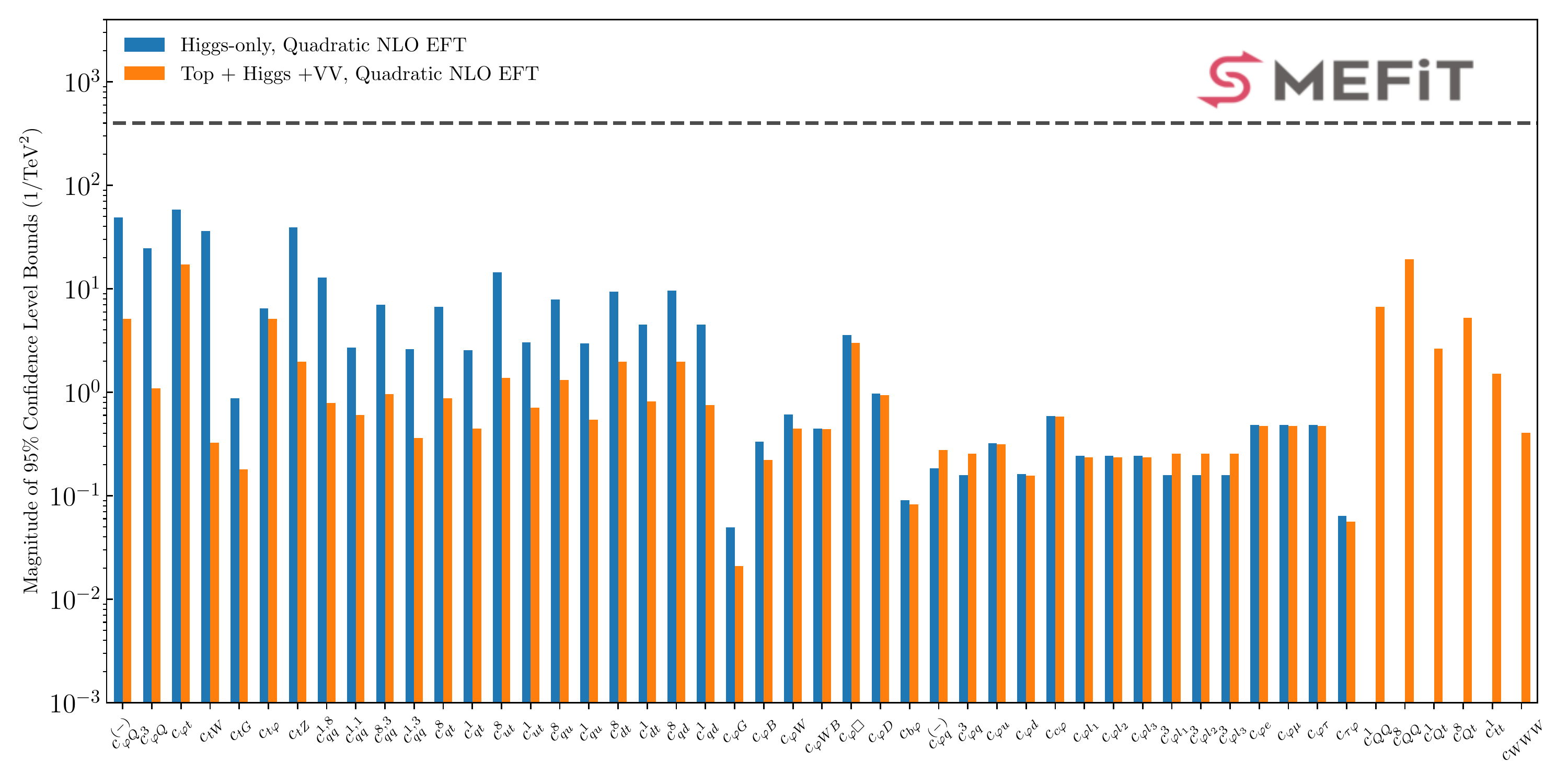}
     \includegraphics[width=0.8\linewidth]{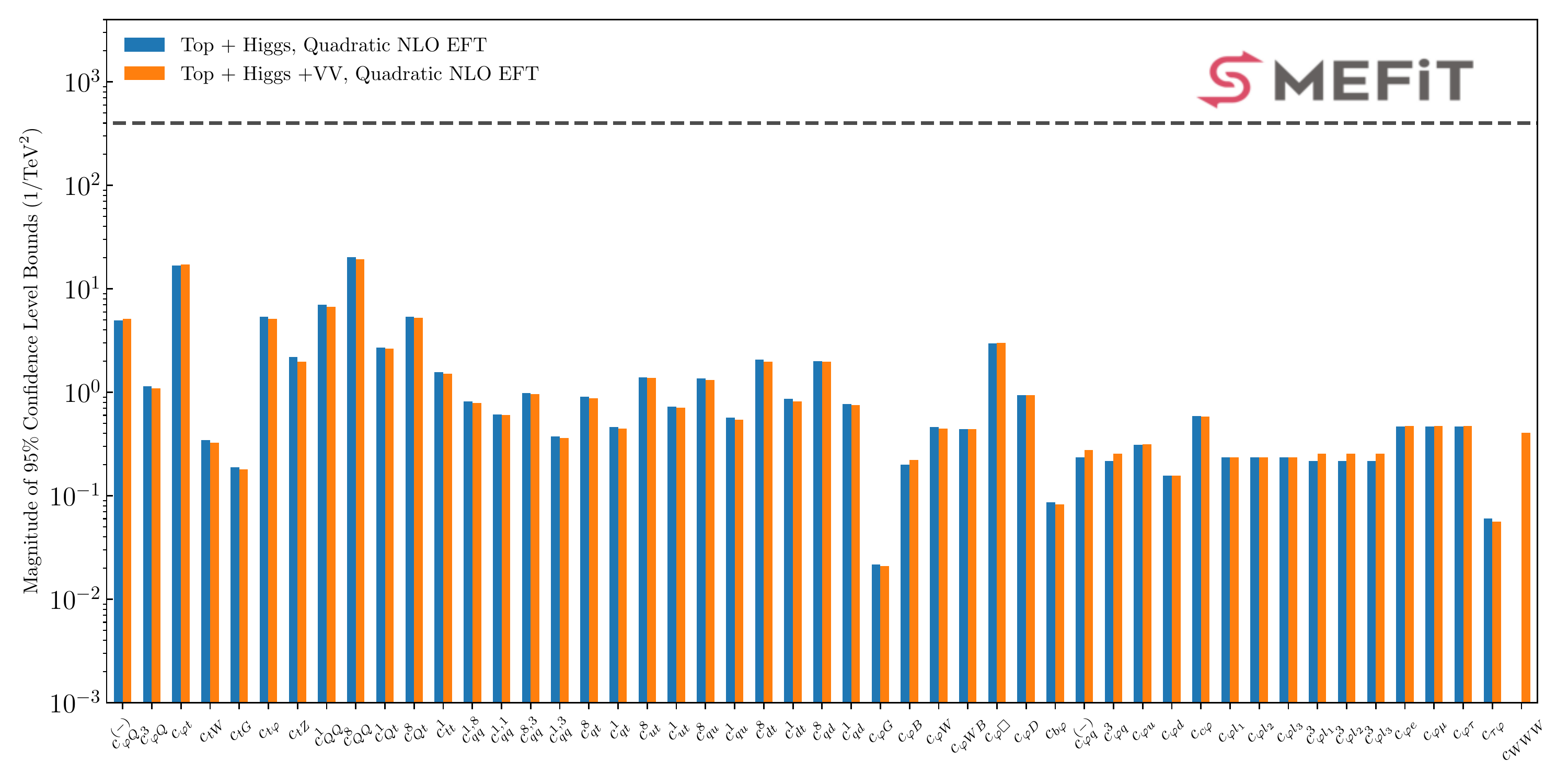}
     \caption{\label{fig:global_vs_toponly} Comparison of the magnitude of the 95\% CL bounds
       on the Wilson coefficients considered in the SMEFiT~\cite{Ethier:2021bye} analysis obtained in the global fit
     with those of fits to restricted datasets: a top-only, a Higgs-only, and a no-diboson fit.}
  \end{center}
\end{figure}

Finally, we should mention that ultimately the cleanest method to quantify the impact
of a specific dataset or group of processes is to repeat the fit by removing them
and studying what are the differences at the level of EFT coefficients.\footnote{One can
also deploy the methods of Bayesian reweighting~\cite{vanBeek:2019evb} to achieve the same goal.}
In this Section, we illustrate the method with results of the SMEFiT global fit reported in Ref.~\cite{Ethier:2021bye} 
without producing new results or recommending any EFT analysis. 
For example, Fig.~\ref{fig:global_vs_toponly} shows a 
comparison of the size of the 95\% CL bounds
on the Wilson coefficients considered in this analysis obtained in the global fit
with those of fits to restricted datasets: a top-only, a Higgs-only, and a no-diboson fit.
This way one can identify which coefficients are most sensitive to which datasets
or groups of processes.
Some observations that can be derived from these plots are that diboson data
provides the unique handle on $c_{WWW}$; that Higgs processes provide some sensitivity
on the top electroweak couplings $c_{tW}$ and $c_{tZ}$ but much less than the top data itself;
and that most of the purely bosonic operators are completely unconstrained unless
Higgs measurements are included in the analysis.

\clearpage

\section{Summary}
\label{sect:summary}

This note serves as a guide to experimental measurements and observables leading to EFT fits, but does not establish 
authoritative guidelines how those measurements should be performed. There is a spectrum of approaches to 
experimental measurements and observables, each approach has its own stronger and weaker sides, 
and none of the approaches has been established as the universally best approach to perform the EFT measurements. 
Analysis of LHC data in the framework of global EFT approach is a relative young field at this point. It has evolved 
from the less demanding task of interpreting of LHC data in the framework of standard model and of testing for
its deviations in individual measurements. The field is still actively evolving, and the numbers of parameters and
channels to be analyzed in an optimal and unbiased way are enormous. This poses both technical and 
conceptual challenges in performing the measurements, which will be resolved only gradually in close
collaboration of experimental and theoretical particle physicists. 
One of the goals of ongoing effort is bridging the gap between theory and experimental communities working on EFT, 
and in particular concerning optimised analyses that so far are only accessible within the experiment.

The types of observables range from the usual quantities calculated for the SM measurements, 
to the EFT-sensitive observables, and to the optimized observables utilizing advanced matrix-element 
and machine-learning techniques. Good sensitivity to EFT effects is desired when constructing an observable. 
The single-step measurement approach allows extraction of EFT parameters from the data directly without intermediate results. 
This approach is conceptually straightforward, unbiased, can be an optimal approach, but suffers from complexity 
and the lack of re-interpretation capability. The two-step measurement approach allows wide dissemination and 
preservation of the data in the form of differential truth-level distribution (first step) for later re-interpretation
(second step), but potentially this may lead to a non-optimal and/or biased analysis if care is not taken to 
address these concerns. Reporting the measurement results may represent a challenge on its own, 
as reporting the full likelihood function would be the ideal outcome and would allow interpretation 
and combination with other results. However, treatment of parameters with weak constraints and 
of systematic uncertainties, which may be correlated with other measurements, may complicate this process. 
This note does not provide specific guidelines how these issues should be technically addressed, 
but points to their importance.

In the second part of the note, we presented the main aspects of the mapping between observables and operators 
in the context of global SMEFT interpretations, in the specific cases of two global fit efforts. 
The set of operators considered in different flavour assumptions has been presented as well as a discussion of the 
set of measurements which have been used to constrain these operators. The sensitivity of different observables 
to different operators in the flavour universal scenario is quantified by examining the linear dependencies of Higgs, 
diboson, and electroweak measurements obtained in the \verb|fitmaker| analysis. 
This allows the results of individual or marginalised constraints from global fits to be understood more intuitively. 
Similarly, a detailed analysis based on the Fisher information formalism can be used to explore the relative sensitivity 
of various observables to the EFT operators as presented in the SMEFiT analysis. 
In the results of a global fit, the question also naturally arises as to which measurement is responsible for setting 
the strongest constraints. In this note this decomposition is shown for individual constraints. 
It has been observed that in order to understand the relation between operators and measurements, 
a single tool is often not sufficient, and that the whole picture only comes together by comparing the output 
of complementary tools. For example, the Fisher information is powerful, but in some cases only a direct fit 
with some specific datasets omitted provides the necessary insight. 
Finally results of the global fit have been shown. 

\bigskip

\noindent
{\bf Acknowledgments}:
This work was done on behalf of the LHC EFT WG and 
we would like to thank members of the LHC EFT WG for stimulating discussion of experimental measurements
and observables which led to this document. 
%



\providecommand{\href}[2]{#2}\begingroup\raggedright\endgroup

\end{document}